\documentclass{article}
\usepackage{rotcapt}

\renewcommand{\arraystretch}{1.3}
\makeatletter
\newdimen\normalarrayskip              
\newdimen\minarrayskip                 
\normalarrayskip\baselineskip \minarrayskip\jot
\newif\ifold             \oldtrue            \def\new{\oldfalse}
\def\arraymode{\ifold\relax\else\displaystyle\fi} 
\def\eqnumphantom{\phantom{(\theequation)}}     
\def\@arrayskip{\ifold\baselineskip\z@\lineskip\z@
     \else
     \baselineskip\minarrayskip\lineskip2\minarrayskip\fi}
\def\@arrayclassz{\ifcase \@lastchclass \@acolampacol \or
\@ampacol \or \or \or \@addamp \or
   \@acolampacol \or \@firstampfalse \@acol \fi
\edef\@preamble{\@preamble
  \ifcase \@chnum
     \hfil$\relax\arraymode\@sharp$\hfil
     \or $\relax\arraymode\@sharp$\hfil
     \or \hfil$\relax\arraymode\@sharp$\fi}}
\def\@array[#1]#2{\setbox\@arstrutbox=\hbox{\vrule
     height\arraystretch \ht\strutbox
     depth\arraystretch \dp\strutbox
     width\z@}\@mkpream{#2}\edef\@preamble{\halign
\noexpand\@halignto
\bgroup \tabskip\z@ \@arstrut \@preamble \tabskip\z@ \cr}%
\let\@startpbox\@@startpbox \let\@endpbox\@@endpbox
  \if #1t\vtop \else \if#1b\vbox \else \vcenter \fi\fi
  \bgroup \let\par\relax
  \let\@sharp##\let\protect\relax
  \@arrayskip\@preamble}
\def\eqnarray{\stepcounter{equation}%
              \let\@currentlabel=\theequation
              \global\@eqnswtrue
              \global\@eqcnt\z@
              \tabskip\@centering
              \let\\=\@eqncr
 \halign to \displaywidth\bgroup
    \eqnumphantom\@eqnsel\hskip\@centering
    $\displaystyle \tabskip\z@ {##}$%
    \global\@eqcnt\@ne \hskip 2\arraycolsep
         $\displaystyle\arraymode{##}$\hfil
    \global\@eqcnt\tw@ \hskip 2\arraycolsep
         $\displaystyle\tabskip\z@{##}$\hfil
         \tabskip\@centering
    &{##}\tabskip\z@\cr}
\begingroup\ifx\undefined\newsymbol \else\def\input#1 {\endgroup}\fi

\catcode`\@=11
\def\marginnote#1{}

\newcount\hour
\newcount\minute
\newtoks\amorpm
\hour=\time\divide\hour by60 \minute=\time{\multiply\hour by60
\global\advance\minute by-\hour}
\edef\standardtime{{\ifnum\hour<12 \global\amorpm={am}%
        \else\global\amorpm={pm}\advance\hour by-12 \fi
        \ifnum\hour=0 \hour=12 \fi
        \number\hour:\ifnum\minute<10 0\fi\number\minute\the\amorpm}}
\edef\militarytime{\number\hour:\ifnum\minute<10 0\fi\number\minute}

\def\draftlabel#1{{\@bsphack\if@filesw {\let\thepage\relax
      \xdef\@gtempa{\write\@auxout{\string
          \newlabel{#1}{{\@currentlabel}{\thepage}}}}}\@gtempa \if@nobreak
    \ifvmode\nobreak\fi\fi\fi\@esphack} \gdef\@eqnlabel{#1}}
    \def\@eqnlabel{}
\def\@vacuum{}
\def\draftmarginnote#1{\marginpar{\raggedright\scriptsize\tt#1}}

\def\draft{
  \oddsidemargin -.5truein
  \def\@oddfoot{\footnotesize \sl preliminary draft \hfil
    \rm\thepage\hfil\sl\today\quad\militarytime}
  \let\@evenfoot\@oddfoot \overfullrule 3pt
    \let\label=\draftlabel
    \let\marginnote=\draftmarginnote
  \def\@eqnnum{(\theequation)\rlap{\kern\marginparsep\tt\@eqnlabel}%
    \global\let\@eqnlabel\@vacuum}

  }

\def\nn{\nonumber}
\def\beq{\begin{equation}}
\def\eeq{\end{equation}}

\def\ba{\beq\new\begin{array}{c}}
\def\ea{\end{array}\eeq}
\def\be{\ba}
\def\ee{\ea}

\newfont{\alef}{msbm10 at 12pt}
\newfont {\goth}{eufm10 at 11pt}
\def\mathbb#1{\hbox{{\alef #1}}}

\def\12{{\footnotesize \frac{1}{2}}}
\def\32{{\footnotesize \frac{3}{2}}}
\def\52{{\footnotesize \frac{5}{2}}}

\let\@@savethanks\thanks
\def\thanks#1{\gdef\thefootnote{\alph{footnote}}\@@savethanks{#1}}

\def\theequation{\arabic{section}.\arabic{equation}}

\title{{\bf Virasoro constraints for Kontsevich-Hurwitz
partition function} \vspace{.5cm}}
\author{{\bf A. Mironov}\footnote{E-mail: \ mironov@itep.ru; mironov@lpi.ru}
\date{ } \\
{\small {\it Lebedev Physics Institute}
and {\it ITEP, Moscow, Russia}}\\ \\
{\bf A. Morozov}\thanks{E-mail: \ morozov@itep.ru}
\date{ } \\ {\small {\it ITEP, Moscow, Russia}}
}

\begin{document}

\maketitle

\vspace{-8.cm}

\begin{center}
\hfill FIAN/TD-17/08\\
\hfill ITEP/TH-33/08\\
\end{center}

\vspace{6.0cm}

\begin{abstract}
\noindent In \cite{LaKa,Kaz} M.Kazarian and S.Lando found a 1-parametric
interpolation between Kontsevich and Hurwitz partition
functions, which entirely lies within the space of KP
$\tau$-functions.
In \cite{BM} V.Bouchard and M.Marino suggested that
this interpolation satisfies some deformed Virasoro constraints.
However, they described the constraints in a somewhat
sophisticated form of AMM-Eynard  equations \cite{AMM.I,AMM.IM,Ey,EO}
for the rather involved Lambert spectral curve.
Here we present the relevant family of Virasoro constraints
{\it explicitly}.
They differ from the conventional continuous Virasoro constraints
in the simplest possible way:
by a  constant shift $\frac{u^2}{24}$ of the $\hat L_{-1}$
operator, where $u$ is an interpolation parameter between
Kontsevich and Hurwitz models.
This trivial modification of the string equation
gives rise to the entire deformation which
is a conjugation
of the Virasoro constraints
$\hat L_m \rightarrow \hat U \hat L_m \hat U^{-1}$
and "twists" the partition function,
${\cal Z}_{KH} = \hat U Z_K$.
The conjugation $\hat U =
\exp \left\{\frac{u^2}{3}(\hat N_1-\hat L_1)+O(u^6)\right\}
= \exp\left\{\frac{u^2}{12}\left(
\sum_k T_k\partial/\partial T_{k+1} -\frac{g^2}{2}\,
\partial^2/\partial T_0^2\right)+O(u^6)\right\}$
is expressed through the previously unnoticed operators like
$\hat N_1 = \sum_k (k+1)^2 T_k \partial/\partial T_{k+1}$
which annihilate the quasiclassical (planar)
free energy $F_K^{(0)}$
of the Kontsevich model, but do not belong to the symmetry
group $GL(\infty)$ of the universal Grassmannian.
\end{abstract}

\bigskip

\def\thefootnote{\arabic{footnote}}

\newpage

\tableofcontents

\section{Introduction}

Modern quantum field theory, nicknamed string theory \cite{UFN2},
looks for a unified approach to seemingly different
problems in different branches of sciences.
In some areas it is already quite successful, for example,
in the field of enumerative geometry.
String theory usually formulates combinatorial problems
in terms of generating functions and then interpret them
as partition functions, i.e. as elements of certain
$D$-modules, satisfying some sets of differential equations
and usually possessing various ("dual") integral representations,
often matrix or even functional.
Therefore, these partition functions acquire a "hidden symmetry",
with respect to change of integration variables,
which manifests itself through
rich integrability properties, which are already revealed in many examples.
These examples begin from matrix models \cite{UFN3},
in particular, from the celebrated Kontsevich model
\cite{Ko,GKM}, and from that on spread in many different
directions. Hidden integrability is now found,
as was originally predicted, in a vast variety of problems,
both in physics and mathematics, and today it is accepted
as an important and universal phenomenon.
However, the underlying $D$-module structure, i.e. the
set of constraints imposed on partition functions,
is often ignored and not enough effort is given to
identify and investigate it in each concrete example,
what obscures the common Lie algebra origin of all these
seemingly different situations.

In this paper, we take as an example the currently popular
deformation of the Kontsevich model, used to describe the
Hurwitz numbers (characterizing combinatorics of certain
ramified coverings of a Riemann sphere)
and Hodge integrals over the moduli space of complex curves.
This Kontsevich-Hurwitz model is very interesting and a
number of spectacular results is already obtained about it.
In particular, the KP integrability of the model is already
established. We are not going to enter any details about
the model but one: our goal is to describe an
underlying deformation of the "continuous Virasoro constraints"
\cite{covirco,MMM}, which
control the original Kontsevich model and should undoubtedly
control its Kontsevich-Hurwitz generalization.
In \cite{BM} it was actually suggested that the constraints
remain Virasoro and bilinear, however, a somewhat
sophisticated machinery of the AMM-Eynard equations
\cite{AMM.I,AMM.IM,Ey,EO} was used to discuss the issue.
In our opinion, this formalism is very useful for a variety
of purposes, both conceptual and technical
(see \cite{AMM.I,AMM.IM} for explanation
of our views on this issue), but the Ward identities in
a given model should be better formulated in a more direct
and straightforward form, and we provide some evidence
that such a form is indeed available: see eq.(\ref{defvirco})
below.

\section{The main statement: eqs.(\ref{defvirco})-(\ref{N1annih})
}
\setcounter{equation}{0}

According to \cite{Kaz}, the
Kontsevich-Hodge free energy is the double expansion
\be
{\cal F}(T) = \sum_{q=0}^\infty u^{2q}F_q(T), \nn \\
F_q(T) = \sum_{p\geq q}^\infty g^{2p}F_q^{(p)}(T)
\ee
where each
\be
F_q^{(p)}(T) = (-)^q
\sum_{n=0}^\infty \frac{1}{n!}\sum_{k_1,\ldots,k_n=0}^\infty
\delta\left(\sum_{i=1}^n(k_i-1)
- (3p-3-q)\right) I^{(p)}_q(k_1,\ldots,k_n)\,
T_{k_1}\ldots T_{k_n}
\label{KHfe}
\ee
is a generating function for the Hodge integrals \cite{G1,Hodge}
\be
I^{(p)}_q(k_1,\ldots,k_n) =
\int_{{\cal M}_{p,n}}\lambda_q \psi_1^{k_1}\ldots \psi_n^{k_n}
\ee
This definition, together with the ELSV formula \cite{ELSV},
allowed M.Kazarian in \cite{Kaz}
to relate ${\cal F}(T)$ to the generating function
$H(p)$ of the Hurwitz numbers \cite{Hurwitz},
which has a simple alternative
representation \cite{G} in terms of discrete Virasoro and
$W$-operators \cite{virc,disVW}
\be
e^{H(p)} = e^{u^3\hat W_0}e^{p_1},
\label{Hp}
\ee
where
\be
\hat W_0 = \sum_{m=0}^\infty p_m \hat V_m
= \frac{1}{2}\sum_{i,j\geq 1}\left(
(i+j)p_ip_j\frac{\partial}{\partial p_{i+j}}
+ ijp_{i+j}\frac{\partial^2}{\partial p_i\partial p_j}\right)
\ee
and $\hat V_m$ are the discrete Virasoro operators ($p_k=kt_k$)
\be
\hat V_m = \sum_{k=0}^\infty
(k+m)p_k\frac{\partial}{\partial p_{k+m}}
+ \sum_{i+j=m} ij\frac{\partial^2}{\partial p_i\partial p_j}
\ee
This allows one to call ${\cal F}(T)$ the Kontsevich-Hodge-Hurwitz
or simply Kontsevich-Hurwitz free energy.
We return to Kazarian's construction in s.\ref{Ka},
and now switch to an alternative description.

Its starting point is the fact that $F_0(T)$,
the ordinary Kontsevich free energy,
satisfies the "continuous Virasoro" constraints
\be
\hat L_m Z_0 = 0 \ \ \ \ {\rm for}\ m\geq -1,
 \nn \\
\hat L_m = -\frac{\partial}{\partial \tau_{m+1}}
+ \sum_{k=\delta_{m,-1}}^\infty \left(k+\frac{1}{2}\right)\tau_{k}
\frac{\partial}{\partial \tau_{k+m}} +
\frac{g^2}{8} \sum_{k=0}^{m-1}
\frac{\partial^2}{\partial \tau_{k}\partial \tau_{m-1-k}}
+ \frac{\tau_0^2}{2g^2}\,\delta_{m,-1}
+ \frac{1}{16}\,\delta_{m,0},
\nn \\
Z_0 = \exp\left(\frac{1}{g^2}F_0(T)\right),
\ \ \ \ \ \ \ \ \ \ \ \ \ \
T_k = \frac{(2k+1)!!}{2^k}\tau_{k} =
\frac{\Gamma(k+\frac{3}{2})}{\Gamma(\frac{3}{2})}\tau_{k}
\ee
It is well known that, while the Virasoro constraints look
more elegant in terms of $\tau$-variables, their
solutions such as $Z$ and $F$, are seriously simplified if
expressed through $T$-variables. Sometimes one denotes
$\tau_k = \check\tau_{2k+1}$ to emphasize that $\{\tau_k\}$
are only half of all the time-variables in the Generalized
Kontsevich Model (GKM) \cite{GKM}: this is also reflected
in the fact that $Z_0$ is a KdV $\tau$-function \cite{GKM},
while the generic $Z_{GKM}$ belongs to the KP family.
According to \cite{Kaz}, this is also true for ${\cal Z}$:
in appropriate variables
(called $\check q_k$ in s.\ref{Ka} below)
${\cal Z}$ is a KP $\tau$-function, and reduces to the KdV one only
for $u=0$.
As usual, we often denote the first two terms
(the linear-in-derivatives piece) in the Virasoro
operator $\hat L_m$ through $\hat l_m$.
The deformation of the Kontsevich model is described by a parameter
$u$ so that the Kontsevich model corresponds to $u=0$, while
the Hurwitz partition function \cite{Hurwitz},
analyzed in some detail in \cite{BM}, corresponds to $u=1$.

Our claim, parallel to a rather implicit
suggestion of \cite{BM},
is that the full Kontsevich-Hurwitz partition function,
\be
{\cal Z}=\prod_{q=0}^\infty Z_q =
\exp\left(\frac{1}{g^2}\sum_{q}^\infty F_q\right)
= \exp \left(\frac{1}{g^2}\sum_{p\geq q\geq 0}^\infty
u^{2q} g^{2p}F_q^{(p)}(T)\right) =
\exp\left(\frac{1}{g^2}{\cal F}(T)\right)
\ee
satisfies the deformed continuous Virasoro constraints.
We claim that the relevant deformation is actually
a {\it conjugation} \cite{AMM.IM}
\be
\hat{\cal L}_m {\cal Z}= 0\ \ \ \ {\rm for}\ m\geq -1, \nn\\
\hat{\cal L}_m = \hat U \hat L_m \hat U^{-1} \label{defvirco} \ee
which obviously preserves structure of the Virasoro (sub)algebra. It
follows that \be {\cal Z} = \hat U Z_0 \label{defpartfun} \ee is
obtained by a simple twisting of the Kontsevich partition function
(note also that, according to \cite{Kaz}, ${\cal Z}$ in proper
variables is a KP $\tau$-function at any given value of $u$, for
$Z_0$ this was originally proved in \cite{GKM} and for ${\cal Z}$
this follows from (\ref{Hp}) and the theory of equivalent
hierarchies \cite{Shiota,equivhi,Kh}). The operator $\hat U$ is
explicitly given by\footnote{The full operator $\hat U$ with higher
order terms included is, in fact, of the form
\be
\hat U=\exp\left(\sum_k s_ku^{4k+2}\hat M_{2k+1}\right),\ \ \ \ \ \ \
\hat M_{2k+1}\equiv\sum_l \hat T_l{\partial\over\partial T_{l+k}}-{1\over 2}
\sum_{a+b=2k}(-)^a{\partial^2\over\partial T_a\partial T_b}
\ee
where the coefficients $s_k={B_{k+1}\over k(k+1)}$ are expressed through the Bernoulli
numbers, $\sum_{k=1}^\infty {B_{2k}x^{2k}\over (2k)!}\equiv {x\over 1-e^{-x}}
-1-{x\over 2}$. This formula is found in \cite{Giv,Kpc}. We are indebted to M.Kazarian
for pointing out reference \cite{Giv} which, in its turn, is based on
refs.\cite{Mumford,FP} and for providing us with the text of his unpublished paper
\cite{Kpc}.}
\be \hat U = \exp\left\{-\frac{u^2}{3}\Big(\hat L_1 - \hat N_1
\Big)+O(u^6)\right\} = \exp \left\{\frac{u^2}{12}\left( \sum_k
\tilde T_k\frac{\partial}{\partial T_{k+1}} -
\frac{g^2}{2}\frac{\partial^2}{\partial T_0^2}\right)+O(u^6)\right\}
\label{Uop} \ee where $\tilde T_k = T_k - \delta_{k,1}$ are
''shifted times'' and \be \hat N_1 = \sum_{k=0}^\infty (k+1)^2
\tilde T_k\frac{\partial} {\partial T_{k+1}} \ee is a new (for
Kontsevich-model theory) operator, which annihilates the genus-zero
free energy, \be \hat N_1 F^{(0)}_0 = 0 \label{N1annih} \ee and
gives rise to an infinite family of such annihilators. Commutation
relations between $\hat L_m$ and $\hat N_1$ imply that the lowest
deformed Virasoro constraints act on ${\cal Z}$ as \be \hat{\cal
L}_{-1} = \hat L_{-1} - \frac{u^2}{24} \label{calL-1} \ee and \be
\hat{\cal L}_0 = \hat L_0 + u^2\frac{\partial}{\partial u^2}
\label{calL0} \ee

In what follows we recursively define coefficients
in the free energy expansion from our suggested Virasoro
constraints: see eqs.(\ref{F0})-(\ref{F4}) below.
In particular, we reproduce in this way all the terms,
explicitly calculated in \cite{BM}. Note that non-trivial is
already the property that conjugation operator
(\ref{Uop}) generates only contributions with $p\geq q$
to the logarithm of ${\cal Z}$.

After that we perform the transformation \cite{Kaz}
from $T$ to $p$ variables,
\be
T_k = u^{2k+1}\sum_{n=1}^\infty \frac{n^{n+k}}{n!}u^{3n} p_n
\label{Tp}
\ee
and demonstrate that the answer coincides with $H(p)$
in (\ref{Hp}) modulo necessary subtractions of certain
$p$-linear and $p$-quadratic terms (the rooted and double-rooted
tree contributions to $H(p)$):
\be\label{H2}
{\cal F}\big(T(p)\big)
= H(p) - H_{01}(p) - H_{02}(p) =
H(p) - \sum_{n\geq 1} \frac{n^{n-2}}{n!}p_n u^{3(n-1)}
- \frac{1}{2}\sum_{m,n\geq 1}\frac{m^mn^n}{(m+n)m!n!}
p_mp_nu^{3(m+n)}
\ee
Note that (\ref{Hp}) and (\ref{Tp}) involve {\it odd}
powers of $u$, i.e. semi-integer powers of the deformation
parameter $u^2$, but all odd powers drop away from the
answer. Finally, we explain the relation between
the AMM-Eynard equations of \cite{BM} and our conjugation
of the continuous Virasoro algebra.

\section{Solving Virasoro constraints}
\setcounter{equation}{0}

This is a standard procedure, so we do not go into
too many details.

First, the deformed
$\hat {\cal L}_0$ constraint (\ref{calL0})
implies that the $T_1$ dependence
can be defined exactly:
from
\be
(\hat l_0 + q)F_q = -\frac{1}{16}\delta_{q,0}
\label{L0co}
\ee
it follows that
\be
F_q^{(p)} = \sum_{m=0}^\infty \sum_{k_1,\ldots,k_m=0}^\infty
\delta\left(\sum_{i=1}^m(k_i-1) - (3p-3-q)\right)
I_q^{(p)}[k_1,\ldots,k_m]\frac{T_{k_1}\ldots T_{k_m}}
{(1-T_1)^{2p-2+m}}
\label{fpq}
\ee
The r.h.s. in (\ref{L0co}) is taken into account by a
peculiar contribution to $F_0^{(1)}$, which deviates from
(\ref{fpq}): see the first term in eq.(\ref{F1}) below.

Second, the string equation
(i.e. the $\hat {\cal L}_{-1}$ constraint)
is satisfied separately by every constituent free energy
$F^{(p)}_q$:
\be
\hat l_{-1} {F}^{(p)}_q =
\left((T_1-1)\frac{\partial}{\partial T_0} +
\sum_{k=1}^\infty T_{k+1}\frac{\partial}{\partial T_k}\right)
{F}^{(p)}_q = \beta^{(p)}_q(T)
\label{seq}
\ee
Moreover, together with $p,q$-dependent selection rules in
(\ref{fpq}) this equation defines the series $F^{(p)}_q(T)$
up to a finite number of free coefficients
(we call them $\gamma$-parameters), one per each
$T_0$-independent monomial allowed by selection rules.

The role of all remaining Virasoro constraints is just to
fix these remaining undefined coefficients in front of
different solutions of (\ref{seq}).

\subsection{Solving string equation}

\subsubsection{$F^{(0)}_0$ -- the genus-zero component of
Kontsevich free energy \label{K0}}

Since $\beta^{(0)}_0 = -\frac{1}{2}T_0^2$, we have,
after explicit resolution of the selection rules,
\be
F^{(0)}_0 = \frac{1}{6}\cdot\frac{T_0^3}{1-T_1}
+ \sum_{s=1}^\infty\left(
\sum_{k_1,\ldots,k_s\geq 2}^\infty J^{(0)}_0[k_1,\ldots,k_s]\,
\frac{ T_0^{3+\sum_{i=1}^s (k_i-1)}T_{k_1}\ldots T_{k_s} }
{(1-T_1)^{1+\sum_{i=1}^s k_i}}\right) = \nn \\
= \frac{1}{6}\cdot\frac{T_0^3}{1-T_1} +
\sum_{k\geq 2} J[k]\,\frac{T_0^{k+2}T_k}{(1-T_1)^{k+1}}
+ \sum_{k_1,k_2\geq 2}J[k_1,k_2]\,
\frac{T_0^{k_1+k_2+1}T_{k_1}T_{k_2}}
{(1-T_1)^{k_1+k_2+1}} + \nn \\ +
\sum_{k_1,k_2,k_3\geq 2}J[k_1,k_2,k_3]\,
\frac{T_0^{k_1+k_2+k_3}T_{k_1}T_{k_2}T_{k_3}}
{(1-T_1)^{k_1+k_2+k_3+1}} + \ldots
\label{fK0}
\ee
(for simplicity we omitted the labels $p,q=0,0$ in the
last two lines).
Substitution into (\ref{seq}) gives
$$
0\ \  =\ \ \frac{1}{6}\,\frac{T_0^3T_2}{(1-T_1)^2}
- \sum_{k\geq 2}(k+2)J[k]\,\frac{T_0^{k+1}T_k}{(1-T_1)^k}
+ \sum_{k\geq 2}J[k]\,\frac{T_0^{k+2}T_{k+1}}{(1-T_1)^{k+1}}\ +
$$
{\footnotesize
$$
+ \sum_{k\geq 2}(k+1)J[k]\,\frac{T_0^{k+2}T_2T_k}{(1-T_1)^{k+2}}\
- \sum_{k_1,k_2\geq 2}(k_1+k_2+1)J[k_1,k_2]\,
\frac{T_0^{k_1+k_2}T_{k_1}T_{k_2}}{(1-T_1)^{k_1+k_2}}\ +
\sum_{k_1,k_2\geq 2}J[k_1,k_2]\,
\frac{T_0^{k_1+k_2+1}(T_{k_1+1}T_{k_2}+T_{k_1}T_{k_2+1})}
{(1-T_1)^{k_1+k_2+1}}\ +
$$}
{\footnotesize
$$
+ \sum_{k_1,k_2\geq 2}(k_1+k_2+1)J[k_1,k_2]\,
\frac{T_0^{k_1+k_2}T_2T_{k_1}T_{k_2}}{(1-T_1)^{k_1+k_2+2}}\ +
\sum_{k_1,k_2,k_3\geq 2}J[k_1,k_2,k_3]
\left(-(k_1+k_2+k_3)\frac{T_0^{k_1+k_2+k_3-1}T_{k_1}T_{k_2}T_{k_3}}
{(1-T_1)^{k_1+k_2+k_3}}\ +
\right.
$$ $$ \ \ \ \ \ \ \ \ \ \ \ \ \ \ \ \ \ \ \ \ \ \ \ \ \ \ \
\left. +  \frac{T_0^{k_1+k_2+k_3}
(T_{k_1+1}T_{k_2}T_{k_3} +T_{k_1}T_{k_2+1}T_{k_3}
+ T_{k_1}T_{k_2}T_{k_3+1})}{(1-T_1)^{k_1+k_2+k_3+1}}
\right) \ \ \ + \ldots
$$}
The structure of recurrent relations is obvious from these formulas.
The first line implies that $J[2]=\frac{1}{24}$ and
\be
J[k] = \frac{1}{k+2}J[k-1]
= \frac{1}{(k+2)!}\ \ \ {\rm for}
\ \ \ k\geq 2
\ee
The second line implies that
\be
J[2,2] = \frac{3}{5}J[2] = \frac{1}{40}, \ \ \ \
J[2,k] = \frac{1}{2(k+3)}\Big((k+1)J[k] + 2J[2,k-1]\Big)
= \frac{(k-1)(k+4)}{2(k+3)!}
\ \ \ {\rm for} \ \ \ k\geq 3,\nn \\
J[k_1,k_2] = \frac{1}{k_1+k_2+1}\Big(J[k_1-1,k_2]+
J[k_1,k_2-1]\Big)\ \ \ {\rm for} \ \ \ k_1,k_2\geq 3
\ee
and so on.

\subsubsection{Generic $F^{(p)}_q$}

Now we are ready to proceed to the case of generic $q$ and $p$.
Generic version of (\ref{fK0}) is
\be
F^{(p)}_q =  \sum_{s=1}^\infty\left(
\sum_{\stackrel{k_1,\ldots,k_s\geq 2}{\sum_{i=1}^s (k_i-1)\geq 3p-3-q}
}^\infty J^{(p)}_q[k_1,\ldots,k_s]\,
\frac{ T_0^{q+3-3p\,+\sum_{i=1}^s (k_i-1)}T_{k_1}\ldots T_{k_s} }
{(1-T_1)^{q+1-p+\sum_{i=1}^s k_i}}\right) =
\label{Fpq}
\ee
$$
= \!\!\!\!\!\! \sum_{k\geq 3p-2-q} \!\!\!\!\!\!
J[k]\,\frac{T_0^{k+2+q-3p\,}T_k}{(1-T_1)^{k+1+q-p}}\
+ \!\!\!\!\!\!\!\!\!\!\!\!
\sum_{\stackrel{k_1,k_2\geq 2}{k_1+k_2\geq 3p-1-q}}
\!\!\!\!\!\!\!\!\!\!\!\! J[k_1,k_2]\,
\frac{T_0^{k_1+k_2+1+q-3p\,}T_{k_1}T_{k_2}}
{(1-T_1)^{k_1+k_2+1+q-p}}\ +
\!\!\!\!\!\!\!\!\!\!\!\!
\sum_{\stackrel{k_1,k_2,k_3\geq 2}{k_1+k_2+k_3\geq 3p-q}}
\!\!\!\!\!\!\!\!\!\!\!\!J[k_1,k_2,k_3]\,
\frac{T_0^{k_1+k_2+k_3+q-3p\,}T_{k_1}T_{k_2}T_{k_3}}
{(1-T_1)^{k_1+k_2+k_3+1+q-p}} + \ldots
$$
Substitution into the string equation (\ref{seq}) gives
$$
0\ \ = \ \
- \sum_{k\geq 3p-2-q}(k+2+q-3p)J[k]\,\frac{T_0^{k+1+q-3p\,}T_k}
{(1-T_1)^{k+q-p}}
+ \!\!\!\!\!\!\sum_{k\geq 3p-2-q}\!\!\!\!\!\!
J[k]\,\frac{T_0^{k+2+q-3p\,}T_{k+2+q-3p\,}}
{(1-T_1)^{k+1+q-p}}\ +
$$
{\footnotesize
$$
+ \!\!\!\!\!\!\sum_{k\geq 3p-2-q}\!\!\!\!\!\!
(k+1+q-p)J[k]\,\frac{T_0^{k+2+q-3p\,}T_2T_k}{(1-T_1)^{k+2+q-p}}\
- \!\!\!\!\!\!\!\!\!\!\!\!
\sum_{\stackrel{k_1,k_2\geq 2}{k_1+k_2\geq 3p-1-q}}
\!\!\!\!\!\!\!\!\!\!\!\!(k_1+k_2+1+q-3p)J[k_1,k_2]\,
\frac{T_0^{k_1+k_2+q-3p\,}T_{k_1}T_{k_2}}{(1-T_1)^{k_1+k_2+q-p}}\ +
\ \ \ \ \ \ \ \ \ \ \ \ \ \ \ \ \ \ \ \
$$ $$ \ \ \ \ \ \ \ \ \ \ \ \ \ \ \ \ \ \ \ \ \ \ \ \ \ \ \ \
 \ \ \ \ \ \ \ \ \ \ \ \ \ \ \ \ \ \ \ \ \ \ \ \ \ \ \ \
  \ \ \ \ \ \ \ \ \ \ \ \ \ \ \ \ \ \ \ \ \ \ \ \ \ \ \ \ +
\!\!\!\!\!\!\!\!\!\!\!\!
\sum_{\stackrel{k_1,k_2\geq 2}{k_1+k_2\geq 3p-1-q}}
\!\!\!\!\!\!\!\!\!\!\!\! J[k_1,k_2]\,
\frac{T_0^{k_1+k_2+1+q-3p\,}(T_{k_1+1}T_{k_2}+T_{k_1}T_{k_2+1})}
{(1-T_1)^{k_1+k_2+1+q-p}}\ +
$$}
{\footnotesize
$$
\!\!\!\!\!\!\!\!\!\!\! + \!\!\!\!\!\!\!\!\!\!\!\!\!
\sum_{\stackrel{k_1,k_2\geq 2}{k_1+k_2\geq 3p-1-q}}
\!\!\!\!\!\!\!\!\!\!\!\!(k_1+k_2+1+q-p)J[k_1,k_2]\,
\frac{T_0^{k_1+k_2+1+q-3p\,}T_2T_{k_1}T_{k_2}}{(1-T_1)^{k_1+k_2+2+q-p}}\ +
\!\!\!\!\!\!\!\!\!\!\!\!
\sum_{\stackrel{k_1,k_2,k_3\geq 2}{k_1+k_2+k_3\geq 3p-q}}
\!\!\!\!\!\!\!\!\!\!\!\!J[k_1,k_2,k_3]
\left(-(k_1+k_2+k_3+q-3p)\frac{T_0^{k_1+k_2+k_3-1+q-3p\,}
T_{k_1}T_{k_2}T_{k_3}}
{(1-T_1)^{k_1+k_2+k_3+q-p}}\ +
\right.
$$ $$ \ \ \ \ \ \ \ \ \ \ \ \ \ \ \ \ \ \ \ \ \ \ \ \ \ \ \
\left. +  \frac{T_0^{k_1+k_2+k_3+q-3p\,}
(T_{k_1+1}T_{k_2}T_{k_3} +T_{k_1}T_{k_2+1}T_{k_3}
+ T_{k_1}T_{k_2}T_{k_3+1})}{(1-T_1)^{k_1+k_2+k_3+1+q-p}}
\right) \ \ \  + \ldots \
$$}
This time the first line implies that
\be
J^{(p)}_q[k] = \frac{1}{k+2+q-3p}J^{(p)}_q[k-1]
= \frac{1}{(k+2+q-3p)!}J^{(p)}_q[3p-2-q]
\ \ \ {\rm for}
\ \ \ k\geq {\rm max}(3p-2-q,2)
\ee
Since $p\geq q$, one has $3p-2-q\geq 2q-2$ and
separate consideration is needed only for a few cases:
$(q,p) = (0,0),\ (0,1)$
and $(1,1)$, these are the only cases when terms
which depend only on $T_0$ and $T_1$ can arise.

The second line implies that
\be
J^{(p)}_q[2,2] = \frac{3+q-p}{(5+q-3p)!}J^{(p)}_q[2]
\ \ \ \ {\rm for} \ \ \ 3p-5-q\leq 0
\ee
Such terms exist only in five cases:
$(q,p) = (0,0),\ (0,1),\ (1,1),\ (1,2)$ and $(2,2)$,
moreover, in the case $(1,2)$ the $T_2^2$ term is independent
of $T_0$ and  enters with an independent coefficient,
which is not fixed by the string equation (only by the higher
Virasoro constraints): there is no $J[2]$ in this case
to constrain $J[2,2]$.

Except for a few exceptional cases, the coefficients $J[2,k]$
are expressed through two independent (at the level of the string
equation) parameters:
$$
J^{(p)}_q[2,k] = \frac{k+1+q-p}{2(k+3+q-3p)}J^{(p)}_q[k] +
\frac{1}{k+3+q-3p}J^{(p)}_q[2,k-1]=
$$ $$
= \frac{(k+1+q-p) + (k+q-p) + \ldots + (k+2+q-p-j)}
{2(k+3+q-3p)!} + \frac{(k+3+q-3p-j)!}{(k+3+q-3p)!}J[2,k-j] =
$$ $$
=\frac{k+q+p}{4(k+2+q-2p)!}J^{(p)}_q[3p-2-q] +
\frac{1}{(k+3+q-3p)!}J^{(p)}_q[2,3p-3-q]
$$
As usual,
\be
J^{(p)}_q[k_1,k_2] = \frac{1}{k_1+k_2+1+q-3p}
\Big(J^{(p)}_q[k_1-1,k_2]+
J^{(p)}_q[k_1,k_2-1]\Big) \ \ \ {\rm for} \ \ \
k_1,k_2\geq {\rm max}(3p-3-q,3)
\ee
and so on.

\subsection{The first terms of $F$ expansion \label{sum}}

Putting different pieces together, one obtains
\be
F_0 = \underbrace{\left\{
\frac{1}{6}\cdot\frac{T_0^3}{1-T_1} + \sum_{k\geq 2}
\frac{1}{(k+2)!}\frac{T_0^{k+2}T_k}{(1-T_1)^{k+1}} +
\frac{1}{40}\cdot\frac{T_0^5T_2^2}{(1-T_1)^5} +
\sum_{k\geq 3}\frac{(k+1)(k+2)}{2(k+3)!}\cdot
\frac{T_0^{k+3}T_2T_k}{(1-T_1)^{k+3}}
+\ldots\right\} }_{{\rm genus}\ 0} +\nn \\
+ \frac{1}{24}\underbrace{\left\{
\Big(1-\log(1-T_1)\Big) + \sum_{k\geq 2}
\frac{1}{(k-1)!}\frac{T_0^{k-1}T_k}{(1-T_1)^{k}} +
\frac{T_0^2T_2^2}{(1-T_1)^4} +
\sum_{k\geq 3}\frac{k^2+k+2}{2k!}
\frac{T_0^{k}T_2T_k}{(1-T_1)^{k+2}}
+\ldots\right\} }_{{\rm genus}\ 1 } + \nn
\ee
{\footnotesize
\centerline{$
+\underbrace{\left(\gamma_{01}^{(2)}= \frac{1}{9\cdot 128}\right)
\left\{
\sum_{k\geq 4}\frac{1}{(k-4)!}
\left(\frac{T_0^{k-4}T_k}{(1-T_1)^{k-1}}+
\frac{k+2}{2}\frac{T_0^{k-3}T_2T_k}{(1-T_1)^{k+1}}\right)+
\ldots\right\} +
\left(\gamma_{02}^{(2)} = \frac{29}{45 \cdot128}\right) \left\{
\sum_{k\geq 3}\frac{1}{(k-3)!}\frac{T_0^{k-3}T_2T_k}{(1-T_1)^{k+1}}
+\ldots\right\} }_{{\rm genus}\ 2} +
$}
$$ \ \ \ \ \ \ \ \ \ \ \ \ \ \ \ \ \ \ \ \ \ \ \
\ \ \ \ \ \ \ \ \ \ \ \ \ \ \ \ \ \ \ \ \ \ \
\ \ \ \ \ \ \ \ \ \ \ \ \ \ \ \ \ \ \ \ \ \ \
\ \ \ \ \ \ \ \ \ \ \ \ \ \ \ \ \ \ \ \ \ \ \
\ \ \ \ \ \ \ \ \ \ \ \ \ \ \ \ \ \ \ \ \ \ \
+\underbrace{\left(\gamma_{03}^{(2)}
=\frac{7}{3\cdot 128}\right)
\left\{\frac{T_2^3}{(1-T_1)^5} + \ldots\right\}
}_{{\rm genus}\ 2} +
$$
$$
+\underbrace{\left( \gamma_{01}^{(3)} =
\frac{1}{9^2\cdot 1024}\right)
\left\{
\sum_{k\geq 7}\frac{1}{(k-7)!}
\left(\frac{T_0^{k-7}T_k}{(1-T_1)^{k-2}}+
\frac{k+3}{2}\cdot\frac{T_0^{k-6}T_2T_k}{(1-T_1)^{k}}\right) +
\ldots\right\} + \gamma_{02}^{(3)}\left\{
\sum_{k\geq 6}\frac{1}{(k-6)!}\frac{T_0^{k-6}T_2T_k}{(1-T_1)^{k}}
+\ldots\right\} }_{{\rm genus}\ 3} +
$$ $$
+\underbrace{\gamma_{01}^{(4)}\left\{
\sum_{k\geq 10}\frac{1}{(k-10)!}
\left(\frac{T_0^{k-10}T_k}{(1-T_1)^{k-3}}+
\frac{k+4}{2}\cdot \frac{T_0^{k-9}T_2T_k}{(1-T_1)^{k-1}}\right) +
\ldots\right\} + \gamma_{02}^{(4)}\left\{
\sum_{k\geq 9}\frac{1}{(k-9)!}\frac{T_0^{k-9}T_2T_k}{(1-T_1)^{k-1}}
+\ldots\right\} }_{{\rm genus}\ 4}  +
$$}
\be
+\underbrace{\ldots}_{{\rm higher\ genera}}
\label{F0}
\ee
\be
F_1 = \left(\gamma_1^{(1)}=-\frac{1}{24}\right)\underbrace{\left\{
\frac{T_0}{1-T_1} +
\sum_{k\geq 2}\frac{1}{k!}\frac{T_0^{k}T_k}{(1-T_1)^{k+1}}+
\frac{1}{2}\cdot\frac{T_0^3T_2^2}{(1-T_1)^5} +
\sum_{k\geq 3}\frac{k+2}{2k!}\frac{T_0^{k+1}T_2T_k}{(1-T_1)^{k+3}}
+\ldots\right\} }_{{\rm genus}\ 1} + \nn
\ee
{\footnotesize
\centerline{$
+\underbrace{\left(\gamma_{11}^{(2)} = -\frac{1}{15\cdot 32}\right)
\left\{
\sum_{k\geq 3}\frac{1}{(k-3)!}\left(\frac{T_0^{k-3}T_k}{(1-T_1)^{k}}
+ \frac{k+3}{2}\cdot\frac{T_0^{k-2}T_2T_k}{(1-T_1)^{k+2}}\right)
+\ldots\right\} +
\left(\gamma_{12}^{(2)} = -\frac{5}{9\cdot 128}\right)\left\{
\sum_{k\geq 2}\frac{1}{(k-2)!}\frac{T_0^{k-2}T_2T_k}{(1-T_1)^{k+2}}
+\ldots\right\} }_{{\rm genus}\ 2} +
$} \centerline{$
+\underbrace{
\left(\gamma_{11}^{(3)}= -\frac{7}{5\cdot 27 \cdot 1024}\right)
\left\{
\sum_{k\geq 6}\frac{1}{(k-6)!}
\left(\frac{T_0^{k-6}T_k}{(1-T_1)^{k-1}}+
\frac{k+4}{2}\cdot\frac{T_0^{k-5}T_2T_k}{(1-T_1)^{k+1}}\right)
+\ldots\right\} + \gamma_{12}^{(3)}\left\{
\sum_{k\geq 5}\frac{1}{(k-5)!}
\frac{T_0^{k-5}T_2T_k}{(1-T_1)^{k+1}}
+\ldots\right\} }_{{\rm genus}\ 3} +
$} $$
+\underbrace{\gamma_{11}^{(4)}\left\{
\sum_{k\geq 9}\frac{1}{(k-9)!}
\left(\frac{T_0^{k-9}T_k}{(1-T_1)^{k-2}}+
\frac{k+5}{2}\cdot\frac{T_0^{k-8}T_2T_k}{(1-T_1)^{k}}\right)
+\ldots\right\} + \gamma_{12}^{(4)}\left\{
\sum_{k\geq 8}\frac{1}{(k-8)!}
\frac{T_0^{k-8}T_2T_k}{(1-T_1)^{k}}
+\ldots\right\} }_{{\rm genus}\ 4} +
$$}
\be
+\underbrace{\ldots}_{{\rm higher\ genera}}
\label{F1}
\ee
\be\label{F2}
F_2 = \left(\gamma_2^{(2)} = \frac{7}{45\cdot 128}\right)
\underbrace{\left\{
\sum_{k\geq 2}\frac{1}{(k-2)!}\frac{T_0^{k-2}T_k}{(1-T_1)^{k+1}}+
3\frac{T_0T_2^2}{(1-T_1)^5} +
\sum_{k\geq 3}\frac{(k+1)(k+2)}{2(k-1)!}
\frac{T_0^{k-1}T_2T_k}{(1-T_1)^{k+3}}
+\ldots\right\} }_{{\rm genus}\ 2} + \nn
\ee
{\footnotesize
$$
+\underbrace{
\left(\gamma_{21}^{(3)} = \frac{41}{7\cdot 81\cdot 1024}\right)
\left\{
\sum_{k\geq 5}\frac{1}{(k-5)!}
\left(\frac{T_0^{k-5}T_k}{(1-T_1)^{k}}+
\frac{k+5}{2}\cdot \frac{T_0^{k-4}T_2T_k}{(1-T_1)^{k+2}}\right)
+\ldots\right\} + \gamma_{22}^{(3)}\left\{
\sum_{k\geq 4}\frac{1}{(k-4)!}
\frac{T_0^{k-4}T_2T_k}{(1-T_1)^{k+2}}
+\ldots\right\} }_{{\rm genus}\ 3} +
$$ $$
+\underbrace{\gamma_{21}^{(4)}\left\{
\sum_{k\geq 8}\frac{1}{(k-8)!}
\left(\frac{T_0^{k-8}T_k}{(1-T_1)^{k-1}}+
\frac{k+6}{2}\cdot \frac{T_0^{k-7}T_2T_k}{(1-T_1)^{k+1}}\right)
+\ldots\right\} + \gamma_{22}^{(4)}\left\{
\sum_{k\geq 7}\frac{1}{(k-7)!}
\frac{T_0^{k-7}T_2T_k}{(1-T_1)^{k+1}}
+\ldots\right\} }_{{\rm genus}\ 4} +
$$}
\be
+\underbrace{\ldots}_{{\rm higher\ genera}}
\ee
{\footnotesize
\centerline{$
F_3 = \underbrace{
\left(\gamma_{31}^{(3)} = -\frac{31}{5\cdot 27 \cdot 7\cdot 1024}
\right)\left\{
\sum_{k\geq 4}\frac{1}{(k-4)!}
\left(\frac{T_0^{k-4}T_k}{(1-T_1)^{k+1}}+
\frac{k+6}{2}\cdot \frac{T_0^{k-3}T_2T_k}{(1-T_1)^{k+3}}\right)
+\ldots\right\} + \gamma_{32}^{(3)}\left\{
\sum_{k\geq 3}\frac{1}{(k-3)!}
\frac{T_0^{k-3}T_2T_k}{(1-T_1)^{k+3}}
+\ldots\right\} }_{{\rm genus}\ 3} +
$} $$
+\underbrace{\gamma_{31}^{(4)}\left\{
\sum_{k\geq 7}\frac{1}{(k-7)!}
\left(\frac{T_0^{k-7}T_k}{(1-T_1)^{k}}+
\frac{k+7}{2} \cdot \frac{T_0^{k-6}T_2T_k}{(1-T_1)^{k+2}}\right)
+\ldots\right\} + \gamma_{32}^{(3)}\left\{
\sum_{k\geq 6}\frac{1}{(k-6)!}
\frac{T_0^{k-6}T_2T_k}{(1-T_1)^{k+2}}
+\ldots\right\} }_{{\rm genus}\ 4} +
$$}
\be
+\underbrace{\ldots}_{{\rm higher\ genera}}
\ee
{\footnotesize
$$
F_4 = \underbrace{\gamma_{41}^{(4)}\left\{
\sum_{k\geq 6}\frac{1}{(k-6)!}
\left(\frac{T_0^{k-6}T_k}{(1-T_1)^{k+1}}+
\frac{k+8}{2}\cdot \frac{T_0^{k-5}T_2T_k}{(1-T_1)^{k+3}}\right)
+\ldots\right\} + \gamma_{42}^{(3)}\left\{
\sum_{k\geq 5}\frac{k+1}{(k-5)!}
\frac{T_0^{k-5}T_2T_k}{(1-T_1)^{k+3}}
+\ldots\right\} }_{{\rm genus}\ 4} +
$$}
\be
+\underbrace{\ldots}_{{\rm higher\ genera}}
\label{F4}
\ee
$$
+ \ldots
$$
We omitted the factors $g^{2p}$ in these formulas,
they can be immediately restored.

These expressions satisfy the string equation (\ref{seq})
for arbitrary values of $\gamma$-parameters, provided
\be
\beta^{(0)}_0 = -\frac{1}{2}T_0^2\ \ \ \ \ {\rm and}\ \ \ \ \
\beta^{(1)}_1 = \frac{1}{24} = -\gamma^{(1)}_1
\ee
Actual values of $\gamma$'s are given in brackets
in the above formulas.
There are many more $\gamma$-parameters than
can be seen in these lines: they appear in front of
$T_0$-independent terms which have powers in $T$ higher
than shown in these formulas. $\gamma$-parameters can be defined in
different ways. In our approach, they are dictated by higher Virasoro
constraints (\ref{defvirco}).

In practice, {\bf we derived (\ref{F1})-(\ref{F4})
with all the proper values of $\gamma$-parameters
with the help of (\ref{defpartfun})}:
by acting with the explicitly known operator (\ref{Uop})
on the known expression (\ref{F0}) for the Kontsevich
partition function $Z_0$
(see, for example, the second paper of \cite[Appendix A1.2]{AMM.IM}).
Now we are going to demonstrate that the free energy
(\ref{F0})-(\ref{F4}) is indeed the same as that
considered in \cite{LaKa,Kaz} and \cite{BM}.

\subsection{Consistency with \cite{LaKa,Kaz}
\label{introg}}

It is an easy MAPLE exercise to check that substitution of
$T(p)$ from (\ref{Tp}) into ${\cal F}(T)$
reproduces $H(p,u)$ in (\ref{Hp}):
\be
H(p,u) - H_{01}(p,u) - H_{02}(p,u) = \sum_{p\geq q\geq 0}
u^{2q}F^{(p)}_q\Big(T(p)\Big)
\label{HFrel}
\ee
Of course, this demonstration is not a conceptual proof,
which should be based on relating the Virasoro constraints
(\ref{defvirco}) to the ones imposed on $\exp\Big(H(p)\Big)$.
Such a proof  seems straightforward, but it
is left beyond the scope of the present paper. For some more details
see s.4 below.

One comment, is, however, necessary already at this point.
The Hurwitz function (\ref{Hp}) and, thus, relation (\ref{HFrel})
are so far defined for $g^2=1$.
One can restore the $g^2$-dependence, making use of the homogeneity
property
\be
F^{(p)}_q(\lambda^{2k-2}T_k) = \lambda^{6p-6-2q} F^{(p)}_q(T_k)
\ee
It follows that
\be
\sum_{p\geq q\geq 0} \lambda^{6(p-1)}u^{2q}F^{(p)}_q(T_k) =
\sum_{p\geq q\geq 0} (\lambda u)^{2q}F^{(p)}_q( \lambda^{2k-2}T_k)
= \nn \\
= H\left(\frac{p_n}{\lambda^{3+3n}},\lambda u\right)
- H_{01}\left(\frac{p_n}{\lambda^{3+3n}},\lambda u\right)
- H_{02}\left(\frac{p_n}{\lambda^{3+3n}},\lambda u\right)
\label{Hg}
\ee
since
\be
T_k \ \stackrel{(\ref{Tp})}{=}\
u^{2k+1}\sum_{n=1}^\infty \frac{n^{n+k}}{n!}u^{3n} p_n
\label{Tp1}
\ee
is equivalent to
\be
\lambda^{2k-2}T_k =
(\lambda u)^{2k+1}\sum_{n=1}^\infty \frac{n^{n+k}}
{n!}(\lambda u)^{3n} \frac{p_n}{\lambda^{3+3n}}
\label{Tp2}
\ee
It remains to put $\lambda^6 = g^2$.

\subsection{Consistency with \cite{BM}
\label{cons} }

It is also easy to compare our ${\cal F}(T)$ with its smaller
fragments, explicitly evaluated in \cite{BM} from
the AMM-Eynard equation on the Lambert curve.
To this end, one should interpret multidensities from
that paper as
\be
\rho^{(p|m)}(y_1,\ldots,y_m) = W_p(y_1,\ldots,y_m)
= \hat\nabla(y_1)\ldots\hat\nabla(y_m) {\cal F}(T), \nn \\
\hat\nabla(y) = \sum_{m=0}^\infty \zeta_m(y)
\frac{\partial}{\partial T_m}
\ee
With this interpretation it is easy to extract from
eqs.(2.44)-(2.47) of \cite{BM}:
\be
F_0 = \underbrace{\frac{1}{6}T_0^3 + \frac{1}{6}T_0^3T_1 +
\ldots}_{{\rm genus}\ 0} +
\underbrace{\frac{1}{24}T_1 + \frac{1}{48}T_1^2 +
\frac{1}{24}T_0T_2}_{{\rm genus}\ 1} + \nn \\
+\underbrace{\frac{1}{9\cdot 128}T_4 + \frac{1}{9\cdot 128}T_0T_5
+ \frac{1}{3\cdot 128}T_1T_4 + \frac{29}{45\cdot 128}T_2T_3
+ \ldots}_{{\rm genus}\ 2}
+ \underbrace{\frac{1}{9^2\cdot 1024}T_7 + \ldots}_{{\rm genus}\ 3}
+ \ldots
\ee
\be
F_1 = \underbrace{-\frac{1}{24}T_0 - \frac{1}{24}T_0T_1
- \ldots}_{{\rm genus}\ 1}
\ \underbrace{-\frac{1}{15\cdot 32}T_3 - \frac{1}{15\cdot 32}T_0T_4
- \frac{1}{5\cdot 32}T_1T_3 - \frac{5}{9\cdot 128}T_2^2
- \ldots}_{{\rm genus}\ 2} - \nn \\
\underbrace{-\frac{7}{5\cdot 27\cdot 1024}T_6
- \ldots}_{{\rm genus}\ 3} - \ldots
\ee
\be
F_2 = \underbrace{\frac{7}{45\cdot 128}T_2
+\frac{7}{45\cdot 128}T_0T_3 + \frac{7}{15\cdot 128}T_1T_2
+ \ldots}_{{\rm genus}\ 2} +
\underbrace{\frac{41}{7\cdot 81\cdot 1024}T_5 + \ldots}_{{\rm
genus}\ 3} + \ldots
\ee
\be
F_3 = \underbrace{-\frac{31}{5\cdot 27\cdot 7\cdot 1024}T_4
- \ldots}_{{\rm genus}\ 3} - \ldots,  \\
\ldots \nn
\ee
what obviously coincides with formulas in s.\ref{sum} above.
These formulas from \cite{BM} are written for $u=1$, but
one can easily restore the $u$-dependence.
We return to discussion of this approach in the
special section \ref{BMap} below.

\subsection{A few comments}

\paragraph{1.} In order to avoid possible confusion about our notation
and normalization conditions we explicitly list a few first
terms in the lowest Virasoro constraints:
\be
\left((T_1-1)\frac{\partial}{\partial T_0}+\ldots\right)
F^{(0)}_0 = - \frac{1}{2}T_0^2,  \nn \\
\left(T_0 \frac{\partial}{\partial T_0}+
3(T_1-1)\frac{\partial}{\partial T_1}+\ldots\right)
F^{(0)}_0 = 0 \nn\\
\left(3T_0 \frac{\partial}{\partial T_1}+
15(T_1-1)\frac{\partial}{\partial T_2}+\ldots\right)
F^{(0)}_0 +
\frac{1}{2}\left(\frac{\partial F^{(0)}_0}{\partial T_0}
\right)^2=0,\nn\\
\nn\\ \ldots \nn \\
\left((T_1-1)\frac{\partial}{\partial T_0}+\ldots\right)
F^{(1)}_0  = 0,  \nn \\
\left(T_0 \frac{\partial}{\partial T_0}+
3(T_1-1)\frac{\partial}{\partial T_1}+\ldots\right)
F^{(1)}_0 = - \frac{1}{8}  \nn\\
\left(
\frac{\partial F^{(0)}_0}{\partial T_0}
\frac{\partial}{\partial T_0} +
3T_0 \frac{\partial}{\partial T_1}+
15(T_1-1)\frac{\partial}{\partial T_2}+\ldots
\right)
F^{(1)}_0
+ \frac{1}{2}\frac{\partial^2 F^{(0)}_0}{\partial T_0^2} = 0
\nn \\ \nn \\ \ldots \nn \\
\left((T_1-1)\frac{\partial}{\partial T_0}+\ldots\right)
F^{(1)}_1 = \frac{1}{24},  \nn \\
\left(T_0 \frac{\partial}{\partial T_0}+
3(T_1-1)\frac{\partial}{\partial T_1}+\ldots\right)
F^{(1)}_1 = - F^{(1)}_1  \nn\\
\left(
\frac{\partial F^{(0)}_0}{\partial T_0}
\frac{\partial}{\partial T_0} +
3T_0 \frac{\partial}{\partial T_1}+
15(T_1-1)\frac{\partial}{\partial T_2}+\ldots\right)
F^{(1)}_1  = 0
\nn \\
\ldots
\ee

\paragraph{2.} The general form of  terms,
explicitly shown in (\ref{F0})-(\ref{F4}), is
\be
F^{(p)}_q = \sum_k \frac{1}{(k+2+q-3p)!}
\frac{T_0^{k+2+q-3p}T_k}{(1-T_1)^{k+1+q-p}} +
\sum_k \frac{(k+1+q-p)(k+2+q-p)+\beta^{(p)}_q}{2(k+3+q-3p)!}
\frac{T_0^{k+3+q-3p}T_2T_k}{(1-T_1)^{k+3+q-p}} + \ldots
\nn
\ee
The values of $\beta^{(p)}_q$ are not constrained by
$\hat L_0$ and $\hat L_1$ conditions: these values are
examples of $\gamma$-parameters.

\paragraph{3.} It is interesting to note that,
if all $\beta$ are vanishing, one would have
\be
F_q^{(p+q)} \approx \partial_0^{2q}F_0^{(p)}
\label{d2rec}
\ee

In order to understand this, note that, using (\ref{calL-1}), one obtains
\be\label{1}
\hat l_{-1}F^{(p)}_q={\delta_{q,1}\delta_{p,1}\over 24}-
{T_0^2\over 2}\delta_{p,0}\delta_{q,0}
\ee
and, therefore,
\be
\hat l_{-1}\partial^2_0F^{(p)}_q=-\delta_{p,0}\delta_{q,0}
\ee
Similarly, using (\ref{L0co}), one obtains
\be
(\hat l_0+q)F^{(p)}_q=-{\delta_{q,0}\delta_{p,1}\over 16}
\ee
and, therefore,
\be\label{4}
(\hat l_0+q+1)\partial^2_0F^{(p)}_q=0
\ee
For $q,p>1$ this means that $F^{(p+1)}_{q+1}$ and $\partial^2_0F^{(p)}_q$ satisfy
the same first two Virasoro constraints and the same selection rules
(otherwise one could take $F^{(k)}_{q+1}$ with any $k>1$, since it
satisfies the same two Virasoro constraints). This does not mean that these two
functions coincide, since there are higher constraints, which fix the ambiguity
expressed in terms of arbitrary coefficients $\gamma$. For instance,
in most cases $\beta\neq 0$, and (\ref{d2rec})
acquires corrections (i.e. the $\gamma$-parameters are all
different). However, in those cases
when there is no freedom in solutions of the two first constraints (there are no
many $\gamma$'s), relation (\ref{d2rec}) is correct. For instance,
\be
F_2^{(2)} \sim \partial_0^{2}F_1^{(1)}
\ee
since, as it follows from (\ref{F1}), (\ref{F2}), there is no $\gamma$-freedom
in these free energies but the general coefficient.

Moreover, even the relative coefficient is fixed in the combination
$F^{(1)}_1 +\frac{1}{24}
\frac{\partial^2}{\partial T_0^2}F^{(0)}_0
$
which is canceled by the first two Virasoro constraints (see (\ref{1})-(\ref{4})),
which means that
\be
F^{(1)}_1 = -\frac{1}{24}
\frac{\partial^2}{\partial T_0^2}F^{(0)}_0
\ee

\bigskip

Thus, we provided a decisive evidence that the Kontsevich-Hurwitz partition function is,
indeed, given by (\ref{defpartfun}), i.e. is a solution to the conjugated Virasoro
constraints (\ref{defvirco}). This means that it is one of the phases in the M-theory
of matrix models \cite{AMM.IM}. From here, the reader can directly proceed to our
conclusions in s.6.
Still we find the claims of \cite{LaKa,Kaz,BM} so interesting, that we devote the next
two sections s.4 and s.5 to deeper discussion about the claims of these papers.

\section{Hurwitz partition function $H(p)$}
\setcounter{equation}{0}

We do not go into details of this very interesting story,
which is nicely presented in numerous papers.
Only some facts of direct relevance for our consideration
are briefly reviewed in this section.

\subsection{Hurwitz numbers}

Hurwitz numbers count ramified coverings of a Riemann sphere.
Relevant for our considerations are coverings with $N$ sheets,
connected only pairwise (double ramifications) except for
at a single point (usually posed at infinity),
where one can glue together $m_1, m_2,
\ldots, m_n$ sheets, with $\sum_{i=1}^n m_i = N$ and some
$m_i\geq 1$.
The number of double  ramification
(i.e. of simple critical) points is then equal to
\be
M = 2p-2 + \sum_{i=1}^n (m_i+1)
\label{RH}
\ee
where $p$ is the genus of the covering.
Positions of ramification points (moduli) are
not taken into account, only combinatorics.

Fig.\ref{covering} illustrates the setting in the simplest
possible case of the covering, $y\rightarrow x$ described
by the equation $Q_N(y) = x$, $Q_N$ being a polynomial
of degree $N$. The function $y(x)$ has $N$
branches, and its Riemann surface is an $N$-fold covering
of the Riemann sphere, parameterized by $x$.
The covering is ramified at $N-1$ zeroes of the derivative
$Q_N^\prime(y_i)=0$, i.e. at $x_i = Q_N(y_i)$, which are all
assumed different (condition that the critical/ramification
points are simple/double) and at $x=\infty$, where all the
$N$ sheets of the Riemann surface are glued together.
Thus, in this case $n=1$, $m_1=N$, $M=N-1$ and, obviously,
$p=0$ -- in accordance with the Riemann-Hurwitz formula
(\ref{RH}).

\begin{figure}
\begin{center}
\unitlength 1mm 
\linethickness{0.4pt}
\ifx\plotpoint\undefined\newsavebox{\plotpoint}\fi
\begin{picture}(156.539,70.887)(0,20)
\multiput(113.443,32.049)(-.07448,-.033634286){175}
{\line(-1,0){.07448}}
\put(100.409,26.163){\line(1,0){42.886}}
\multiput(143.295,26.163)(.073224719,.033657303){178}
{\line(1,0){.073224719}}
\put(156.329,32.154){\line(-1,0){42.675}}
\multiput(113.653,41.614)(-.07448,-.033634286){175}
{\line(-1,0){.07448}}
\put(100.619,35.728){\line(1,0){42.886}}
\multiput(143.505,35.728)(.073224719,.033657303){178}
{\line(1,0){.073224719}}
\put(156.539,41.719){\line(-1,0){42.675}}
\multiput(113.443,50.549)(-.07448,-.033634286){175}
{\line(-1,0){.07448}}
\put(100.409,44.663){\line(1,0){42.886}}
\multiput(143.295,44.663)(.073224719,.033657303){178}
{\line(1,0){.073224719}}
\put(156.329,50.654){\line(-1,0){42.675}}
\multiput(113.653,58.958)(-.07448,-.033634286){175}
{\line(-1,0){.07448}}
\put(100.619,53.072){\line(1,0){42.886}}
\multiput(143.505,53.072)(.073224719,.033657303){178}
{\line(1,0){.073224719}}
\put(156.539,59.063){\line(-1,0){42.675}}
\multiput(113.653,67.367)(-.07448,-.033634286){175}
{\line(-1,0){.07448}}
\put(100.619,61.481){\line(1,0){42.886}}
\multiput(143.505,61.481)(.073224719,.033657303){178}
{\line(1,0){.073224719}}
\put(156.539,67.472){\line(-1,0){42.675}}
\put(148.139,48.614){\line(0,-1){18.385}}
\put(148.139,39.775){\circle*{1.118}}
\put(148.316,48.79){\circle*{1.118}}
\put(148.139,30.229){\circle*{1.061}}
\put(147.962,23.511){\makebox(0,0)[cc]{$\infty$}}
\put(148.139,48.614){\line(0,1){16.971}}
\put(148.316,65.231){\circle*{1.118}}
\put(148.139,57.276){\circle*{1.118}}
\qbezier(70.534,68.589)(1.856,61.872)(51.619,55.154)
\qbezier(51.972,55.154)(78.842,51.53)(38.537,46.846)
\qbezier(38.537,46.846)(6.718,43.045)(52.679,35.355)
\qbezier(52.679,35.355)(103.149,26.87)(15.38,24.749)
\put(8.132,20.683){\vector(0,1){50.205}}
\put(8.132,20.506){\vector(1,0){75.66}}
\put(13.081,67.882){\makebox(0,0)[cc]{$y$}}
\put(81.494,24.395){\makebox(0,0)[cc]{$x$}}
\put(25.102,20.683){\vector(0,-1){.07}}
\multiput(25.386,42.886)(-.01537,-.96843){24}{{\rule{.4pt}{.4pt}}}
\put(30.229,20.86){\vector(0,-1){.07}}
\multiput(30.512,60.564)(-.008839,-.994369){41}{{\rule{.4pt}{.4pt}}}
\put(62.225,20.683){\vector(0,-1){.07}}
\multiput(62.509,52.256)(-.011049,-.988845){33}{{\rule{.4pt}{.4pt}}}
\put(70.711,21.036){\vector(0,-1){.07}}\multiput(71.171,29.982)(-.05303,-.90156){11}{{\rule{.4pt}{.4pt}}}
\put(141.598,38.714){\line(0,-1){9.546}}
\put(137.002,56.392){\line(0,-1){8.485}}
\put(120.385,64.523){\line(0,-1){8.132}}
\put(115.082,47.906){\line(0,-1){8.839}}
\end{picture}
\caption{{\footnotesize
The covering $y\rightarrow x$ of the Riemann sphere in
the simplest case of the curve $Q_N(y) = x$.
Left picture: the real section. Right picture:
symbolical complex view. All critical points (zeroes
of $Q'(y)$ are assumed different. The $N$ sheets merge
together at infinity. }}
\label{covering}
\end{center}
\end{figure}
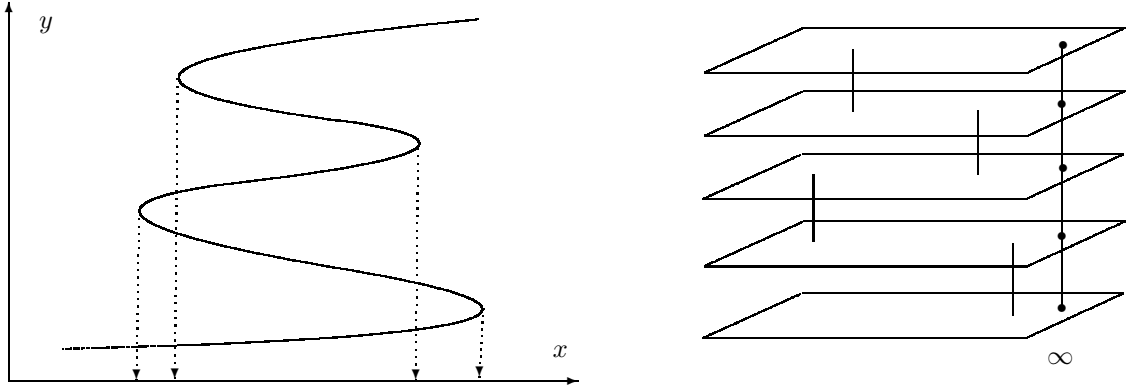

\begin{figure}
\begin{center}
\unitlength 1mm 
\linethickness{0.4pt}
\ifx\plotpoint\undefined\newsavebox{\plotpoint}\fi
\begin{picture}(156.539,47.472)(0,20)
\multiput(31.218,42.36)(-.07448,-.033634286){175}
{\line(-1,0){.07448}}
\put(18.184,36.474){\line(1,0){42.886}}
\multiput(61.07,36.474)(.073224719,.033657303){178}
{\line(1,0){.073224719}}
\put(74.104,42.465){\line(-1,0){42.675}}
\multiput(31.008,51.294)(-.07448,-.033634286){175}
{\line(-1,0){.07448}}
\put(17.974,45.408){\line(1,0){42.886}}
\multiput(60.86,45.408)(.073224719,.033657303){178}
{\line(1,0){.073224719}}
\put(73.894,51.399){\line(-1,0){42.675}}
\multiput(31.218,59.703)(-.07448,-.033634286){175}
{\line(-1,0){.07448}}
\put(18.184,53.817){\line(1,0){42.886}}
\multiput(61.07,53.817)(.073224719,.033657303){178}
{\line(1,0){.073224719}}
\put(74.104,59.808){\line(-1,0){42.675}}
\multiput(113.443,32.049)(-.07448,-.033634286){175}
{\line(-1,0){.07448}}
\put(100.409,26.163){\line(1,0){42.886}}
\multiput(143.295,26.163)(.073224719,.033657303){178}
{\line(1,0){.073224719}}
\put(156.329,32.154){\line(-1,0){42.675}}
\multiput(113.653,41.614)(-.07448,-.033634286){175}
{\line(-1,0){.07448}}
\put(100.619,35.728){\line(1,0){42.886}}
\multiput(143.505,35.728)(.073224719,.033657303){178}
{\line(1,0){.073224719}}
\put(156.539,41.719){\line(-1,0){42.675}}
\multiput(113.443,50.549)(-.07448,-.033634286){175}
{\line(-1,0){.07448}}
\put(100.409,44.663){\line(1,0){42.886}}
\multiput(143.295,44.663)(.073224719,.033657303){178}
{\line(1,0){.073224719}}
\put(156.329,50.654){\line(-1,0){42.675}}
\multiput(113.653,58.958)(-.07448,-.033634286){175}
{\line(-1,0){.07448}}
\put(100.619,53.072){\line(1,0){42.886}}
\multiput(143.505,53.072)(.073224719,.033657303){178}
{\line(1,0){.073224719}}
\put(156.539,59.063){\line(-1,0){42.675}}
\multiput(113.653,67.367)(-.07448,-.033634286){175}
{\line(-1,0){.07448}}
\put(100.619,61.481){\line(1,0){42.886}}
\multiput(143.505,61.481)(.073224719,.033657303){178}
{\line(1,0){.073224719}}
\put(68.962,35.511){\makebox(0,0)[cc]{$\infty$}}
\put(156.539,67.472){\line(-1,0){42.675}}
\put(29.345,56.745){\line(0,-1){8.132}}
\put(29.345,56.745){\circle*{1.118}}
\put(29.345,48.613){\circle*{1.118}}
\put(35.532,57.806){\line(0,-1){7.601}}
\put(35.532,57.806){\circle*{1.118}}
\put(35.532,50.205){\circle*{1.118}}
\put(41.719,55.508){\line(0,-1){17.324}}
\put(41.719,55.508){\circle*{1.118}}
\put(41.719,38.184){\circle*{1.118}}
\put(46.846,57.983){\line(0,-1){17.147}}
\put(46.846,57.983){\circle*{1.118}}
\put(46.846,40.836){\circle*{1.118}}
\put(56.745,48.083){\line(0,-1){8.662}}
\put(56.745,48.083){\circle*{1.118}}
\put(56.745,39.421){\circle*{1.118}}
\put(62.402,49.851){\line(0,-1){8.485}}
\put(62.402,49.851){\circle*{1.118}}
\put(62.402,41.366){\circle*{1.118}}
\put(147.609,64.877){\line(0,-1){7.955}}
\put(148.139,48.614){\line(0,-1){18.385}}
\put(148.139,39.775){\circle*{1.118}}
\put(148.316,48.79){\circle*{1.118}}
\put(148.139,30.229){\circle*{1.118}}
\put(147.785,56.922){\circle*{1.118}}
\put(147.609,64.877){\circle*{1.118}}
\put(147.962,23.511){\makebox(0,0)[cc]{$\infty$}}
\put(113.137,63.993){\line(0,-1){16.263}}
\put(113.137,63.993){\circle*{1.118}}
\put(113.137,47.733){\circle*{1.118}}
\put(119.678,56.392){\line(0,-1){8.132}}
\put(119.678,56.392){\circle*{1.118}}
\put(119.678,48.262){\circle*{1.118}}
\put(125.335,47.73){\line(0,-1){8.662}}
\put(125.335,47.73){\circle*{1.118}}
\put(125.335,39.068){\circle*{1.118}}
\put(138.239,57.452){\line(0,-1){17.324}}
\put(138.239,57.452){\circle*{1.118}}
\put(138.239,40.128){\circle*{1.118}}
\put(132.406,64.7){\line(0,-1){35.179}}
\put(132.406,64.7){\circle*{1.118}}
\put(132.406,29.521){\circle*{1.118}}
\end{picture}
\caption{{\footnotesize
The covering $y\rightarrow x$ of the Riemann sphere in
the case of generic $P_N(x,y)=0$.
Left picture: a fully reducible symbol, no branching at infinity.
Actually in the picture $N=3$ and $M=6$, so that $p=1$
(this is the cubic representation of a torus,
like $x^3+y^3+\alpha xy = 0$).
Right picture: generic branching at infinity, with $n$
groups of merging $m_1$, $m_2$, $\ldots$, $m_n$ sheets.
Actually in the picture $n=2$, $m_1=2$, $m_2=3$, $M=5$,
$p = 0$.
}}
\label{coveringgen}
\end{center}
\end{figure}
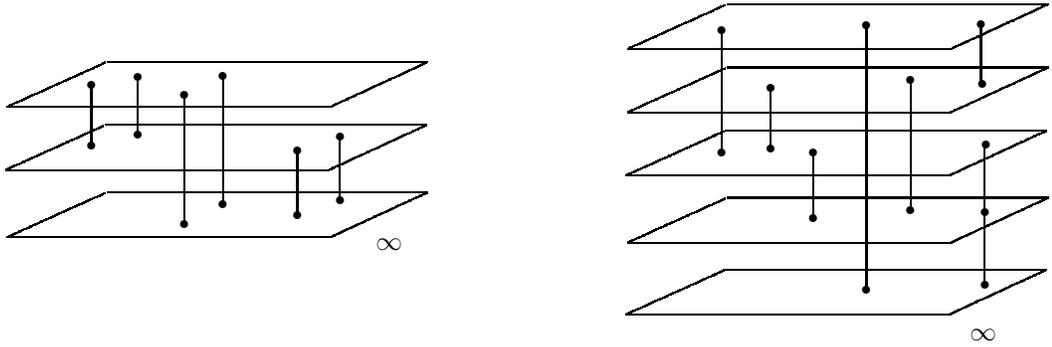

Another familiar case is the hyperelliptic covering
$y^2 = P_{2p+1}(x)$ where all the ramification points,
including the one at infinity are simple (double).
In this case $n=1$, $m_1=2$ and $M=2p+1$.

In the opposite extreme case of {\it generic} irreducible
polynomial of degree $N$, $P_N(x,y) = y^N + \sum_{k+l\leq N}y^kx^l =0$,
the function $y(x)$ has $N$ branches, i.e. its Riemann surface has
$N$ sheets of the corresponding
and is ramified at $N(N-1)$ points, where
$\frac{\partial P_N}{\partial y}=0$
(discriminant ${\rm Disc}_y P_n(x,y)$ is a polynomial
of degree $N(N-1)$ in $x$), which are
(generically) all different, and at $x=y=\infty$.
The branching at infinity is controlled by the
{\it homogeneous} part of $P_N(x,y)$
 (a "symbol" of $P_N$), which is
fully reducible,
$P_N(x,y) \sim y^N + \sum_{k+l=N} p_{kl}y^kx^l
= \prod_{i=1}^N(y-\lambda_i x)$
as $x,y\rightarrow \infty$.
This means that there is actually
no branching at infinity, thus $n=N$,
$m_1=\ldots=m_N=1$
and genus $p = \frac{(N-1)(N-2)}{2}$.

However, if our polynomial has different degrees $N$
and $n\leq N$ in $y$ and $x$ respectively and behaves as
$P(x,y) \sim \prod_{i=1}^n(y^{m_i}-\lambda_i x)$
at $x,y\rightarrow \infty$ then things are different:
non-trivial branching structure occurs over $x=\infty$,
and it is characterized by partition
$N=\sum_{i=1}^n m_i$ of $N$,
see Fig.\ref{coveringgen} for a simple example.
However, partition does not characterize the
covering unambiguously:
what remains not fixed, is combinatorics
of pairwise gluing of sheets, and Hurwitz number
$h(p\,|m_1,\ldots,m_n)$
counts the number of different possibilities
(modulo {\it location} of the critical points --
if they are taken into account we get a whole
continuous  moduli space of coverings and
Hurwitz number counts the number of {\it its} sheets).

Hurwitz numbers are simple to define, but not so
easy to calculate.
As usual, the problem is drastically simplified by
passing to generating functions --
this is one of the ideas, put into the basis of string theory.
Moreover, not a single, but a number of various (dual)
descriptions immediately arise in this way.
Most straightforwardly,
the Hurwitz free energy is the generating function
\be
H(p) = \frac{1}{g^2}
\sum_{n=1}^\infty \frac{1}{n!} \sum_{p;m_1,\ldots,m_n;M}
\delta\left( \sum_{i=1}^n (m_i+1) + 2p-2 - M\right)
\frac{u^{3M}g^{2p}}{M!}
h(p\,|m_1,\ldots,m_n)p_{m_1}\ldots p_{m_n}
\label{Hufe}
\ee
The parameters $u$ and $g$ serve to separate Hurwitz numbers
for different numbers $M$ of simple ramification points
and different genera. As already mentioned in (\ref{Hg}),
eq.(\ref{RH}) allows one to absorb $g^2$ into rescalings of
$u$ and $p_m$: $u \rightarrow gu$, $p_m \rightarrow p_m/g^{m+1}$,
and we do not keep $g^2$ dependence explicitly in this section.

\subsection{$e^H$ as KP $\tau$-function}

According to \cite{G}, the exponential of $H(p|u)$ can be alternatively
represented by eq.(\ref{Hp}):
\be
e^{H(p)} = e^{u^3\hat W_0(p)} e^{p_1}
\label{Hp1}
\ee
what immediately implies that it is a KP $\tau$-function \cite{Ok},
simply because $e^{p_1}$ is, and
all the $W_\infty$ generators belong to ${GL(\infty)}$
which is the symmetry group of the Universal Grassmannian \cite{UGr,W}.
Eq.(\ref{Hp1}) is motivated by relation to partitions
and characters, see also \cite{MMS,part}, but comments
on its derivation are beyond the scope of this paper.
We concentrate instead on its implications.

\subsubsection{Diagram technique}

\begin{figure}
\begin{center}
\unitlength 1mm 
\linethickness{0.4pt}
\ifx\plotpoint\undefined\newsavebox{\plotpoint}\fi
\begin{picture}(100,36.151)(0,0)
\multiput(30.052,16.882)(.057733429,-.03372041){173}
{\line(1,0){.057733429}}
\multiput(40.04,11.049)(-.063297462,-.033644597){155}
{\line(-1,0){.063297462}}
\put(29.964,27.931){\line(1,0){9.988}}
\put(49.939,33.941){\vector(3,2){.07}}
\multiput(40.04,27.842)(.054693342,.033695005){181}
{\line(1,0){.054693342}}
\put(50.028,22.097){\vector(2,-1){.07}}
\multiput(40.04,27.754)(.059451686,-.033671751){168}
{\line(1,0){.059451686}}
\put(49.939,11.049){\vector(1,0){.07}}\put(40.217,10.96)
{\line(1,0){9.7227}}
\put(39.068,30.229){\makebox(0,0)[cc]{$\frac{1}{2}ij$}}
\put(46.934,36.151){\makebox(0,0)[cc]{$p_i$}}
\put(46.315,19.534){\makebox(0,0)[cc]{$p_j$}}
\put(30.052,25.102){\makebox(0,0)[cc]{$p_{i+j}$}}
\put(45,14.142){\makebox(0,0)[cc]{$\frac{1}{2}(i+j)$}}
\put(28.903,18.738){\makebox(0,0)[cc]{$p_i$}}
\put(28.991,3.801){\makebox(0,0)[cc]{$p_j$}}
\put(48.083,7.601){\makebox(0,0)[cc]{$p_{i+j}$}}
\end{picture}
\caption{{\footnotesize
The two vertices in diagram technique which describes
the action of $\hat W_0$ on $e^{p_1}$. Arrows denote
derivatives with respect to $p$, ends without arrows
carry $p$ themselves. Vertices contain factors of $ij$
and $i+j$. In what follows we often write $i$ instead
of $p_i$.
}}
\label{vertices}
\end{center}
\end{figure}
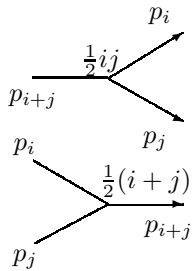

According to (\ref{Hp1}),
$H(p)$ is obtained by the diagram technique with two triple-vertex
elements (similar to the one analyzed in \cite{DCh}),
see Fig.\ref{vertices}.
Direction of arrows is important, lines with different
orientation are different, since the weights of two
vertices, $ij$ and $i+j$ do not coincide, i.e.
these are rather Heitler than Feynman diagrams.
All possible diagrams describe the r.h.s. of (\ref{Hp1}),
and its logarithm, $H(p)$ contains only {\it connected}
diagrams.  The power of $u^3$ is the number of vertices,
however, this does not immediately provide the power of
$u$ in ${\cal F}(T)$, because an $u$-dependence is also
contained in $T(p)$ (moreover, the diagrams with odd number
of vertices contain odd powers of $u$, which are all
converted into even powers after the transformation
from $p$ to $T$).
Remarkably, despite $q$ does not have a
direct diagrammatic meaning,
$p$ has: the diagram with $p\ $ loops contributes
only to components $F^{(p)}_q$ of the free energy.
$H_{01}$ is the sum of all rooted tree diagrams,
and $H_{02}$ of all double-rooted
trees.

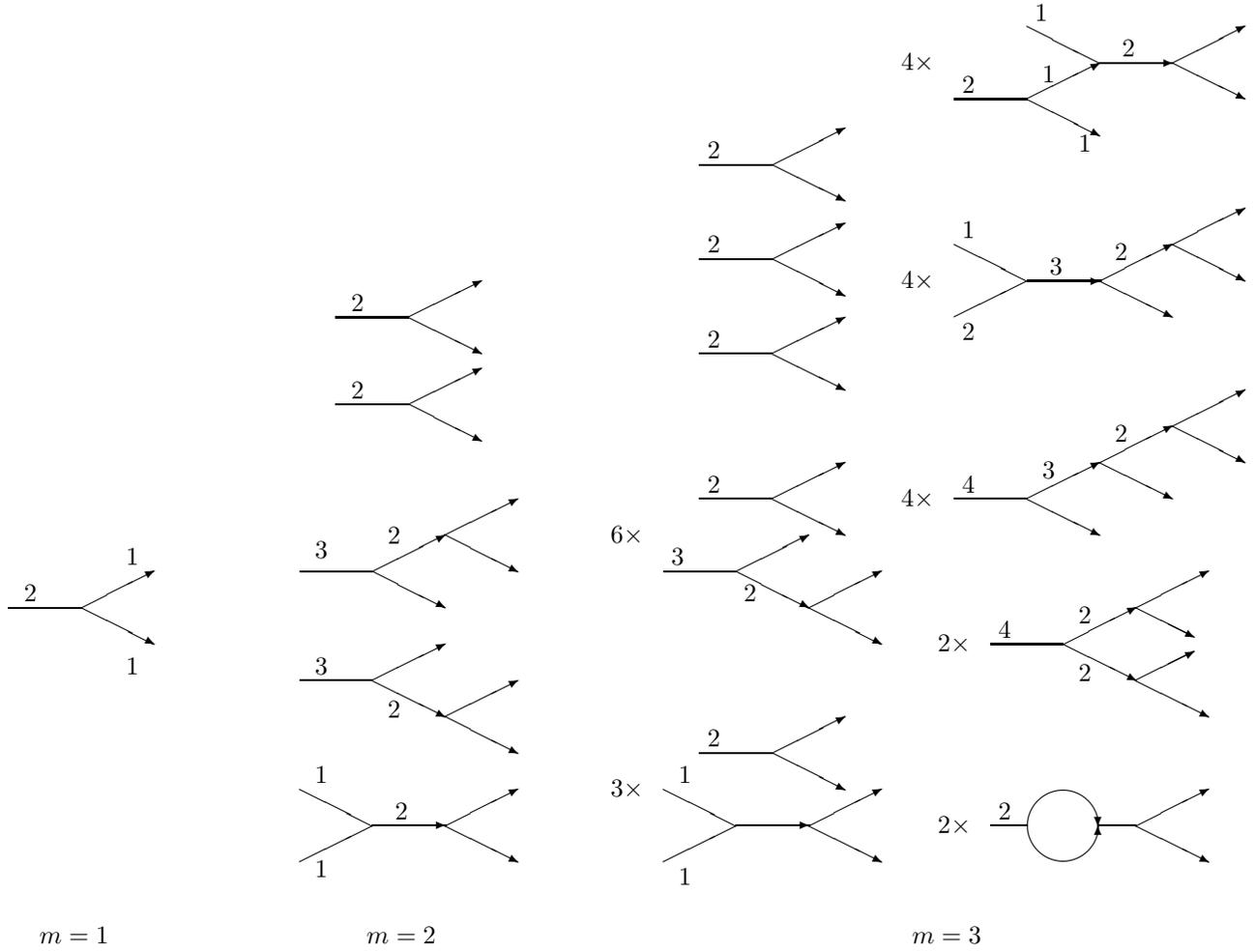
\begin{figure}
\begin{center}
\unitlength 1mm 
\linethickness{0.4pt}
\ifx\plotpoint\undefined\newsavebox{\plotpoint}\fi
\begin{picture}(273.5,130.007)(10,0)
%
%
\put(10,50){\line(1,0){9.95}}
\put(20,50){\line(2,1){10}}   \put(30,55){\vector(2,1){.2}}
\put(20,50){\line(2,-1){10}} \put(30,45){\vector(2,-1){.2}}
\put(13,52){\makebox(0,0)[cc]{$2$}}
\put(27,57){\makebox(0,0)[cc]{$1$}}
\put(27,42){\makebox(0,0)[cc]{$1$}}
\put(19,5){\makebox(0,0)[cc]{$m=1$}}
%
%
\put(55,90){\line(1,0){9.95}}
\put(58,92){\makebox(0,0)[cc]{$2$}}
\put(65,90){\line(2,1){10}}   \put(75,95){\vector(2,1){.2}}
\put(65,90){\line(2,-1){10}} \put(75,85){\vector(2,-1){.2}}
\put(55,78){\line(1,0){9.95}}
\put(58,80){\makebox(0,0)[cc]{$2$}}
\put(65,78){\line(2,1){10}}   \put(75,83){\vector(2,1){.2}}
\put(65,78){\line(2,-1){10}} \put(75,73){\vector(2,-1){.2}}
\put(50,55){\line(1,0){9.95}}
\put(53,58){\makebox(0,0)[cc]{$3$}}
\put(60,55){\line(2,1){10}}   \put(70,60){\vector(2,1){.2}}
\put(63,60){\makebox(0,0)[cc]{$2$}}
\put(60,55){\line(2,-1){10}} \put(70,50){\vector(2,-1){.2}}
\put(70,60){\line(2,1){10}}   \put(80,65){\vector(2,1){.2}}
\put(70,60){\line(2,-1){10}} \put(80,55){\vector(2,-1){.2}}
\put(50,40){\line(1,0){9.95}}
\put(53,42){\makebox(0,0)[cc]{$3$}}
\put(60,40){\line(2,1){10}}   \put(70,45){\vector(2,1){.2}}
\put(60,40){\line(2,-1){10}} \put(70,35){\vector(2,-1){.2}}
\put(63,36){\makebox(0,0)[cc]{$2$}}
\put(70,35){\line(2,1){10}}   \put(80,40){\vector(2,1){.2}}
\put(70,35){\line(2,-1){10}} \put(80,30){\vector(2,-1){.2}}
\put(50,25){\line(2,-1){10}} \put(53,27){\makebox(0,0)[cc]{$1$}}
\put(50,15){\line(2,1){10}}  \put(53,14){\makebox(0,0)[cc]{$1$}}
\put(60,20){\line(1,0){9.95}}\put(70,20){\vector(2,0){.2}}
\put(64,22){\makebox(0,0)[cc]{$2$}}
\put(70,20){\line(2,1){10}}   \put(80,25){\vector(2,1){.2}}
\put(70,20){\line(2,-1){10}} \put(80,15){\vector(2,-1){.2}}
\put(64,5){\makebox(0,0)[cc]{$m=2$}}
%
%
%
%
\put(105,111){\line(1,0){9.95}} \put(107,113){\makebox(0,0)[cc]{$2$}}
\put(115,111){\line(2,1){10}}   \put(125,116){\vector(2,1){.2}}
\put(115,111){\line(2,-1){10}} \put(125,106){\vector(2,-1){.2}}
\put(105,98){\line(1,0){9.95}} \put(107,100){\makebox(0,0)[cc]{$2$}}
\put(115,98){\line(2,1){10}}   \put(125,103){\vector(2,1){.2}}
\put(115,98){\line(2,-1){10}} \put(125,93){\vector(2,-1){.2}}
\put(105,85){\line(1,0){9.95}} \put(107,87){\makebox(0,0)[cc]{$2$}}
\put(115,85){\line(2,1){10}}   \put(125,90){\vector(2,1){.2}}
\put(115,85){\line(2,-1){10}} \put(125,80){\vector(2,-1){.2}}
\put(105,65){\line(1,0){9.95}} \put(107,67){\makebox(0,0)[cc]{$2$}}
\put(115,65){\line(2,1){10}}   \put(125,70){\vector(2,1){.2}}
\put(115,65){\line(2,-1){10}} \put(125,60){\vector(2,-1){.2}}
\put(100,55){\line(1,0){9.95}} \put(102,57){\makebox(0,0)[cc]{$3$}}
\put(110,55){\line(2,1){10}}   \put(120,60){\vector(2,1){.2}}
\put(110,55){\line(2,-1){10}} \put(120,50){\vector(2,-1){.2}}
\put(112,52){\makebox(0,0)[cc]{$2$}}
\put(120,50){\line(2,1){10}}   \put(130,55){\vector(2,1){.2}}
\put(120,50){\line(2,-1){10}} \put(130,45){\vector(2,-1){.2}}
\put(95,60){\makebox(0,0)[cc]{$6\times$}}
\put(105,30){\line(1,0){9.95}} \put(107,32){\makebox(0,0)[cc]{$2$}}
\put(115,30){\line(2,1){10}}   \put(125,35){\vector(2,1){.2}}
\put(115,30){\line(2,-1){10}} \put(125,25){\vector(2,-1){.2}}
\put(100,25){\line(2,-1){10}} \put(103,27){\makebox(0,0)[cc]{$1$}}
\put(100,15){\line(2,1){10}}\put(103,13){\makebox(0,0)[cc]{$1$}}
\put(110,20){\line(1,0){9.95}}\put(120,20){\vector(2,0){.2}}
\put(120,20){\line(2,1){10}}   \put(130,25){\vector(2,1){.2}}
\put(120,20){\line(2,-1){10}} \put(130,15){\vector(2,-1){.2}}
\put(95,25){\makebox(0,0)[cc]{$3\times$}}
\put(139,5){\makebox(0,0)[cc]{$m=3$}}
%
%
%
\put(150,120){\line(2,1){10}}\put(160,125){\vector(2,1){.2}}
\put(153,123.5){\makebox(0,0)[cc]{$1$}}
\put(150,130){\line(2,-1){10}} \put(152,132){\makebox(0,0)[cc]{$1$}}
\put(160,125){\line(1,0){9.95}} \put(170,125){\vector(2,0){.2}}
\put(164,127){\makebox(0,0)[cc]{$2$}}
\put(170,125){\line(2,1){10}}   \put(180,130){\vector(2,1){.2}}
\put(170,125){\line(2,-1){10}} \put(180,120){\vector(2,-1){.2}}
\put(140,120){\line(1,0){9.95}} \put(142,122){\makebox(0,0)[cc]{$2$}}
\put(150,120){\line(2,-1){10}} \put(160,115){\vector(2,-1){.2}}
\put(158,114){\makebox(0,0)[cc]{$1$}}
\put(135,125){\makebox(0,0)[cc]{$4\times$}}
\put(140,90){\line(2,1){10}}  \put(142,88){\makebox(0,0)[cc]{$2$}}
\put(140,100){\line(2,-1){10}} \put(142,102){\makebox(0,0)[cc]{$1$}}
\put(150,95){\line(1,0){9.95}} \put(160,95){\vector(2,0){.2}}
\put(154,97){\makebox(0,0)[cc]{$3$}}
\put(160,95){\line(2,1){10}}   \put(170,100){\vector(2,1){.2}}
\put(163,99){\makebox(0,0)[cc]{$2$}}
\put(160,95){\line(2,-1){10}} \put(170,90){\vector(2,-1){.2}}
\put(170,100){\line(2,1){10}}   \put(180,105){\vector(2,1){.2}}
\put(170,100){\line(2,-1){10}} \put(180,95){\vector(2,-1){.2}}
\put(135,95){\makebox(0,0)[cc]{$4\times$}}
\put(140,65){\line(1,0){9.95}}\put(142,67){\makebox(0,0)[cc]{$4$}}
\put(150,65){\line(2,1){10}}   \put(160,70){\vector(2,1){.2}}
\put(153,69){\makebox(0,0)[cc]{$3$}}
\put(150,65){\line(2,-1){10}} \put(160,60){\vector(2,-1){.2}}
\put(160,70){\line(2,1){10}}   \put(170,75){\vector(2,1){.2}}
\put(163,74){\makebox(0,0)[cc]{$2$}}
\put(160,70){\line(2,-1){10}} \put(170,65){\vector(2,-1){.2}}
\put(170,75){\line(2,1){10}}   \put(180,80){\vector(2,1){.2}}
\put(170,75){\line(2,-1){10}} \put(180,70){\vector(2,-1){.2}}
\put(135,65){\makebox(0,0)[cc]{$4\times$}}
\put(145,45){\line(1,0){9.95}} \put(147,47){\makebox(0,0)[cc]{$4$}}
\put(155,45){\line(2,1){10}}   \put(165,50){\vector(2,1){.2}}
\put(158,49){\makebox(0,0)[cc]{$2$}}
\put(155,45){\line(2,-1){10}} \put(165,40){\vector(2,-1){.2}}
\put(158,41){\makebox(0,0)[cc]{$2$}}
\put(165,50){\line(2,1){10}}   \put(175,55){\vector(2,1){.2}}
\put(165,50){\line(2,-1){8}} \put(173,46){\vector(2,-1){.2}}
\put(165,40){\line(2,1){8}}   \put(173,44){\vector(2,1){.2}}
\put(165,40){\line(2,-1){10}} \put(175,35){\vector(2,-1){.2}}
\put(140,45){\makebox(0,0)[cc]{$2\times$}}
\put(145,20){\line(1,0){5}} \put(147,22){\makebox(0,0)[cc]{$2$}}
\put(155,20){\circle{10}}  \put(159.8,20){\vector(0,-1){.2}}
\put(159.8,20){\vector(0,1){.2}}
\put(160,20){\line(1,0){5}}
\put(165,20){\line(2,1){10}}   \put(175,25){\vector(2,1){.2}}
\put(165,20){\line(2,-1){10}} \put(175,15){\vector(2,-1){.2}}
\put(140,20){\makebox(0,0)[cc]{$2\times$}}
\end{picture}
\caption{{\footnotesize
The lowest-order diagrams for Hurwitz function $H(p)$.
All free arrows at the right hand are supposed to act on
$e^{p_1}$, i.e. they carry index $1$ (from $p_1$)
and come with weight $1$. Each diagram is a monomial in
$p_k$'s, where relevant values of $k$ are indices
of the incoming lines at the left. The sum of $k$'s
is equal to the number of free arrows in the diagram.
Expression for diagram is made out of $ij/2$ and $(i+j)/2$ factors
at the vertices and
$\frac{u^{3m}}{m!}$ where $m$ is the total number of vertices.
Diagrams with $p$ loops
contribute only to $F^{(p)}$. Selection rule for $q$
is more complicated, because $u$ enters not only through
$u^{3m}$ but also through the $T(p)$ dependence.
For $m=3$ we do not draw identical diagrams,
instead their multiplicity are shown.
}}
\label{firstdia}
\end{center}
\end{figure}

\subsubsection{Low-order terms in $u$}

The lowest-order diagrams in Fig.\ref{firstdia}
describe the first terms of the $p$-series $H(p)$:
\be
\hat W_0 e^{p_1} = \frac{1}{2}p_2e^{p_1}, \nn \\
\hat W_0^2 e^{p_1} = \frac{1}{4}\Big(p_2^2+2p_1^2+4p_3\Big)e^{p_1},
\nn\\
\hat W_0^3 e^{p_1} = \frac{1}{8}\Big(
p_2^3 + 6p_1^2p_2 + 12p_2p_3 + 32p_1p_2 + 32p_4 + 4p_2
\Big)e^{p_1},\nn\\
\hat W_0^4 e^{p_1} = \frac{1}{16}\Big(
12p_1^4+12p_1^2p_2^2+48p_1^2p_3+64p_1^3+8p_1^2+p_2^4+24p_3p_2^2+\nn\\
+128p_2^2p_1+208p_2^2+128p_4p_2+48p_3^2+432p_3p_1+144p_3+400p_5
\Big)e^{p_1},\nn\\
\ldots
\ee
Thus
\be
H = p_1 + \frac{1}{2}u^3p_2 + \frac{1}{4}u^6(2p_3+p_1^2)
+\frac{1}{12}u^9(8p_1p_2 + 8p_4 + p_2) + \ldots
\ee
\be
H_{01} = p_1 + \frac{1}{2}u^3p_2 + \frac{1}{2}u^6p_3 +
\frac{2}{3}u^9p_4 + \ldots
\ee
\be
H_{02} = \frac{1}{4}u^6p_1^2 + \frac{2}{3}u^9 p_1p_2 +
u^{12}\left({1\over 2}p_2^2+{9\over 8}p_1p_3\right)+
u^{15}\left({32\over 15}p_1p_4+{9\over 5}p_2p_3\right)+\ldots
\ee
and
\be
H - H_{01} - H_{02} = \frac{1}{12}u^9p_2 +
u^{12}\left(\frac{3}{8}p_3+\frac{1}{6}p_1^3+\frac{1}{48}p_1^2\right)
+ u^{15}\left(\frac{4}{3}p_4 + p_1^2p_2 + \frac{1}{3}p_1p_2
+\frac{1}{240}p_2\right)
+\nn\\
+u^{18}\left(\frac{625}{144}p_5 + \frac{9}{4}p_1^2p_3+2p_2^2p_1 +
\frac{27}{16}p_1p_3 +\frac{2}{3}p_2^2+\frac{1}{6}p_1^4
+ \frac{9}{80}p_3 + \frac{1}{18}p_1^3 + \frac{1}{1440}p_1^2\right)
+ O(u^{21})
\ee

\subsubsection{ Linear contributions to ${\cal F}$}

Especially instructive is to compare the linear contributions to
$H(p)$ and to ${\cal F}(T)$.
The first $T$-linear terms in ${\cal F}$ are
\be
\begin{array}{ccccccccc}
{\rm lin}\Big({\cal F}\Big) &=&
\gamma_0^{(1)}T_1 &+& \gamma_0^{(2)}T_4 &+& \gamma_0^{(3)}T_7
&+& \ldots \\
&& -\gamma_1^{(1)}T_0 &-& \gamma_1^{(2)}T_3 &-& \gamma_1^{(3)}T_6
&-& \ldots \\
&& &+& \gamma_2^{(2)}T_2 &+& \gamma_2^{(3)}T_5
&+& \ldots \\
&&&& &-& \gamma_3^{(3)}T_4
&-& \ldots \\
&&&&&&&+&  \ldots
\end{array}
\ee
and they all enter with different $\gamma$-factors,
which are not fixed by the string equation ($L_{-1}$-constraint),
only by the higher Virasoro constraints.
Substituting $T(p)$ from (\ref{Tp3}), one obtains
\be
{\rm lin}\Big({\cal F}\Big)
= \frac{1}{24}\sum_{n=1}^\infty u^{3+3n}p_n\frac{n^n}{n!}(n-1) +
\sum_{n=1}^\infty u^{6+3n} p_n\frac{n^n}{n!}\Big(\gamma_0^{(2)}n^4 -
\gamma_1^{(2)}n^3 + \gamma_2^{(2)}n^2\Big) + \ldots =\nn \\
= \sum_{p,n=1}^\infty u^{3p+3n}p_n\frac{n^n}{n!}
\Big(\gamma_0^{(p)}n^{3p-2} -
\gamma_1^{(p)}n^{3p-3} + \ldots + (-)^p\gamma_p^{(p)}n^{2p-2}\Big)
\ee
We explicitly substituted
$\gamma_1^{(1)} = \gamma_0^{(1)} = \frac{1}{24}$
in the first term of the first relation
in order to demonstrate that these two $\gamma$-factors
are related so that $p_1$ drops away from ${\cal F}$.
In fact this is true more generally: there are no $p_1$-linear
terms in ${\cal F}\Big(T(p)\Big)$ at all, and this provides
a set of relations for $\gamma$-factors (not a complete one,
of course, moreover, the coefficients in front of $T$-linear
terms do not exhaust the full set of $\gamma$-parameters).

Turning now to the Hurwitz free energy $H(p)$, its $p$-linear
part is provided by "rooted" diagrams, with just one free
leg at the left. Let us begin with the rooted trees.
If $r_m$ is the sum of all rooted tree diagrams with $m$
vertices, then one has an obvious recurrent relation:
\be\label{4.11}
r_{m+1} = \frac{1}{2(m+1)}\sum_{i+j=m+2} is_{i-1}\cdot jr_{j-1}
\ee
i.e. $r(t) = \sum_{m=0}^\infty r_mt^m$ satisfies the differential
equation
\be\label{4.12}
\partial_t r(t) =
\frac{1}{2}\left\{\partial_t \Big(t\cdot r(t)\Big)\right\}^2
\ee
This gives $r = 1 + \frac{1}{2}t + \frac{1}{2}t^2
+ \frac{2}{3}t^3 + \ldots =
\sum_{n=1}^\infty \frac{n^{n-2}}{n!}t^{n-1}$
i.e.  $\sum_{n=1}^\infty r_{n-1}p_n$
is exactly $H_{01}(p)$.
Thus, we see that by subtracting $H_{01}$ from $H(p)$ we throw
away all the $p$-linear terms, coming from the tree diagrams.
This fully eliminates the $p_1$-linear terms, because they
can not come from loops, but other terms $p_{k\geq 2}$
can and do arise in ${\cal F}\Big(T(p)\Big)$.

Thus, one observes the appearance of peculiar series of the form
$\sum_n \frac{n^{n+\alpha}}{n!}x^n$ (with $\alpha=-2$
in this particular case).
Such series arise as inverse to Lambert-like functions.
They are also important ingredient of the ELSV formula.

\subsubsection{Developing diagram calculus}

For future considerations (beyond the scope of the present paper)
it is instructive to elaborate a little more on the diagram
formalism.
Summation of rooted trees is equivalent to
evaluating $R(t)$,
\be
e^{R(t)} = e^{t\hat W_0^{-}}e^{p_1}
\ee
where $t=u^3$ and $\hat W_0^- = \frac{1}{2}\sum_{i,j=1}^\infty
ijp_{i+j}\partial^2_{ij}$ is a "half" of the $\hat W_0$
operator.
Diagram analysis implies that $r(t)$ is a sum of
{\it connected} diagrams, i.e. is linear in $p$-variables,
so that $\partial_{ij}^2 r = 0$,
moreover, conservation of "momentum" $i$ at all vertices
implies the selection rule
\be
R(t) = \sum_{m=0}^\infty t^m r_m p_{m+1} = \oint x r(tx) dp\,(x)
\ \ \ \ \ {\rm with}\ \ \ \
r(x) = \sum_{m=0}^\infty r_mx^m \ \ \ {\rm and} \ \ \
dp\,(x) = \sum_{k=1}^\infty \frac{p_k dx}{x^{k+1}}
\label{selr1}
\ee
Therefore,
\be
\dot R = e^{-R} \hat W_0^{-} e^R =
\frac{1}{2}\sum_{i,j} ij p_{i+j} \partial_i R \partial_j R
\ee
Substituting (\ref{selr1}) and picking up
the coefficient of $t^m$ or, equivalently, of $p_{m+1}$,
one reproduces eq.(\ref{4.11}):
\be
mr_m = \frac{1}{2}\sum_{i+j=m+1} ij r_{i-1}r_{j-1}
\label{srecr}
\ee
and  (\ref{4.12}):
\be
\dot r = \frac{1}{2}\Big[(t\cdot r)\dot{\phantom.}\Big]^2
\label{dotreq}
\ee

After this reformulation one can easily do much more.
For example, we can act with $\hat W_0^{-}$ not only on
$e^{p_1}$, but, for example, on
$e^{p_1+\alpha_n p_n}$ and pick up the $\alpha$-linear
contribution. This would allow us to get an expression for the
rooted tree, with exactly one of the outcoming arrows
carrying index $n$ (while all the rest still carry $1$).
The only thing to change in this case is the selection
rule (\ref{selr1}):
\be
R(t,\alpha) =
\sum_{m=0}^\infty t^m \Big(r_m p_{m+1}
+ \alpha_n r_m^{(n)} p_{m+n} + O(\alpha^2)\Big) =
\oint \Big(x r(tx)+\alpha_n x^n r'(tx)^{(n)} + O(\alpha^2)\Big)
dp\,(x)
\label{selrn}
\ee
where we introduced the evident notation $r^{(n)}_m$ and keep to denote
$r_m^{(1)}$ as $r_m$ and so for $R^{(n)}(t)$ below.
This immediately implies in addition to (\ref{srecr})
\be
mr_m^{(n)} = \sum_{i+j=m+n} ij r_{i-1}r_{j-n}^{(n)}
\label{s'recr}
\ee
-- an already linear equation for $r(t)^{(n)}$ once
$r(t)$ is known:
\be
t^{n-1}\dot r^{(n)} = (t\cdot r^{(n)})
\dot{\phantom.}\big(t^nr^{(n)}\big)\dot{\phantom.}
\label{dotr'eq}
\ee

These equations are solved in terms of the peculiar
special function $w(t)$, which belongs to the Lambert family and
satisfies
\be
t\dot w = w(1+w)^2
\label{wdiff}
\ee
and is the first member $w(t) = w_0(t)$
in the family of series
\be\label{wm}
w_m(t) = \sum_{k=1}^\infty \frac{k^{k+m}}{k!} t^k
\ee
From (\ref{wdiff}) it is easy to find expressions for
all $w_m(t)$:
\be\label{wmr}
w_1 = t\dot w_0 = t\dot w = w(1+w)^2,\nn \\
w_2 = t\dot w_1 = w(1+w)^2(1+4w + 3w^2),
\nn \\
w_3 = t\dot w_2 = w(1+w)^4(1+10w + 15w^2),\nn \\
\ldots \nn \\
w_{m+1} = t\dot w_m = w(1+w)^2\frac{dw_m}{dw}
\ee
These functions will be used in s.\ref{Ka} below
to define the transformations
\be
T_m(p) = t^{\frac{2m+1}{3}}\oint w_m(tx)dp(x)
\ee

One can also consider $w_m(t)$ with negative values of $m$.
Since $t\dot w_{-1} = w_0=w$,
one obtains
\be
w_{-1} = \frac{w}{1+w}
\label{w-1}
\ee
because $\frac{d}{dw}\frac{w}{1+w} = \frac{w}{w(1+w)^2}$
and all the $w$ series begin from $t^{1}$
(the absence of the $t^0$ term fixes the integration constants).
This $w_{-1}$ is the Lambert function {\it per se}.
Similarly,
\be
w_{-2} = \frac{w(2+w)}{2(1+w)^2} = w_{-1}-\frac{1}{2}w_{-1}^2
\label{w-2}
\ee
and so on.

We can now return to $r(t)$ and $r^{(n)}(t)$.
Comparing the second formula for $w_{-2}$ with
eq.(\ref{dotreq}), we obtain:
\be
r(t) = t^{-1}w_{-2}(t)=
\sum_{k=1}^\infty \frac{k^{k-2}}{k!}t^{k-1},
\ \ \ \ {\rm i.e.} \ \ \ r_m = \frac{(m+1)^{m-1}}{(m+1)!},
\ m\geq 0
\label{rw-2}
\ee
Indeed,
$$\big(t^{-1}w_{-2}\big)\!\dot{\phantom 5}
= t^{-2} (t\dot w_{-2} - w_{-2}) = t^{-2}(w_{-1} - w_{-2})
= \frac{1}{2t^2}w_{-1}^2 = \frac{1}{2}\dot w_{-2}^2 $$
and this is exactly eq.(\ref{dotreq}) for $r = t^{-1}w_{-2}$.

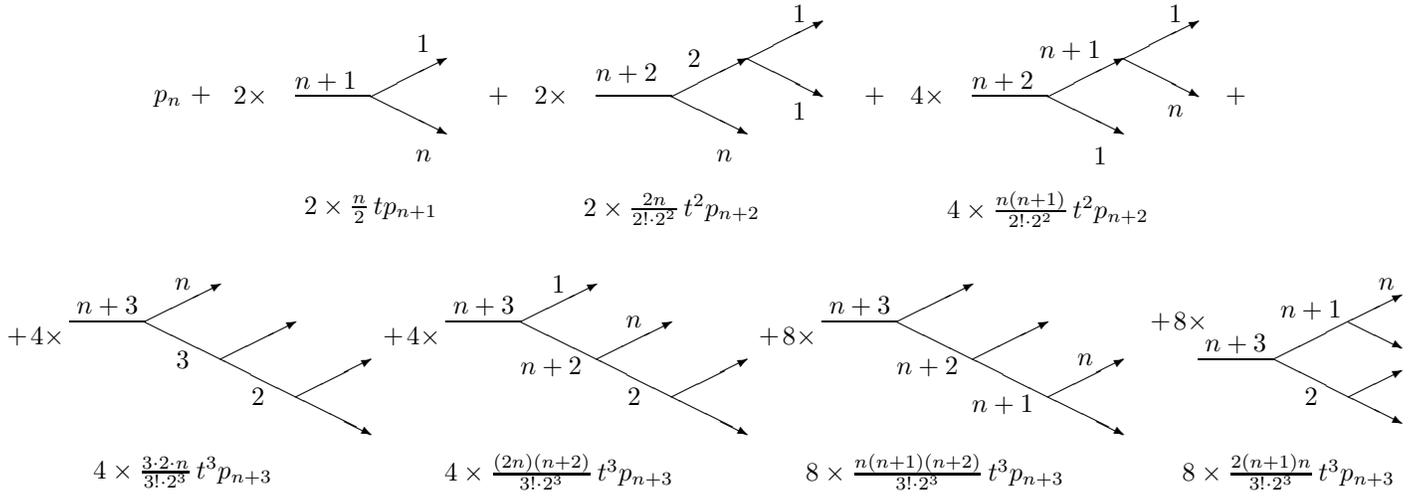
\begin{figure}
\begin{center}
\unitlength 1mm 
\linethickness{0.4pt}
\ifx\plotpoint\undefined\newsavebox{\plotpoint}\fi
\begin{picture}(273.5,70)(-20,0)
%
%
\put(10,55){\line(1,0){9.95}}
\put(20,55){\line(2,1){10}}   \put(30,60){\vector(2,1){.2}}
\put(20,55){\line(2,-1){10}} \put(30,50){\vector(2,-1){.2}}
\put(14,57){\makebox(0,0)[cc]{$n+1$}}
\put(27,62){\makebox(0,0)[cc]{$1$}}
\put(27,47){\makebox(0,0)[cc]{$n$}}
\put(-5,55){\makebox(0,0)[cc]{$p_n\  +$}}
\put(4,55){\makebox(0,0)[cc]{$2\times$}}
\put(20,40){\makebox(0,0)[cc]{$2\times\frac{n}{2}\,tp_{n+1}$}}
\put(50,55){\line(1,0){9.95}}
\put(54,58){\makebox(0,0)[cc]{$n+2$}}
\put(60,55){\line(2,1){10}}   \put(70,60){\vector(2,1){.2}}
\put(63,60){\makebox(0,0)[cc]{$2$}}
\put(60,55){\line(2,-1){10}} \put(70,50){\vector(2,-1){.2}}
\put(70,60){\line(2,1){10}}   \put(80,65){\vector(2,1){.2}}
\put(70,60){\line(2,-1){10}} \put(80,55){\vector(2,-1){.2}}
\put(67,47){\makebox(0,0)[cc]{$n$}}
\put(77,66){\makebox(0,0)[cc]{$1$}}
\put(77,53){\makebox(0,0)[cc]{$1$}}
\put(37,55){\makebox(0,0)[cc]{$+$}}
\put(44,55){\makebox(0,0)[cc]{$2\times$}}
\put(60,40){\makebox(0,0)[cc]{$2\times
\frac{2n}{2!\cdot 2^2}\,t^2p_{n+2}$}}
\put(100,55){\line(1,0){9.95}}
\put(104,57){\makebox(0,0)[cc]{$n+2$}}
\put(110,55){\line(2,1){10}}   \put(120,60){\vector(2,1){.2}}
\put(113,61){\makebox(0,0)[cc]{$n+1$}}
\put(110,55){\line(2,-1){10}} \put(120,50){\vector(2,-1){.2}}
\put(120,60){\line(2,1){10}}   \put(130,65){\vector(2,1){.2}}
\put(120,60){\line(2,-1){10}} \put(130,55){\vector(2,-1){.2}}
\put(117,47){\makebox(0,0)[cc]{$1$}}
\put(127,66){\makebox(0,0)[cc]{$1$}}
\put(127,53){\makebox(0,0)[cc]{$n$}}
\put(87,55){\makebox(0,0)[cc]{$+$}}
\put(94,55){\makebox(0,0)[cc]{$4\times$}}
\put(110,40){\makebox(0,0)[cc]{$4\times
\frac{ n(n+1)}{2!\cdot 2^2}\,t^2p_{n+2}$}}
\put(135,55){\makebox(0,0)[cc]{$+$}}
%
%
%
\put(-20,25){\line(1,0){9.95}}
\put(-10,25){\line(2,1){10}}   \put(0,30){\vector(2,1){.2}}
\put(-10,25){\line(2,-1){30}} \put(20,10){\vector(2,-1){.2}}
\put(0,20){\line(2,1){10}}   \put(10,25){\vector(2,1){.2}}
\put(10,15){\line(2,1){10}}   \put(20,20){\vector(2,1){.2}}
\put(-5,30){\makebox(0,0)[cc]{$n$}}
\put(-15,27){\makebox(0,0)[cc]{$n+3$}}
\put(-5,20){\makebox(0,0)[cc]{$3$}}
\put(5,15){\makebox(0,0)[cc]{$2$}}
\put(-27,23){\makebox(0,0)[cc]{$+$}}
\put(-23,23){\makebox(0,0)[cc]{$4\times$}}
\put(-5,5){\makebox(0,0)[cc]{$4\times\frac{3\cdot 2\cdot n}
{3!\cdot 2^3}\,t^3p_{n+3}$}}
\put(30,25){\line(1,0){9.95}}
\put(40,25){\line(2,1){10}}   \put(50,30){\vector(2,1){.2}}
\put(40,25){\line(2,-1){30}} \put(70,10){\vector(2,-1){.2}}
\put(50,20){\line(2,1){10}}   \put(60,25){\vector(2,1){.2}}
\put(60,15){\line(2,1){10}}   \put(70,20){\vector(2,1){.2}}
\put(55,25){\makebox(0,0)[cc]{$n$}}
\put(35,27){\makebox(0,0)[cc]{$n+3$}}
\put(45,30){\makebox(0,0)[cc]{$1$}}
\put(44,19){\makebox(0,0)[cc]{$n+2$}}
\put(55,15){\makebox(0,0)[cc]{$2$}}
\put(23,23){\makebox(0,0)[cc]{$+$}}
\put(27,23){\makebox(0,0)[cc]{$4\times$}}
\put(45,5){\makebox(0,0)[cc]{$4\times\frac{(2n)(n+2)}
{3!\cdot 2^3}\,t^3p_{n+3}$}}
\put(80,25){\line(1,0){9.95}}
\put(90,25){\line(2,1){10}}   \put(100,30){\vector(2,1){.2}}
\put(90,25){\line(2,-1){30}} \put(120,10){\vector(2,-1){.2}}
\put(100,20){\line(2,1){10}}   \put(110,25){\vector(2,1){.2}}
\put(110,15){\line(2,1){10}}   \put(120,20){\vector(2,1){.2}}
\put(115,20){\makebox(0,0)[cc]{$n$}}
\put(85,27){\makebox(0,0)[cc]{$n+3$}}
\put(94,19){\makebox(0,0)[cc]{$n+2$}}
\put(104,14){\makebox(0,0)[cc]{$n+1$}}
\put(73,23){\makebox(0,0)[cc]{$+$}}
\put(77,23){\makebox(0,0)[cc]{$8\times$}}
\put(95,5){\makebox(0,0)[cc]{$8\times\frac{n(n+1)(n+2)}
{3!\cdot 2^3}\,t^3p_{n+3}$}}
\put(130,20){\line(1,0){9.95}}
\put(140,20){\line(2,1){17}}   \put(157,28.5){\vector(2,1){.2}}
\put(140,20){\line(2,-1){17}} \put(157,11.5){\vector(2,-1){.2}}
\put(150,25){\line(2,-1){7}}   \put(157,21.5){\vector(2,-1){.2}}
\put(150,15){\line(2,1){7}}   \put(157,18.5){\vector(2,1){.2}}
\put(155,30){\makebox(0,0)[cc]{$n$}}
\put(135,22){\makebox(0,0)[cc]{$n+3$}}
\put(145,26.5){\makebox(0,0)[cc]{$n+1$}}
\put(145,15){\makebox(0,0)[cc]{$2$}}
\put(125,25){\makebox(0,0)[cc]{$+$}}
\put(129,25){\makebox(0,0)[cc]{$8\times$}}
\put(142,5){\makebox(0,0)[cc]{$8\times\frac{2(n+1)n}
{3!\cdot 2^3}\,t^3p_{n+3}$}}
\end{picture}
\caption{{\footnotesize
The simplest diagrams contributing to $r^{(n)}$:
connected rooted trees
with exactly one external arrow carrying index $n$.
The coupling constant is $u^3=t$.
The sum of all diagrams is
$R^{(n)} = p_n + np_{n+1}t + \frac{n(n+2)}{2}p_{n+2}t^2 +
\frac{n(n+3)^2}{6}p_{n+3}t^3 + \ldots$
The zeroth-order contribution $p_n$ corresponds to
the diagram with no vertices, not shown on the picture.
The combinatorial coefficient $8$ in the last diagram is made
from a "naive" factor $4$, counting the places to attach the
outgoing $n$ arrow (or, what is the same, the "up-down"
orientations of two vertices with different incoming lines)
and an extra, perhaps less familiar $2$, counting the
"time"-ordering of the two most right vertices:
a phenomenon illustrated also by appearance of $B$ and $C$
diagrams in Fig.\ref{p1p3} below.
}}
\label{ndia}
\end{center}
\end{figure}

Eq.(\ref{dotr'eq}) now acquires the form
\be
t^n \dot r^{(n)} = w_{-1}\cdot(t^n r^{(n)})\dot{\phantom.}
\ \ \ \ \ {\rm or} \ \ \ \ \
t\cdot\dot r^{(n)} = nr^{(n)}w
\ee
which implies that
\be
r^{(n)} = e^{nw_{-1}} = e^{\frac{nw}{{1+w}}} =
1 + nt + \frac{n(n+2)}{2}t^2 + {n(n+3)^2\over 6}t^3+\ldots=1+n\sum_{k=1}^\infty
{(n+k)^{k-1}\over k!}t^k, \nn\\
R^{(n)} = \oint x^n r^{(n)}(tx)dp(x) =
p_n + np_{n+1}t + \frac{n(n+2)}{2}p_{n+2}t^2 + \ldots
= p_n + n\sum_{k=1}^\infty
{(n+k)^{k-1}\over k!}t^kp_{n+k}
\ee
The first terms can be easily reproduced by direct
evaluation of diagrams, see Fig.\ref{ndia}.

\begin{figure}
\begin{center}
\unitlength 1mm 
\linethickness{0.4pt}
\ifx\plotpoint\undefined\newsavebox{\plotpoint}\fi
\begin{picture}(200,150)(-15,-20)
\put(0,115){\makebox(0,0)[cc]{$(\hat W_0^-)^2\hat W_0^+\hat W_0^+$}}
\put(0,60){\makebox(0,0)[cc]{$\hat W^-_0\hat W^+_0 (\hat W_0^-)^2$}}
\put(0,5){\makebox(0,0)[cc]{$\hat W^+_0 (\hat W_0^-)^3$}}
\put(50,130){\makebox(0,0)[cc]{${\bf A}$}}
\put(20,100){\line(2,1){30}}\put(35,110){\makebox(0,0)[cc]{$2$}}
\put(25,105){\makebox(0,0)[cc]{$3$}}
\put(30,105){\line(2,-1){10}}\put(40,100){\vector(2,-1){.2}}
\put(40,110){\line(2,-1){10}}\put(50,105){\vector(2,-1){.2}}
\put(40,120){\line(2,-1){10}}
\put(50,115){\line(1,0){10}}\put(55,117){\makebox(0,0)[cc]{$2$}}
\put(60,115){\line(2,1){10}}\put(70,120){\vector(2,1){.2}}
\put(60,115){\line(2,-1){10}}\put(70,110){\vector(2,-1){.2}}
\put(50,90){\makebox(0,0)[cc]{$4\cdot{\underline 8}\,p_1p_3=32p_1p_3$}}
\put(45,117.5){\vector(0,1){.07}}\put(45,112.5){\vector(0,-1){.07}}
\put(45,112.5){\vector(0,1){.07}}\put(45,107.5){\vector(0,-1){.07}}
\multiput(45,108)(0,.95588){9}{{\rule{.4pt}{.4pt}}}
\put(30,105){\circle{2}}
\put(120,130){\makebox(0,0)[cc]{${\bf B}$}}
\put(90,100){\line(2,1){30}}\put(95,105){\makebox(0,0)[cc]{$3$}}
\put(105,100){\makebox(0,0)[cc]{$2$}}
\put(107.5,110.5){\makebox(0,0)[cc]{$1$}}
\put(100,105){\line(2,-1){20}}\put(120,95){\vector(2,-1){.2}}
\put(110,100){\line(2,1){10}}\put(120,105){\vector(2,1){.2}}
\put(110,120){\line(2,-1){10}}
\put(120,115){\line(1,0){10}}\put(125,117){\makebox(0,0)[cc]{$2$}}
\put(130,115){\line(2,1){10}}\put(140,120){\vector(2,1){.2}}
\put(130,115){\line(2,-1){10}}\put(140,110){\vector(2,-1){.2}}
\put(120,90){\makebox(0,0)[cc]{$4\cdot{\underline 4}\,p_1p_3=16p_1p_3$}}
\put(110,120){\vector(0,1){.07}}\put(110,110){\vector(0,-1){.07}}
\multiput(110,110)(0,.95588){10}{{\rule{.4pt}{.4pt}}}
\put(100,105){\circle{2}}
\put(50,80){\makebox(0,0)[cc]{${\bf C}$}}
\put(30,55){\line(2,1){20}}\put(35,60){\makebox(0,0)[cc]{$3$}}
\put(50,52){\makebox(0,0)[cc]{$2$}}
\put(43,63.7){\makebox(0,0)[cc]{$1$}}
\put(40,60){\line(2,-1){30}}\put(70,45){\vector(2,-1){.2}}
\put(60,50){\line(2,1){10}}\put(70,55){\vector(2,1){.2}}
\put(40,70){\line(2,-1){10}}
\put(50,65){\line(1,0){10}}\put(55,67){\makebox(0,0)[cc]{$2$}}
\put(60,65){\line(2,1){10}}\put(70,70){\vector(2,1){.2}}
\put(60,65){\line(2,-1){10}}\put(70,60){\vector(2,-1){.2}}
\put(50,40){\makebox(0,0)[cc]{$4\cdot{\underline 8}\,p_1p_3=32p_1p_3$}}
\put(60,65){\vector(0,1){.07}}\put(60,50){\vector(0,-1){.07}}
\multiput(60,50)(0,.95588){15}{{\rule{.4pt}{.4pt}}}
\put(45,67.5){\vector(0,1){.07}}\put(45,62.5){\vector(0,-1){.07}}
\multiput(45,62.5)(0,.95588){5}{{\rule{.4pt}{.4pt}}}
\put(40,60){\circle{2}}
\put(120,80){\makebox(0,0)[cc]{${\bf D}$}}
\put(90,55){\line(2,1){20}}\put(95,60){\makebox(0,0)[cc]{$3$}}
\put(127,63.8){\makebox(0,0)[cc]{$2$}}
\put(103,63.8){\makebox(0,0)[cc]{$2$}}
\put(100,60){\line(2,-1){10}}\put(110,55){\vector(2,-1){.2}}
\put(100,70){\line(2,-1){10}}
\put(110,65){\line(1,0){10}}\put(115,67){\makebox(0,0)[cc]{$3$}}
\put(120,65){\line(2,1){10}}\put(130,70){\vector(2,1){.2}}
\put(120,65){\line(2,-1){20}}\put(140,55){\vector(2,-1){.2}}
\put(130,60){\line(2,1){10}}\put(140,65){\vector(2,1){.2}}
\put(120,40){\makebox(0,0)[cc]{$12\cdot{\underline 8}\,p_1p_3=96p_1p_3$}}
\put(100,60){\circle{2}}
\put(110,65){\circle{2}}
\put(120,65){\circle{2}}
\put(50,25){\makebox(0,0)[cc]{${\bf E}$}}
\put(30,0){\line(2,1){10}}\put(35,-0.5){\makebox(0,0)[cc]{$3$}}
\put(30,10){\line(2,-1){10}}\put(35,10.5){\makebox(0,0)[cc]{$1$}}
\put(40,5){\line(1,0){10}}\put(45,7){\makebox(0,0)[cc]{$4$}}
\put(50,5){\line(2,1){30}}\put(80,20){\vector(2,1){.2}}
\put(55,10.5){\makebox(0,0)[cc]{$3$}}
\put(65,15.5){\makebox(0,0)[cc]{$2$}}
\put(50,5){\line(2,-1){10}}\put(60,0){\vector(2,-1){.2}}
\put(60,10){\line(2,-1){10}}\put(70,5){\vector(2,-1){.2}}
\put(70,15){\line(2,-1){10}}\put(80,10){\vector(2,-1){.2}}
\put(50,-15){\makebox(0,0)[cc]{$24\cdot{\underline 8}\,p_1p_3
=192p_1p_3$}}
\put(40,5){\circle{2}}
\put(50,5){\circle{2}}
\put(60,10){\circle{2}}
\put(120,25){\makebox(0,0)[cc]{${\bf F}$}}
\put(100,0){\line(2,1){10}}\put(105,-0.5){\makebox(0,0)[cc]{$3$}}
\put(100,10){\line(2,-1){10}}\put(105,10.5){\makebox(0,0)[cc]{$1$}}
\put(110,5){\line(1,0){10}}\put(115,7){\makebox(0,0)[cc]{$4$}}
\put(120,5){\line(2,1){20}}\put(140,15){\vector(2,1){.2}}
\put(120,5){\line(2,-1){20}}\put(140,-5){\vector(2,-1){.2}}
\put(125,10.5){\makebox(0,0)[cc]{$2$}}
\put(125,-0.5){\makebox(0,0)[cc]{$2$}}
\put(130,10){\line(2,-1){7}}\put(137,6.5){\vector(2,-1){.2}}
\put(130,0){\line(2,1){7}}\put(137,3.5){\vector(2,1){.2}}
\put(120,-15){\makebox(0,0)[cc]{$16\cdot{\underline 4}\,p_1p_3
=64p_1p_3$}}
\put(130,10){\vector(0,1){.07}}\put(130,0){\vector(0,-1){.07}}
\multiput(130,0)(0,.95588){10}{{\rule{.4pt}{.4pt}}}
\put(110,5){\circle{2}}
\put(0,-15){\makebox(0,0)[cc]{$\frac{t^4}{4!}
\left(\frac{1}{2}\right)^4\times $}}
\end{picture}
\caption{{\footnotesize
Six diagrams, contributing to the double-rooted tree $p_1p_3$
in eq.(\ref{drt}).
Under each diagram its total contribution is written,
-- to be further multiplied by $\frac{1}{2^4\cdot 4!}$ --
and the coefficient in front of $p_1p_3$ is split into
the "obvious" part, given by a product of $ij$ and $i+j$
factors in the vertices and into "combinatorial factor"
(underlined), counting the multiplicity of the diagram.
When evaluating combinatorial factors one should also remember
that vertices with $i<j$ should be counted twice,
since they enter twice in the sum over $i,j$ in the definition
of $\hat W_0$. Such vertices are marked by a circle in the
picture.
Another contribution to combinatorial factor come from
different possibilities to form the same diagram due to
permutations of connections between vertices. Such
permutations are shown by dotted lines. Each dotted line
and each circle contribute a $2$ to the combinatorial factor.
Note that, since the two most right vertices in {\bf F} are
identical, no "time-ordering" factor of $2$ contributes
in this case.
The sum of all the six diagrams is
$\ \frac{16(2+1+2+6+12+4)}{2^4\cdot 4!}
p_1p_3 = \frac{27}{24}p_1p_3 = \frac{9}{8}p_1p_3\ $
and exactly reproduces the corresponding term in $H_{02}$.
}}
\label{p1p3}
\end{center}
\end{figure}
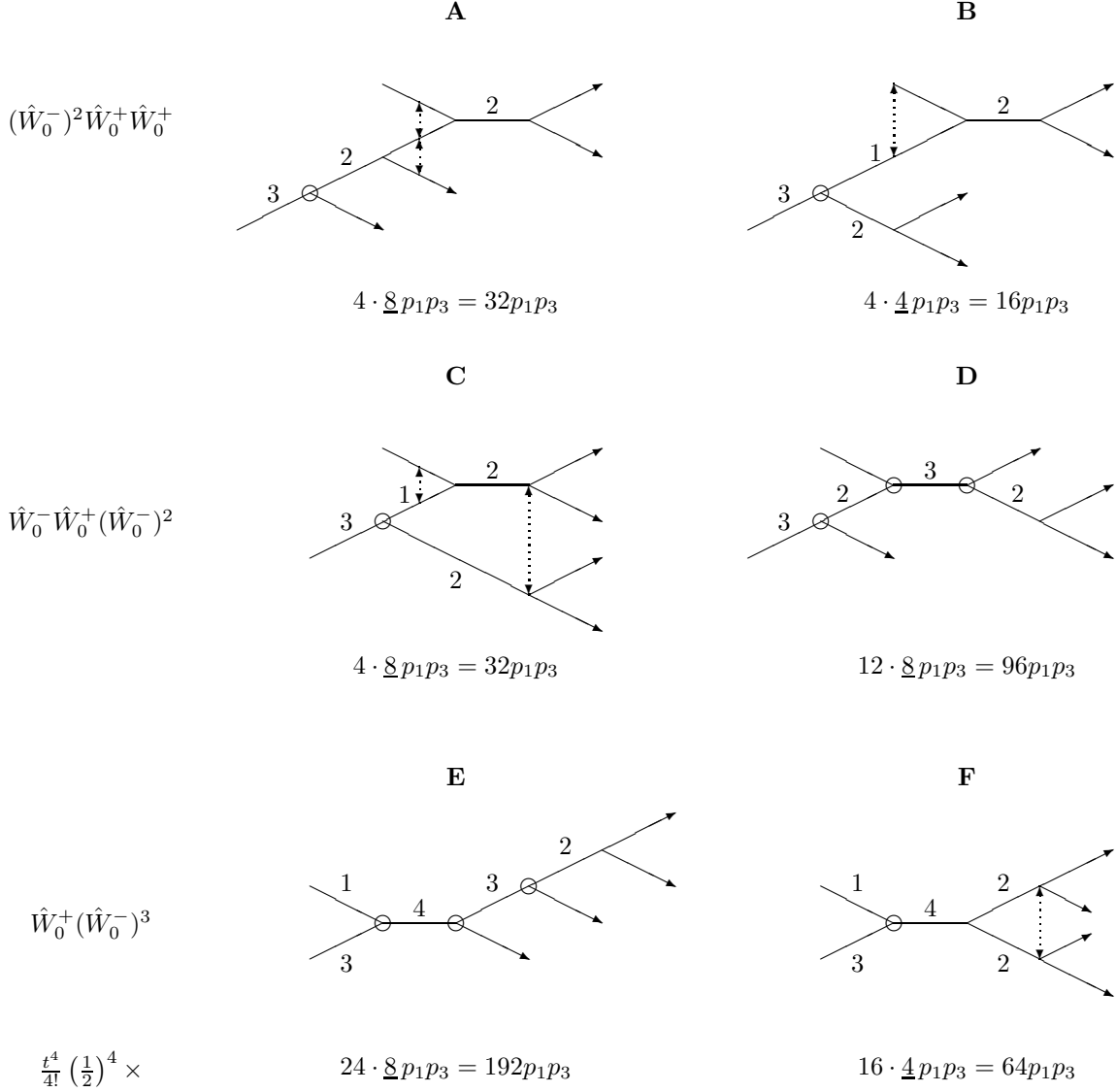

Similarly one can introduce and evaluate $R^{(m,n)}$,
the sum of rooted trees with {\it two} outgoing arrows
carrying indices $m$ and $n$, and more generally,
\be
R^{(n_1,\ldots,n_\nu)} = \sum_{k\geq 0}
r^{(n_1,\ldots,n_\nu)}_kt^k
p_{k+1+\sum_{i=1}^\nu(n_i-1)} =
\oint x^{1+\sum_{i=1}^\nu(n_i-1)}
r^{(n_1,\ldots,n_\nu)}(xt)dp(x), \nn \\
r^{(n_1,\ldots,n_\nu)}_k =
\frac{n_1\ldots n_\nu m^{k-1}}{(k+1-\nu)!}, \ \ \ \
m = k+1+\sum_{i=1}^\nu(n_i-1), \nn \\
\hat R^{(n_1,\ldots,n_\nu)} = \sum_{k,m=0}^\infty
\frac{n_1n_2\ldots n_\nu m^{k-1}t^k}{(k+1-\nu)!}
\delta\left(k+1+\sum_{i=1}^\nu (n_i-1) - m\right)
p_m\partial_{n_1}\partial_{n_2}\ldots\partial_{n_\nu}
\ee
where $p_m$ is the momentum at the root, $k$ -- the number
of vertices and the sum actually runs from $k=\nu-1$.
Of course, $R(t)$ and $R^{(n)}(t)$ are particular
cases of this formula for $\nu=0$ and $\nu=1$ respectively.
The corresponding generating functions are
\be \nn
\begin{array}{cl}
\nu=0: & r(t) = \sum_{k=0}^\infty \frac{(k+1)^{k-1}}{(k+1)!}\,t^k
= t^{-1}w_{-2}, \\
\nu=1: & r^{(n)}(t) = 1 + n\sum_{k=1}^\infty
\frac{(n+k)^{k-1}}{k!}\,t^k = e^{nw_{-1}},\nn \\
\nu=2: & r^{(n_1,n_2)}(t) = n_1n_2\sum_{k=1}^\infty
\frac{(n_1+n_2+k-1)^{k-1}}{(k-1)!}\,t^k
= t\cdot n_1n_2(1+w)e^{(n_1+n_2)w_{-1}},\nn \\
\nu=3: & r^{(n_1,n_2,n_3)}(t) = n_1n_2n_3\sum_{k=2}^\infty
\frac{(n_1+n_2+n_3+k-2)^{k-1}}{(k-2)!}\,t^k
=t^2\cdot n_1n_2n_3(1+w)^2(N+w)e^{Nw_{-1}},
\nn \\
\nu=4: & r^{(n_1,n_2,n_3,n_4)}(t) = n_1n_2n_3n_4\sum_{k=3}^\infty
\frac{(n_1+n_2+n_3+n_4+k-3)^{k-1}}{(k-3)!}\,t^k
= \\
& \ \ \ \ \ \ \ \ \ \ \ \ \ \ \ \ \ \ \ \ \ \
=t^3\cdot n_1n_2n_3n_4(1+w)^3\Big(N^2+(3N+1)w+3w^2\Big)
e^{Nw_{-1}},
\\
\nu=5: & r^{(n_1,n_2,n_3,n_4,n_5)}(t) = n_1n_2n_3n_4n_5
\sum_{k=4}^\infty
\frac{(N+k-4)^{k-1}}{(k-4)!}\,t^k
= \\
& \ \ \ \ \ \ \ \ \
=t^4\cdot n_1n_2n_3n_4n_5(1+w)^4\Big(N^3 + (6N^2+4N+1)w +
(15N+10)w^2 +15w^3\Big)
e^{Nw_{-1}},
\\
\nu=6: & r^{(n_1,n_2,n_3,n_4,n_5,n_6)}(t) = n_1n_2n_3n_4n_5n_6
\sum_{k=5}^\infty
\frac{(N+k-5)^{k-1}}{(k-5)!}\,t^k
= \\
&
\ \ \ \ \ \ \ \ \ \ \ 
=t^5\cdot n_1n_2n_3n_4n_5n_6(1+w)^5\Big(N^4
+ ( 10N^3 + 10N^2 + 5N + 1)w + \nn \\
& \ \ \ \ \ \ \ \ \ \ \ \ \ \ \ \ \ \ \ \ \ \
\ \ \ \ \ \ \ \ \ \ \ \ \ \ \ \ \ \ \ \ \ \
+(45N^2 + 60N + 25)w^2 +
(105N+105)w^3 + 105w^4\Big)e^{Nw_{-1}},
\\
& \ldots
\end{array}
\ee
with $N = n_1+n_2+\ldots+n_\nu$.
These "$(\nu+1)$-point functions" are the main building blocks
in the diagram technique.
Knowing them, one can proceed to more
complicated problems, for example, to double-root tree
diagrams. They arise when one vertex of another kind,
coming from the operator $\hat W_0^+ = \frac{1}{2}\sum_{i,j}
(i+j)p_ip_j\partial_{i+j}$, is allowed.
Thus, what we need is a linear-in-$\hat W^+_0$ term in
\be
e^{t(\hat W^-_0 + \hat W^+_0)}e^{p_1} =
\left(e^{t\hat W^-_0} + \int_0^1
e^{st\hat W^-_0}t\hat W^+_0 e^{(1-s)t\hat W^-_0}ds + \ldots
\right)e^{p_1} = \nn \\
= \left(R(t) + \frac{t}{2}  \sum_{i,j}(i+j) \int_0^1 ds\
e^{st\hat W^-_0} p_ip_j \frac{\partial}{\partial p_{i+j}}
R\big((1-s)t\big) + \ldots\right)e^{p_1}
\ee
where the Campbell-Hausdorf formula was used for the exponential:
$$
e^{A+B} = e^A + \int_0^1 e^{sA}Be^{(1-s)A} ds +
\int_0^1 ds_1 \int_0^{1-s_1}ds_2
e^{s_1 A} B e^{s_2 A} B e^{1-s_1-s_2} + \ldots
$$
In order to evaluate the action of the remaining operator
we need $R^{(n)}$.
The generating function for the connected double-rooted trees is
\be \frac{1}{2}\sum_{\nu=0}^\infty\sum_{\mu=0}^\nu
\sum_{{i,j=1}\atop{l_1,\ldots,l_\nu=2}}^\infty (i+j)r_{i+j-1} r_{l_1-1}\ldots
r_{l_\nu-1} \int_0^1\left[(1-s)t\right]^{i+j-1+\sum_{i=1}^\nu (l_i-1)}
R^{(i,l_1,\ldots,l_\mu)}(ts)R^{(j,l_{\mu+1},\ldots,l_\nu)}(ts)tds \label{drt} \ee
see Fig.\ref{droot}, where $\nu=2$ additional trees are explicitly
shown. It is straightforward to check that this expression
reproduces $H_{02}$. In particular, Fig.\ref{p1p3} explicitly shows
six diagrams, contributing to the $p_1p_3$ double-rooted tree.
Note that they are all different, for example, $B$ and $C$
differ by the "time" ordering of vertices, i.e. by the
ordering of operators: $(\hat W_0^-)^2\hat W_0^+\hat W_0^+$
and $\hat W^-_0\hat W^+_0 (\hat W_0^-)^2$ respectively,
moreover these two topologically equivalent diagrams enter with
different combinatorial factors!
This explicit example can serve to illustrate the
peculiarities of Heitler diagram technique as compared
to the more familiar Feynman one.

\begin{figure}
\begin{center}
\unitlength 1mm 
\linethickness{0.4pt}
\ifx\plotpoint\undefined\newsavebox{\plotpoint}\fi
\begin{picture}(156.539,60.472)(-30,20)
\put(50,50){\line(1,0){10}}
\put(30,40){\line(2,1){20}}\put(50,50){\vector(2,1){.2}}
\put(30,60){\line(2,-1){20}}\put(50,50){\vector(2,-1){.2}}
\put(60,50){\line(2,1){10}}
\put(60,50){\line(2,-1){10}}
\put(50,20){\line(0,1){60}}
\put(20,35){\line(2,1){10}}
\put(20,60){\line(1,0){10}}
\put(30,60){\line(2,1){40}}\put(60,75){\vector(2,1){.2}}
\put(30,40){\line(2,-1){40}}\put(60,25){\vector(2,-1){.2}}
\put(60,75){\line(2,-1){10}}
\put(60,25){\line(2,1){10}}
\put(40,65){\line(2,-1){7}}
\put(40,35){\line(2,1){7}}
\put(10,35){\line(1,0){10}}
\put(20,35){\line(2,-1){10}}
\put(56,52){\makebox(0,0)[cc]{$i+j$}}
\put(46,54){\makebox(0,0)[cc]{$i$}}
\put(46,46){\makebox(0,0)[cc]{$j$}}
\put(53,75){\makebox(0,0)[cc]{$i'$}}
\put(53,32){\makebox(0,0)[cc]{$j'$}}
%
%
\end{picture}
\caption{{\footnotesize
Schematic representation of a double-rooted graph
contributing to (\ref{drt}).
The vertical line shows the "time moment" of the action of
operator $\hat W^+$. All the vertices to the right of this line
("before") come with the factor $(1-s)$, all the vertices
to the left ("after") come with the factor $s$.
The right parts of such diagrams are described in (\ref{drt})
by $r$-factors, the left parts by $R$'s.
Explicitly shown is the case of $\nu+1=3$ disconnected branches
evolved to the moment of $\hat W^+$ action, in general the
sum over all $\nu$ should be made.
}}
\label{droot}
\end{center}
\end{figure}
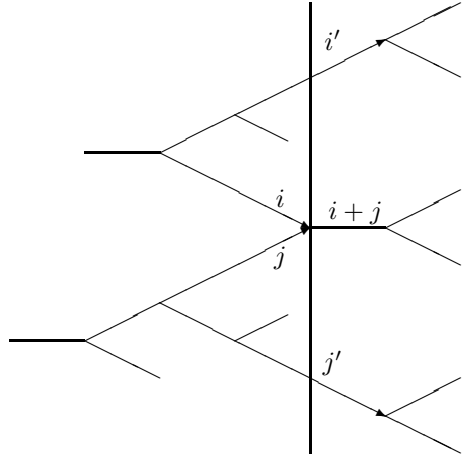

Similarly one can express through $R^{(m,n)}(t)$
the sum of rooted 1-loop diagrams and so on.
This is a straightforward, though somewhat tedious procedure.

\subsection{The claims of \cite{LaKa,Kaz}
\label{Ka}}

\subsubsection{Relation between Hurwitz and
Kontsevich-Hodge free energies}

The Hurwitz free energy (\ref{Hufe}), which is the generating
function of the Hurwitz numbers and
the Kontsevich-Hodge free energy (\ref{KHfe}), which is
the generating function of the Hodge integrals, can be related with the help
of the ELSV formula \cite{ELSV}:
\be
\frac{1}{M!}h(p\,|m_1,\ldots,m_n) =
\prod_{i=1}^n \frac{m_i^{m_i}}{m_i!}
\int_{{\cal M}_{p,n}}\frac{1-\lambda_1 + \lambda_2 -\ldots
\pm \lambda_p}{\prod_{i=1}^n(1-m_i\psi_i)},
\ \ \ \ \ M=2p-2+\sum_{i=1}^n (m_i+1)
\label{ELSV}
\ee
The Hodge integrals
\be
I^{(p)}_q(k_1,\ldots,k_n) =
\int_{{\cal M}_{p,n}}\lambda_q \prod_{i=1}^n \psi_i^{k_i}
\ee
do not vanish only provided
\be
q + \sum_{i=0}^n (k_i-1) = 3p-3
\ee
This implies \cite{Kaz} that, when (\ref{ELSV}) is multiplied
by $u^{3m}$, the $u$-factors can be redistributed as follows:
\be
\frac{u^{3M}}{M!}h(p\,|m_1,\ldots,m_n) =
\prod_{i=1}^n \frac{m_i^{m_i}}{m_i!}u^{3m_i+1}
\int_{{\cal M}_{p,n}}\frac{1-u^2\lambda_1 + u^4\lambda_2
-\ldots \pm u^{2p}\lambda_p}{\prod_{i=1}^n(1-m_i\psi_i u^2)}
\label{ELSV1}
\ee
Indeed, the total power of $u$ in the integral is
$u^{2(j + \sum k_i)} = u^{6p-6 + 2n}$.
Together with $\prod_i u^{3m_i+1}$ this gives
$u^{6p-6 + 3\sum (m_i+1)} = u^{3M}$, as required.

Converting (\ref{ELSV1}) with $p$-variables, one obtains \cite{Kaz}:
$$
H(p) \ \stackrel{(\ref{Hufe})}{=}\
\sum_{n=0}^\infty \frac{1}{n!} \sum_{p;m_1,\ldots,m_n;M}
\frac{u^{3M}}{M!}h(p\,|m_1,\ldots,m_n) p_{m_1}\ldots p_{m_n}
\delta\left( \sum_{i=1}^n (m_i+1) + 2p-2 - M\right)
\ \stackrel{(\ref{ELSV1})}{=}
$$ $$
= \sum_{p,n=0}^\infty \frac{1}{n!}
\int_{{\cal M}_{p,n}}\!\!\Big(1-u^2\lambda_1 + u^4\lambda_2
-\ldots \pm u^{2p}\lambda_p\Big)
\prod_{i=1}^n \left(\sum_{m_i=1}^\infty
\frac{m_i^{m_i}}{m_i!}\frac{u^{3m_i+1}p_{m_i}}
{(1-m_i\psi_i u^2)}\right) =
$$ $$
= \sum_{q=0}^\infty (-)^q u^{2q}
\sum_{p,n=0}^\infty \frac{1}{n!}
\int_{{\cal M}_{p,n}} \!\!\!\!\!\!\lambda_q \prod_{i=1}^n
\left(\sum_{k=0}^\infty T_k\psi_i^k\right) =
$$
\be
= \sum_{q=0}^\infty (-)^q u^{2q}
\sum_{p,n=0}^\infty \frac{1}{n!}\,
\delta\left(\sum_{i=1}^n(k_i-1)
- (3p-3-q)\right) I^{(p)}_q(k_1,\ldots,k_n)\,
T_{k_1}\ldots T_{k_n}
= \sum_{p\geq q\geq 0} u^{2q} F^{(p)}_q(T)
\label{ELSV3}
\ee
with
\be
T_k = u^{2k+1}\sum_{m\geq 1}^\infty \frac{m^m}{m!}u^{3m}p_m
\label{Tp3}
\ee
This explains the relation (\ref{Tp}) between $T$ and $p$
variables.
The factors $g^{2p}$ are introduced straightforwardly,
as was already demonstrated in s.\ref{introg}.

The restriction $p\geq q$ is an important property of the
Hodge integrals.
It is one of non-trivial things to be checked in analyzing
our main claim (\ref{defpartfun}).

\bigskip

\centerline{
\begin{tabular}{|c|c|c|}
\hline
& Hurwitz & Hodge \\
\hline\hline
curve & a ramified covering of Riemann sphere
& arbitrary curve \\
& with all but one (at $\infty$) critical points simple & \\
\hline
$p$ & genus of the covering & genus of the curve \\
\hline
$n$ & number of different preimages of $\infty$
& number of marked points on a complex curve \\
\hline
$\{m_1,\ldots,m_n\}$ & multiplicities of different
preimages & --- \\
\hline
$M$ & number of simple (double) ramification points & --- \\
\hline
$\psi_i$ & --- & ${\rm Chern}_1$ of the bundle of cotangent lines \\
&& at $i$-th marked point \\
\hline
$\lambda_j$ & --- & ${\rm Chern}_j$ of the bundle of holomorphic
1-forms\\
\hline
\end{tabular}
}

\bigskip

The r.h.s. of eq.(\ref{ELSV3}) can be also rewritten in terms
of Witten's topological correlators \cite{Wico}
\be
I^{(p)}_q =
\int_{{\cal M}_{p,n}}\lambda_q \prod_{i=1}^n \psi_i^{k_i}
= \ \Big<\lambda_q \sigma_{k_1}\ldots \sigma_{k_n}\Big>,
\ \ \ \ \ \
\sum_{i=0}^n (k_i-1) = 3p-3-q
\ee
In this notation
\be
{\cal F} = \sum_{q,m_0,m_1,\ldots} u^{2q}
\Big<\lambda_q \sigma_0^{m_1}\sigma_1^{m_1}\ldots\Big>
\frac{T_0^{m_0}}{m_0!}\frac{T_1^{m_1}}{m_1!}\ldots =
\sum_{q=0}^\infty u^{2q} \left<\lambda_q
\exp \left(\sum_{k=0}^\infty  T_k\sigma_k\right)\right>
\ee

\subsubsection{Interplay between the
$T$, $p$ and $q$ time-variables}

We can now elaborate more on the $T\!-\!p\ $
relation (\ref{Tp3}).
First of all, one can rewrite it recursively and, in terms of
the generating functions,
\be
T_k =
u^{2k+4} \sum_{n=1}^\infty \frac{n^{n+k}}{n!} u^{3n} p_n
= u^{2k+1} \left(\frac{u}{3}\frac{\partial}{\partial u}\right)^k
\!\!\left(\frac{T_0}{u}\right)
\ee
where
\be
\frac{1}{u}T_0 = \sum_{n=1}^\infty \frac{n^n}{n!} u^{3n}p_n =
\oint w(x) dp(x)
\ee
with
\be
dp(x) \equiv \sum_n \frac{p_ndx}{x^{n+1}}
\ee
and
\be
w(x) \equiv \sum_{n=1}^\infty \frac{n^n}{n!}(u^3x)^n
\ee
is an inverse of the function
\be
u^3x = \frac{w}{1+w}\exp\left(-\frac{w}{1+w}\right)
\stackrel{(\ref{w-1})}{=}w_{-1}\exp\left(-w_{-1}\right)
\label{xw}
\ee

$T(p)$ is linear, but not a triangular change of time-variables.
It can be decomposed into upper and lower triangular
transformations.

Expanding powers of $x$ into formal series in $w$,
we introduce a set of expansion coefficients $c_{n|k}$:
\be
(u^3x)^n = \sum_{k\geq n} c_{n|k} w^k
\ee
Then
\be
p_n = \oint x^n dp(x) = \sum_{k\geq n} c_{n|k} u^{3(k-n)} q_k
\ee
where
\be
q_k = u^{-3k}\oint w^k(x) dp(x)
\ee
We can now express the $T$-variables through the $q$-variables.
Obviously,
\be
T_0 = u \oint w(x)dp(x) = u^4 q_1
\ee
Next,
\be
 T_1 = \frac{u^4}{3}\oint \frac{\partial w(x)}{\partial u}dp(x)
\ee
The derivative of $w$ is taken at constant $x$ and
can be obtained by differentiating (\ref{xw}),
\be
u^3 x = \left(\frac{u}{3}\frac{\partial w(x)}{\partial u}\right)
\frac{d}{dw}\left\{
\frac{w\exp\left(-\frac{w}{1+w}\right)}{1+w}\right\}
= \frac{\exp\left(-\frac{w}{1+w}\right)}{(1+w)^3}
\left(\frac{u}{3}\frac{\partial w(x)}{\partial u}\right)
\ee
Substituting (\ref{xw}) at the l.h.s., one gets
\be
\frac{u}{3}\frac{\partial w(x)}{\partial u} = w(1+w)^2
\ee
and
\be
 T_1 = u^3\oint w(1+w)^2dp = u^6q_1 + 2u^9 q_2 + u^{12} q_3
\ee
Next,
\be
T_2 = u^5\frac{u}{3}\frac{\partial}{\partial u}\frac{T_1}{u^3}
= u^5\frac{u}{3}\frac{\partial}{\partial u}\oint w(1+w)^2dp
= u^5\oint w(1+w)^2(1+4w+3w^2)dp = \nn \\
= u^8q_1 + 6u^{11}q_2 + 12u^{14}q_3 + 10u^{17}q_4 + 3u^{20}q_5
\ee
In the same way one can deduce expressions
for all other
\be
 T_k = u^{2k+1}\oint w_k(w)dp,\ \ \ \ w_{k+1}
 \stackrel{(\ref{wmr})}{=}w(1+w)^2{dw_k\over dw}
\label{recre}
\ee
They describe a triangular change of variables $T(q)$,
with $T_k$ being a linear combination of $q_1,\ldots,q_{2k+1}$,
and the recurrent relation (\ref{recre}) can be
immediately rewritten in terms of $q$'s \cite{Kaz}:
\be
 T_{k+1} = u^2
\sum_{m=1}^{2k+1} m(q_m + 2u^3q_{m+1} + u^6q_{m+2})
\frac{\partial T_k}{\partial q_m}
\ee

\subsubsection{Comparing $H$ and ${\cal F}$ expressed through
the $q$-variables}

A few lowest transformations $p(q)$ and $T(q)$
look as follows:
\be\label{pq}
p_1 = q_1 - 2u^3q_2 + \frac{7}{2}u^6 q_3 - \frac{17}{3}u^9q_4
+ \frac{209}{24} u^{12}q_5 - \ldots, \nn \\
p_2 = q_2 - 4u^3q_3 + 11u^6q_4 - \frac{76}{3}u^9q_5 + \ldots,\nn\\
p_3 = q_3 - 6u^3q_4 + \frac{45}{2}u^6q_5 -\ldots, \nn \\
p_4 = q_4 - 8 u^3q_5 + \ldots, \nn\\
p_5 = q_5 + \ldots, \nn \\
\ldots
\ee
and
$$
\begin{array}{ccccccc}
T_0 &  =
& u^4\Big(p_1+2u^3p_2 + \frac{9}{2}u^6 p_3 + \frac{32}{3}u^9p_4
+ \ldots\Big) &
=& u^4q_1, \\
 T_1 &=& u^6\Big(p_1 + 4u^3p_2 + \frac{27}{2}u^6p_3
+ \frac{128}{3}u^9p_4 + \ldots\Big) &=&
u^6q_1 + 2u^9q_2 + u^{12}q_3,  \\
T_2 &=& u^8\Big(p_1 + 8u^3p_2 + \frac{81}{2}u^6p_3
+ \frac{512}{3}u^9p_4 + \ldots\Big) &=&
 u^8q_1 + 6u^{11}q_2 + 12u^{14}q_3 + 10u^{17}q_4 + 3u^{20}q_5,  \\
T_3 &=& u^{10}\Big(p_1 + 16u^3p_2 + \frac{243}{2}u^6p_3
+ \frac{2048}{3}u^9p_4 + \ldots\Big) &=&
u^{10}q_1 + 14u^{13}q_2+61\,u^{16}q_3
 +124u^{19}q_4+ \\
 &&&&+131\,u^{22}q_5+ 70u^{25}q_6+15u^{28}q_7,\\
\ldots
\end{array}
$$
Note once again that $T_m$ are infinite series in terms of
$p$, but are {\it finite} linear combinations of $q$'s.

Given these expressions, one can substitute them into the
Hurwitz and Kontsevich-Hurwitz free energies:
\be
{\cal H}(q) = \left. H(p)\right|_{p\rightarrow p(q)} =
\frac{1}{12}u^9q_2 + \frac{1}{48}u^{12}(2q_3+q_1^2+8q_1^3)
+ \frac{1}{240}u^{15}(q_2 + 60q_1q_2) + \nn \\
+ \frac{1}{1440}u^{18}
(120q_2^2+138q_3+80q_1^3+q_1^2+720q_1q_3+240q_1^4)
+ O(u^{21})
\label{calHu}
\ee
while
\be
{\cal F}(u|q) = F_{0,0}u^4q_1 +
\underline{F_{0,1}}(u^6q_1+2u^9q_2+u^{12}q_2)
+ F_{0,2}u^8q_1 + \ldots + \nn \\
+ \frac{1}{2}F_{0,00}u^8q_1^2 +
F_{0,01}u^{10}q_1(q_1+2u^3q_2+u^{6}q_2)
+ \ldots + \nn \\
+ \frac{1}{6}\underline{F_{0,000}}u^{12}q_1^3 + \ldots + \nn \\
+ u^2\underline{F_{1,0}}u^4q_1 + \ldots
\ee
In the last formula we expanded the free energy ${\cal F}(T)$
into Taylor series in $ T$ and made a substitution $T\to T(q)$.
It is important that many terms are actually absent
in Taylor expansions, because the corresponding derivatives
of $F^{(p)}_q$ are vanishing.
Underlined in the above formula are the terms
with non-vanishing $F$-derivatives.
Thus
\be
F_0 = \left<e^{T_k\sigma_k}\right> =
\underbrace{
\frac{1}{6}T_0^3 + \frac{1}{6}T_0^3T_1 + \frac{1}{6}T_0^3T_1^2
+ \frac{1}{24}T_0^4T_2 + \ldots}_{{\rm genus}\ 0} + \nn \\
+ \underbrace{
\frac{1}{24}T_1 + \frac{1}{48}T_1^2 + \frac{1}{24}T_0T_2
+\frac{1}{72}T_1^3+\frac{1}{12}T_0T_1T_2 + \frac{1}{48}T_0^2T_3
+ \ldots}_{{\rm genus}\ 1} + \nn \\
+\underbrace{
\frac{1}{1152}T_4 + \frac{29}{5760}T_2T_3 + \frac{1}{384}T_1T_4
+ \frac{1}{1152}T_0T_5 + \ldots}_{{\rm genus}\ 2} +
\underbrace{\ldots}_{{\rm higer\ genera}} = \nn \\
= \frac{1}{24}u^6q_1 + \frac{1}{12}u^9q_2 +
u^{12}\left(\frac{1}{6}q_1^3 + \frac{1}{16}q_1^2+\frac{1}{1152}q_1
+\frac{1}{24}q_3\right) + \ldots,
\ee
\be
F_1 = \left<\lambda_1 e^{T_k\sigma_k}\right> = \underbrace{
-\frac{1}{24}T_0 -\frac{1}{24} T_0T_1 - \ldots
}_{{\rm genus}\ 1}\
- \underbrace{\frac{1}{15\cdot 32} T_3 + \ldots}_{{\rm genus}\ 2} +
\underbrace{\ldots}_{{\rm higer\ genera}}
\rightarrow \nn\\
\stackrel{\times u^2}{\longrightarrow} -\frac{1}{24}u^6q_1
-\frac{1}{24} u^{12}q_1^2 - \frac{1}{15\cdot 32}u^{12}q_1 - \ldots,
\ee
\be
F_2 = \left<\lambda_2 e^{T_k\sigma_k}\right> = \underbrace{
\frac{7}{45\cdot 128}T_2 + \ldots}_{{\rm genus}\ 2} +
\underbrace{\ldots}_{{\rm higer\ genera}} \rightarrow \nn \\
\stackrel{\times u^4}{\longrightarrow}
\frac{7}{45\cdot 128}u^{12}q_1
+ \ldots,
\ee
Therefore, one has
\be
{\cal F}(u|q)  =
\frac{1}{12}u^9q_2 + \frac{1}{48}u^{12}(2q_3+q_1^2+8q_1^3)
+ O(u^{15}) = \frac{1}{12}u^9q_2 + O(u^{12}),
\ee
in agreement with (\ref{calHu}).

Note that the agreement is based on non-trivial relations
between different components $F^{(p)}_q$ like
$F_{1,0} = -F_{0,1}$ (to cancel the $u^6$ terms) or
$\frac{1}{1152} - \frac{1}{15\cdot 32} +\frac{7}{45\cdot 128} = 0$
etc.

\subsubsection{Kontsevich-Hurwitz partition function
as a KP $\tau$-function: the need to switch from $T$ to $q$}

Substituting $q_m = u^{-4m}\check q_m$ and taking the limit
$u\rightarrow 0$, one gets \be T_k \stackrel{u\rightarrow
0}{\longrightarrow} (2k+1)!!\,\check q_{2k+1} \ee so that one can
identify $2^k\check q_{2k+1}$ with $\tau_k$ in the Kontsevich model.
Since $\tau_k$ are time-variables of the KdV $\tau$-function,
$\check q_k$ provide a doubled set of variables, natural for
description of KP $\tau$-function. Indeed, as proved in \cite{Kaz},
if expressed through the $q$-variables, ${\cal Z}\Big(T(q)\Big)$ becomes
a KP $\tau$-function.

To derive this claim, one has to start from the obvious fact that
$\exp\Big(H(p)\Big)$ is a KP $\tau$-function in $p$-variables --
simply because it is obtained by the action of a $W$-operator from a
trivial $\tau$-function $e^{p_1}$ and all the generators of
$W_\infty$-algebra belong to ${GL(\infty)}$, which acts on
the universal Grassmannian, considered as a universal moduli space
of Riemann surfaces \cite{UGr,Kn}, and maps KP solutions into KP
solutions. This, however, does not imply that ${\cal
Z}\Big(T(p)\Big)$ is a KP $\tau$-function, since the KP hierarchy is
not invariant w.r.t a generic change of variables. Moreover,
${\cal F}\Big(T(p)\Big)$ coincides not with $H(p)$ but with $H(p) -
H_{01}(p) - H_{02}(p)$, and subtraction of the quadratic function
$H_{02}$ would also violate the KP equations.

However, would the change of times be induced by a change of the spectral parameter,
one still can obtain
a KP $\tau$-function. Indeed, as follows from the theory of
equivalent hierarchies \cite{Shiota,equivhi,Kh}, if one makes a change of the
spectral parameter $\mu\to\tilde\mu(\mu)=\mu +\sum_{k\ge 0}\mu^{-k}$ at
the vicinity of infinity, the times are changed by the following
triangle transformation \cite[eq.(16)]{equivhi}
\be
k\tilde t_k=\sum_l \hbox{Res}_{\mu=\infty}{\mu^{l-1}\over\tilde \mu^k(\mu)}lt_l
\ee
while the $\tau$-function in $\tilde t$-variables is multiplied by the exponential
quadratic in times \cite[eq.(46)]{equivhi}:
\be
\tau (t)=
\exp\left(-{1\over 2} \sum_{kl} Q_{kl}\tilde t_k\tilde t_l\right)
\times \tilde\tau (\tilde t)\\
Q_{kl}=\hbox{Res}_{\mu=\infty}\left\{\tilde\mu^k(\mu){d\left[\tilde\mu^l(\mu)\right]_+
\over d\mu}\right\}
\ee
where $[\ldots]_+$ denotes only positive powers of the power series.

As emphasized in \cite{Kaz}, the change of time variables $p\to q$ is
exactly of this type. Indeed, upon
identification of $1/w$ and $1/x$ related by formula (\ref{xw}) with $\mu$
and $\tilde\mu$ respectively, one immediately reproduces the change of time variables
(\ref{pq})
and the proper $Q=H_{02}$ in (\ref{H2}), the times being identified as $q_k=kt_k$,
$p_k=k\tilde t_k$. Note that the linear part $H_{01}$ of (\ref{H2})
is not reproduced by integrability theory arguments,
since the $KP$ $\tau$-function is defined up to an arbitrary exponential factor
linear in times. Note that, although this is the invariance of the Hirota bilinear
equations, if one would like to preserve some reduction of the KP hierarchy, it is
necessary to choose this exponential properly.

From the point of view of the conjugated Virasoro algebra
(\ref{defvirco}) the need to switch from $T$ to $q$-variables
in order to obtain a KP $\tau$-function is related to the fact
that operator $\hat N_1$ does {\it not} belong to the
${GL(\infty)}$
algebra, which acts on the universal Grassmannian \cite{UGr}.

\section{Conjugated Virasoro constraints (\ref{defvirco})
in the BM approach \label{BMap}}
\setcounter{equation}{0}

\subsection{AMM-Eynard equations}

AMM-Eynard equations rewrite Virasoro constraints in terms of a
spectral complex curve $\Sigma$ in the following form \be
\left\{\left(\frac{1}{g^2}V' + g^2\hat\nabla\right)*
\left(\frac{1}{g^2}V' + g^2\hat\nabla\right) \right\}Z = 0
\label{AMME} \ee Here \be \hat\nabla(z) = \sum_{k=0}^\infty
\zeta_k(z) \frac{\partial}{\partial T_k} \label{nabla} \ee \be V'(z)
= \sum_{k=0}^\infty \tilde T_k v_k(z) \label{Vprime} \ee where
$v_k(z)$ and $\zeta_k(z)$ are the full sets of 1-forms on $\Sigma$,
related by the condition \be \hat\nabla(z) V'(z') = B(z,z')
\label{Berg} \ee where $B(z,z')$ is the Bergmann kernel, i.e.
$(1,1)$ Green function $B(z,z') = <\partial\phi(z)
\partial\phi(z')>$ on $\Sigma$. The star product in (\ref{AMME})
denotes a multiplication map $\Omega_\Sigma\times\Omega_\Sigma
\rightarrow \Omega_\Sigma$ on the space of 1-forms $\Omega_\Sigma$,
\be
(\omega_1*\omega_2)(z) =
\sum_i \oint_{a_i} K(z,z') \omega_1(z')\omega_2(\tilde z') \ee which
represents the projection on the "$-$" part of the Virasoro algebra.
For hyperelliptic curves, which are double coverings of the Riemann
sphere, $\{a_i\}$ is a finite set of ramification points and $\tilde
z$ is the counterpart of $z$ on the other sheet. Then the kernel $K$
is actually a differential of the form $\frac{dz}{d\tilde z'}$,
which is a ratio of the $(1,0)$ Green function on $\Sigma$ (which is
the primitive of the Bergmann kernel w.r.t. the second argument
calculated from $z'$ to $\tilde z'$) and the
Seiberg-Witten-Dijkraaf-Vafa differential \footnote{In simplest
case of the sphere, $\Sigma_H:\ y^2_H(z)=z^2-4S$ corresponding to the Hermitean
one-matrix model \cite{AMM.IM}
$$
K(z,z')= \frac{dz}{dz'}\frac{1}{z-z'}\left(\frac{1}{y_H(z)} -
\frac{1}{y_H(z')}\right)
$$}:
\be K(z,z') = \frac{<\partial\phi(z)\, \phi(z')>-<\partial\phi(z)\,
\phi(\tilde z')>}{\Omega_{DV}(z')-\Omega_{SW}(\tilde z')} \ee See
\cite{AMM.IM} and \cite{CMMV,EO} for details.

Substitution of (\ref{nabla}) and (\ref{Vprime}) into (\ref{AMME})
gives: \be \left( \sum_{k,n\geq 0}(v_k*\zeta_n) \tilde
T_k\frac{\partial}{\partial T_n} + \frac{1}{2}\sum_{k,l\geq
0}(\zeta_k*\zeta_l) \frac{\partial^2}{\partial T_k\partial T_l} +
\frac{1}{2}\sum_{k,l\geq 0} (v_k*v_l) \tilde T_k\tilde T_l +{1\over
2}Tr_*B \right) Z=0 \label{AMME2} \ee where, following B.Eynard,
we have corrected
(\ref{AMME})  by adding the $*$-trace of the
Bergmann kernel (note that this prescription is more than just normal ordering
of \ :$V\hat\nabla$:). Expanding the products of 1-forms into linear
combinations of $\zeta$ (no $v$ will arise due to projection
property of the $*$-product), one obtains a one-dimensional set of
constraints on $Z$. They can be also written as recurrent relations
for the multiresolvents \be \rho^{(p|m)}(z_1,\ldots,z_m) = \left.
\hat\nabla(z_1)\ldots\hat\nabla(z_m)F^{(p)}\right|_{T_k=\delta_{k,1}}
\ee in the following form \be \rho^{(p|m+1)}(z,z_1,\ldots,z_m) =
\frac{1}{2}Tr_* B(\bullet,\bullet) \delta_{p,0}\delta_{m,0} +
\sum_{i=1}^m B(\bullet,z_i)*\rho^{(p|m)}(\bullet,z_{I/i}) + \nn \\ +
\sum_{p_1=0}^p\sum_{J\subset I}
\rho^{(p_1|m_J+1)}(\bullet,z_J)*\rho^{(p_2|m_{I/J}+1)}(\bullet,I/J)
+ \frac{1}{2}Tr_* \rho^{(p-1|m+2)}(\bullet,\bullet,z_1,\ldots,z_m)
\ee They are obtained simply by acting with operators $\hat\nabla$
on (\ref{AMME}) or (\ref{AMME2}) and putting $T_k-\delta_{k,1}=0$
afterwards. The terms with the Bergmann kernel come from the action
of $\hat\nabla$ on $V'$, action on the $V'*V'$ term gives rise to
the trace of the Bergmann kernel. The notation here is hopefully
obvious: the bullets, $\bullet$ mark arguments on which the
$*$-product acts, two points are converted into a single $z$. If
both bullets are arguments in the same function, we call the
corresponding product $*$-trace, $Tr_*$: for, say, $H(z_1,z_2) =
\sum_{m,n} H_{mn}\zeta_m(z_1)\zeta_n(z_2)$ the $*$-trace is
$Tr_*H(\bullet,\bullet) = \sum_{m,n} H_{mn}(\zeta_m* \zeta_n)(z)$.

\subsection{*-calculus on Lambert curve}

Bouchard and Marino \cite{BM} suggested to represent the Virasoro
constraints for the Kontsevich-Hurwitz partition function in the
form of the AMM-Eynard equation with the Lambert curve $x =
(z+1)e^{-z}$ with ramification point at $(x,z) = (1,0)$ in the role
of $\Sigma$ with \be B(z,z') = \frac{dz dz'}{(z-z')^2}
\label{BergLam} \ee and $K(z,z')$ given by the 5-dimensional
Seiberg-Witten-Dijkgraaf-Vafa differential\footnote{To compare with
\cite{BM} note that our $z$ is the same as in that paper (i.e. $z =
y-1$), but $x$ differs by a factor of $e$. } $\Omega_{SW}=-\log y
d\log x= -\log (1+z) d\log x$: \be K(z,z') =
\frac{\frac{dz}{z-z'}-\frac{dz}{z-\tilde z'}} {\log (1+z') -
\log(1+\tilde z')} \frac{1+z'}{z'dz'} \ee Accordingly \be \tilde z =
S(z) = -z+\frac{2}{3}z^2-\frac{4}{9}z^3+\frac{44}{135}z^4
-\frac{104}{405}z^5+\frac{40}{189}z^6-\frac{7648}{42525}z^7
+\frac{2848}{18225}z^8 -\frac{31712}{229635}z^9 +\nn\\
+\frac{23429344}{189448875}z^{10} -\frac{89072576}{795685275}z^{11}
+ \frac{1441952704}{14105329875}z^{12}
-\frac{893393408}{9499507875}z^{13}
+\frac{9352282112}{107417512125}z^{14}
- \ldots \ee is
non-trivial solution of $(z+1)e^{-z} = (S+1)e^{-S}$ in the vicinity
of ramification point. Note that in these notations
$\omega_1(z')\omega_2(\tilde z') =
\check\omega_1(z')\check\omega_2\big(S(z')\big)
\frac{dS(z')}{dz'}(dz')^2$ where $\omega(z) = \check\omega(z)dz$,
i.e. \be <\omega_1|\omega_2>_z\ = dz \oint_{z'=0}
\frac{\frac{1}{z-z'}-\frac{1}{z-\tilde z'}} {\log (1+z') -
\log(1+\tilde z')} \frac{1+z'}{z'}\frac{dS(z')}{dz'}\,
\check\omega_1(z')\check\omega_2\big(S(z')\big)dz' =\
<\omega_2|\omega_1>_z \ee Our star product differs by a factor of
$2$ from that in \cite{BM}.

The next suggestion of \cite{BM} is to take \be \zeta_k(z) =
\frac{zdz}{1+z} \left(-\frac{1+z}{z}\frac{d}{dz}\right)^{n+2}\!\!\!z
\ee i.e.
$$
\zeta_{-1} = dz, \ \ \ \zeta_0 = \frac{dz}{z^2}, \ \ \ \zeta_1 =
\frac{2z+3}{z^4}dz, \ \ \ \zeta_2 = \frac{6z^2+20z+15}{z^6}dz, \ \ \
\zeta_3 = \frac{24z^3+130z^2+210z+105}{z^8}dz,
$$ $$
\zeta_4 = \frac{120 z^4  + 924 z^3  + 2380 z^2  + 2520 z + 945}
{z^{10}}dz, \ \ \ \zeta_5 = \frac{720z^5+ 7308z^4+ 26432z^3
+44100z^2+ 34650z+ 10395} {z^{12}}dz,
$$ $$
\zeta_6 = \frac{5040 z^6  + 64224 z^5  + 303660 z^4  + 705320 z^3 +
866250 z^2 + 540540 z + 135135}{z^{14}}dz,
$$ $$
\zeta_7 = \frac{40320 z^7  + 623376 z^6  + 3678840 z^5  + 11098780
z^4  + 18858840 z^3 + 18288270 z^2  + 9459450 z + 2027025}
{z^{16}}dz, $$
$$\ldots$$
\be \zeta_k = (k+1)!\frac{dz}{z^{k+2}}\Big(1 + O(z^{-1})\Big) \ee
Accordingly, from (\ref{Berg}) and (\ref{BergLam}),
$$
v_0 = dz, \ \ \ v_1 = zdz,\ \ \, v_2 = \frac{z^2-z}{2}dz, \ \ \ v_3
= \frac{2z^3-5z^2+5z}{12}dz, \ \ \ v_4 = \frac{1}{120}(5z^4dz -
15v_2-130v_3),
$$ $$\ldots
$$
\be v_k = \frac{z^kdz}{k!}\Big(1 + O(z^{-1})\Big) \ee

It is now easy to evaluate the products of $v$ and $\zeta$
differentials, the lowest products are listed in Table I. Consequently,
\be
\begin{array}{ccl}
v_k*\zeta_{k-l} &=& 0
\ \ \ {\rm for}\ \ \ l\geq 2\ ,\cr
v_{k}*\zeta_{k-1} &=& \zeta_0\ , \cr
v_k*\zeta_k &=& \frac{2k+1}{3}\zeta_1\ , \cr
v_k*\zeta_{k+1} &=& \frac{(2k+1)(2k+3)}{15}\zeta_2 -
\frac{2(k-1)(k+3)}{5\cdot 27}\zeta_1\ , \cr
v_k*\zeta_{k+2} &=& \frac{(2k+1)(2k+3)(2k+5)}{105}\zeta_3 -
\frac{4(k-1)(k+4)(2k+3)}{27\cdot 35} \zeta_2
+ \frac{2(k-1)(k+4)(2k+3)}{3^5\cdot 35}\zeta_1\ ,\cr
v_k*\zeta_{k+3} &=& \frac{(2k+1)(2k+3)(2k+5)(2k+7)}{945}\zeta_4+\ldots\ ,\cr
&\ldots& \label{previr} \end{array}
\ee
Finally, \be \frac{1}{2}Tr_*
B(\bullet,\bullet) = \frac{1}{24}(\zeta_1-\zeta_0) \ee Note that it
can not be represented simply as $\sum_{n=0}^\infty
(v_n*\zeta_n)(z)$: the sum diverges, but contour integral provides a
self-consistent expression for this quantity, which notably includes
$\zeta_0$, not only $\zeta_1$ (!).

\newpage
{\Large\bf Table I}

\bigskip

\hspace{-2.1cm}\rotate{ {\footnotesize $
\begin{array}{|c|c|c|c|c|c|c|}
\hline
A*B& \zeta_0 & \zeta_1 &
\zeta_2 & \zeta_3 & \zeta_4 & \zeta_5\\
\hline\hline
&&&&&&\\
z^5&0&0&35\zeta_0&{245\over 3}\zeta_1+154\zeta_0&147\zeta_2+{2\cdot
7\cdot 93\over 9}
\zeta_1+&231\zeta_3+{2\cdot 7\cdot 1069\over 15}\zeta_2+\\
&&&&&&\\
&&&&&+120\underline{\zeta_0}&+{4\cdot 31\cdot 107\over 3^2\cdot 5}\underline{\zeta_1}\\
&&&&&&\\
\hline
&&&&&&\\
z^4&0&3\zeta_0&5\zeta_1+26\zeta_0 &7\zeta_2+\frac{32\cdot
17}{9}\zeta_1+24\underline{\zeta_0} &9\zeta_3+\frac{2\cdot
163}{3}\zeta 2+
&11\zeta_4+{512\over 3}\zeta_3-\\
&&&&&&\\
&&&&&+ \frac{4\cdot 35\cdot 13}{27}\underline{\zeta_1}
&+{4\cdot 1627\over 45}\underline{\zeta 2}-{8\cdot 337\over 3^4\cdot 5}\zeta_1\\
&&&&&&\\
\hline
&&&&&&\\
z^3&0&5\zeta_0&\frac{25}{3}\zeta_1 + 6\underline{\zeta_0}
&\frac{35}{3}\zeta_2 + \frac{16\cdot 23}{27}\underline{\zeta_1}
&15\zeta_3 + \frac{2\cdot 547}{45}\underline{\zeta_2}-
&\frac{55}{3}\zeta_4+\frac{16\cdot 107}{45}\underline{\zeta_3}-\\
&&&&&&\\
&&&&&-\frac{4\cdot 103}{5\cdot 81}\zeta_1 &-\frac{4\cdot
103}{27\cdot 5}\zeta_2
+\frac{8\cdot 37}{3^5\cdot 5}\zeta_1 \\
&&&&&&\\
\hline
&&&&&&\\
z^2&\zeta_0&\zeta_1 + 2\underline{\zeta_0}
&\zeta_2+\frac{10}{3}\underline{\zeta_1} &\zeta_3 +
\frac{14}{3}\underline{\zeta_2}-\frac{4}{27}\zeta_1 &\zeta_4 +
6\underline{\zeta_3} - \frac{16}{45}\zeta_2+
& \zeta_5+\frac{22}{3}\underline{\zeta_4}-\frac{28}{45}\zeta_3+\\
&&&&&&\\
&&&&&+ \frac{8}{5\cdot 81} \zeta_1
&+\frac{8}{27\cdot 5}\zeta_2+\frac{32}{3^5\cdot 5}\zeta_1 \\
&&&&&&\\
\hline
&&&&&&\\
z&\underline{\zeta_0}&\underline{\zeta_1}&\underline{\zeta_2}
&\underline{\zeta_3}&\underline{\zeta_4}&\underline{\zeta_5}\\
&&&&&&\\
\hline
&&&&&&\\
1&\frac{1}{3}\underline{\zeta_1} &\frac{1}{5}\underline{\zeta_2} +
\frac{2}{45}\zeta_1 &\frac{1}{7}\underline{\zeta_3} +
\frac{16}{35\cdot 9}\zeta_2- &\frac{1}{9}\underline{\zeta_4} +
\frac{10}{7\cdot 27}\zeta_3- &\frac{1}{11}\underline{\zeta_5} +
\frac{16}{27\cdot 11}\zeta_4 &\frac{1}{13}\underline{\zeta_6}
+\frac{70}{99\cdot 13}\zeta_5-\\
&&&&&&\\
&&&- \frac{8}{35\cdot 81}\zeta_1 &- \frac{4}{27\cdot 35}\zeta_2 -
\frac{8}{3^5\cdot 7}\zeta_1 &-\frac{4\cdot 43}{81\cdot 35\cdot
11}\zeta_3 -\frac{8\cdot 179}{81\cdot 25\cdot 77}\zeta_2 &
-\frac{8\cdot 49}{81\cdot 55\cdot13}\zeta_4
-\frac{8\cdot 71\cdot 157}{3^5\cdot 25\cdot 77\cdot 13}\zeta_3\\
&&&&&&\\
&&&&&+\frac{16\cdot 71}{3^6\cdot 25\cdot 77}\zeta_1 & +
\frac{16\cdot 2447}{3^6\cdot 25\cdot 77\cdot 13}\zeta_2
+\frac{32\cdot 67}{3^8\cdot 11\cdot 13}\zeta_1\\
&&&&&&\\
\hline\hline
&&&&&&\\
\zeta_0 &\frac{1}{15}\underline{\zeta_2}-
&\frac{1}{35}\underline{\zeta_3} - \frac{8}{35\cdot 9}\zeta_2 +
&\frac{1}{7\cdot 9}\underline{\zeta_4}
- \frac{16}{27\cdot 35}\zeta_3
&{1\over 3^2\cdot 11}\underline{\zeta_5}-{5\cdot 16\over 3^4\cdot
7\cdot 11}\zeta_4 - &{1\over 11\cdot 13}\underline{\zeta_6}
-{8\cdot 5\over3^3\cdot 11\cdot 13}\zeta_5-&\\
&&&&&&\\
&-\frac{8}{5\cdot 27}\zeta_1&+\frac{4}{35\cdot 81}\zeta_1&+
\frac{16}{35\cdot 81}\zeta_1&-{4\over 3^5\cdot 5\cdot 11}\zeta_3+&
-{512\over 3^5\cdot 5\cdot 7\cdot 11\cdot 13}\zeta_4+{2^3\cdot
17\cdot 167\over
3^6\cdot 5^2\cdot 11\cdot 13}\zeta_3&\hbox{{\Large not}}\\
&&&&&&\\
&&&&{8\cdot 17\cdot 71\over 3^6\cdot 5^2\cdot 7\cdot
11}\zeta_2-{32\cdot 179\over 3^8\cdot 5^2\cdot 7\cdot 11}\zeta_1&
-{2^4\cdot 47\over 3^7\cdot 5\cdot 7\cdot 13}\zeta_2-{2^7\cdot
10099\over
3^9\cdot 5^2\cdot 7\cdot 11\cdot 13}\zeta_1&\\
&&&&&&\\
\hline
&&&&&&\\
\zeta_1&&\frac{1}{105}\underline{\zeta_4} -\frac{2}{9\cdot
35}\zeta_3 + &{1\over 3\cdot 7\cdot 11}\underline{\zeta_5}- {32\over
3^3\cdot 5\cdot 7\cdot 11}\zeta_4&{1\over 3\cdot 11\cdot
13}\underline{\zeta_6}
-{5^2\cdot 2\over 3^3\cdot 7\cdot 11\cdot 13}\zeta_5&&\\
&&&&&&\\
&&+\frac{4}{35\cdot 25}\zeta_2 -\frac{16}{81\cdot
35}\zeta_1&+{8\over 3^4\cdot 7\cdot 11}\zeta_3- {2^4\cdot 67\over
3^5\cdot 5^2\cdot 7\cdot 11}\zeta_2&+{2^3\cdot 139\over 3^5\cdot
5\cdot 7\cdot 11\cdot 13}\zeta_4-{2^4\cdot 3^3\cdot 5\cdot 23\over
3^6\cdot 5^2\cdot 7\cdot 11\cdot 13}\zeta_3&\hbox{{\Large enough}}&
\hbox{{\Large space}}\\
&&&&&&\\
&&&+{2^4\cdot 29\over 3^7\cdot 5^2\cdot 7\cdot 11}\zeta_1&-{2^4\cdot
37\over 3^5\cdot 5^2\cdot 7\cdot 11\cdot 13}\zeta_2+{2^7\cdot
103\over 3^7\cdot 5^2
\cdot7\cdot 13}\zeta_1&&\\
&&&&&&\\
\hline
\end{array}
$}}

\newpage

\noindent

\subsection{AMM-Eynard equations for Lambert curve}

All this implies that the r.h.s. of the AMM-Eynard equations for the
Lambert curve and the BM choice of $\{\zeta_n\}$ has the form \be
\sum_{m=0}^\infty \frac{2^{m-1}}{(2m+1)!!} \zeta_{m}\hat{\cal
M}^{BM}_{m-1} {\cal Z}_{BM} = 0 \ee where \be \hat{\cal M}^{BM}_{-1}
= \sum_{k=1}^\infty \tilde T_k\frac{\partial}{\partial T_{k-1}} +
\frac{1}{2} T_0^2 -\frac{1}{24} =
L^K_{-1} - \frac{1}{24} \nn \\
2\hat{\cal M}^{BM}_{0} =
2\left\{ \hat L^K_0 - \frac{1}{3} \left(\frac{16}{15}\hat L^K_1-\hat
N_1\right) + \frac{2}{5\cdot 81}\left(\frac{8}{7}\hat L^K_2-\hat
N_2\right)
+ \ldots\right\}  \nn \\
4\hat{\cal M}^{BM}_1 = 4\hat L^K_1 - \frac{64}{9\cdot 7} \hat L^K_2
+ \frac{8}{9}\hat N_2= 4\left\{\hat L_1^K -
\frac{2}{9}\left(\frac{8}{7}\hat L^K_2-\hat N_2\right)
+ \ldots\right\} \nn \\
\ldots \ee The constant shift by $1/24$ in ${\cal M}_{-1}^{BM}$
comes from the {\it anomalous} contribution with $\zeta_0$ to
$\frac{1}{2}Tr_*B(\cdot,\cdot)$. The $\zeta_1$-term in the same
trace contributes the usual $3/24=1/8$ to $2L^K_0$. The shifted
times are $\tilde T_k = T_k - \delta_{k,1}$. The new operators at
the r.h.s. are: \be \hat N_1 = \sum_{k=0}^\infty (k+1)^2\tilde T_k
\frac{\partial}{\partial T_{k+1}}, \nn \\
\hat N_2 = \sum_{k=0}^\infty (k+1)(k+3/2)(k+2)\tilde T_k
\frac{\partial}{\partial T_{k+2}},\nn\\
\ldots \ee Higher $\hat N$-operators can will also contain terms
with second derivatives.

Indeed, picking up the terms with $\frac{1}{3}\zeta_1$ in
(\ref{previr}) one obtains
$$
2\hat{\cal M}_{0}^{BM} = \underbrace{\frac{1}{8} + \sum_{k=0}^\infty
(2k+1)\tilde T_k\frac{\partial}{\partial T_k}}_{2\hat L_0^K} -
\frac{2}{45}\sum_{k=0}^\infty (k-1)(k+3)\tilde
T_k\frac{\partial}{\partial T_{k+1}}
-\frac{4}{45}\frac{\partial^2}{\partial T_0^2} + \ldots = $$ $$ =
2\hat L_0^K - \frac{8}{45}\underbrace{\left( \sum_{k=0}^\infty
(2k+1)(2k+3)\tilde T_k\frac{\partial}{\partial T_{k+1}}
+\frac{1}{2}\frac{\partial^2}{\partial T_0^2} \right)}_{4\hat L_1^K}
+ \frac{2}{45}\underbrace{\sum_{k=0}^\infty
\Big(4(2k+1)(2k+3)-(k-1)(k+3)\Big) \tilde
T_k\frac{\partial}{\partial T_{k+1}}}_{15\hat N_1} + \ldots = $$ \be
= 2\hat L_0 - \frac{32}{45}\hat L_1^K + \frac{2}{3}\hat N_1 + \ldots
\ee
Similarly, the next contribution to $2{\cal M}^{BM}_0$ is \be
\frac{1}{35\cdot 81} \left( \sum_k 2(k-1)(k+4)(2k+3) \tilde
T_k\frac{\partial}{\partial T_{k+2}} + 12 \frac{\partial^2}{\partial
T_0\partial T_1}\right) = \nn \\ = \frac{4}{35\cdot
81}\underbrace{\left( \sum_k (2k+1)(2k+3)(2k+5) \tilde T_k
\frac{\partial}{\partial T_{k+2}} + 3 \frac{\partial^2}{\partial
T_0\partial T_1}\right)}_{8\hat L_2^K} - \frac{2}{5\cdot
81}\underbrace{\sum_k (k+1)(k+2)(2k+3)\tilde T_k
\frac{\partial}{\partial T_{k+2}}}_{2\hat N_2} \ee Further, the next
contribution to ${\cal M}_0^{BM}$ (i.e. to the coefficient in front
of $\frac{1}{3}\zeta_1$) will be \be -{2^3\cdot 5\over 35\cdot
81}\tilde T_0\frac{\partial}{\partial T_{3}} + 0\cdot\tilde
T_1\frac{\partial}{\partial T_{4}}+ {2^4\cdot 7\over 35\cdot
81}\tilde T_2\frac{\partial}{\partial T_{5}}+ {2^5\cdot 11\over
35\cdot 81}\tilde T_3\frac{\partial}{\partial T_{6}}+\ldots+ 2\cdot
3\cdot\frac{16}{35\cdot 81}\frac{\partial^2} {\partial T_0\partial
T_2} - 3\cdot\frac{16}{35\cdot 81} \frac{\partial^2}{\partial T_1^2}
\ee while \be 16\hat L_3^K = \sum_k (2k+1)(2k+3)(2k+5)(2k+7) \tilde
T_k\frac{\partial}{\partial T_{k+3}} + 15\frac{\partial^2}{\partial
T_0\partial T_2} + \frac{9}{2}\frac{\partial^2}{\partial T_1^2} \ee
If one arranges to eliminate the $T_0T_{m-1}$ derivative from $\hat
N_m$, then $\hat N_3$ will be a combination of $\hat n_3$ and
$\frac{\partial^2}{\partial T_1^2}$.

Collect now the terms with $\frac{1}{15}\zeta_2$:
$$
4\hat{\cal M}^{BM}_1 = \underbrace{ \sum_k (2k+1)(2k+3)\tilde
T_k\frac{\partial}{\partial T_{k+1}} +
\frac{1}{2}\frac{\partial^2}{\partial T_0^2}}_{4\hat L^K_1} -
\frac{4}{9\cdot 7} \left(\sum_k (k-1)(k+4)(2k+3) \tilde
T_k\frac{\partial}{\partial T_{k+2}} + 6 \frac{\partial^2}{\partial
T_0\partial T_1}\right) + \ldots = $$ \centerline{\footnotesize $ =
4\hat L^K_1 - \frac{8}{9\cdot 7} \underbrace{\left( \sum_k
(2k+1)(2k+3)(2k+5) \tilde T_k \frac{\partial}{\partial T_{k+2}} + 3
\frac{\partial^2}{\partial T_0\partial T_1}\right)}_{8\hat L_2^K}
+\frac{4}{9\cdot 7}\underbrace{ \sum_k
\Big(2(2k+1)(2k+3)(2k+5)-(k-1)(k+4)(2k+3)\Big) \tilde
T_k\frac{\partial}{\partial T_{k+2}} }_{14\hat N_2} + \ldots $ } \be
=4\hat L^K_1 - \frac{64}{9\cdot 7} \hat L^K_2 + \frac{8}{9}\hat N_2
+ \ldots \ee

\subsection{New algebra}

Thus, we are led to a new set of the operator $\hat N_1$ and its
descendants $\hat N_2, ...$ produced by its commuting with the
Virasoro algebra. Together with the Virasoro algebra, they form an
extended algebra of operators $\hat A_m$. Denote the linear (in
derivatives) part of these operators through $\hat a_m =\sum_k
P_m(k) \tilde T_k\frac{\partial}{\partial T_{k+m}}$. The polynomials
$P_m(k)$ are \be
\begin{array}{cl}
\hat L_{-1}: & 1 \cr
\hat L_0: & \frac{1}{2}(2k+1) \cr
\hat L_1: & \frac{1}{4}(2k+1)(2k+3) \cr
\ldots & \cr
\hat N_1: & (k+1)^2 \cr
\hat N_2: & 2(k+1)(k+{3\over 2})(k+2) \cr
\ldots &
\end{array}
\ee One has for the commutator $[\hat a_m,\hat a_n]$ \be P_{m+n}(k)
= P_m(k)P_n(k+m) - P_n(k)P_m(k+n) \ee Using this rule, one obtains
\be
\begin{array}{clc}
\left[\hat N_1, \hat L^K_{-1}\right] &
= 2\left(\hat L^K_0-\frac{1}{16}\right),& \cr
\left[\hat N_1, \hat L^K_{0}\right]  & = \hat N_1,& \cr
\left[\hat N_1, \hat L^K_{1}\right] & = \frac{2}{3}\Big(\hat N_2 -
\hat L_2^K\Big),&
\ \ \ {\rm (quadratic\ pieces\ also\ match)}\cr
& \ldots & \cr
\left[\hat N_2, \hat L^K_{-1}\right] & =
3 \hat N_1 & \cr
\left[\hat N_2, \hat L^K_{0}\right] & = 2\hat N_2,&
\end{array}
\ee The
commutator \be \begin{array}{cll}\left[\hat N_1,\hat N_2\right]& = -2\sum_k
(k+1)(k+2)^2(k+3)\tilde T_k \frac{\partial}{\partial T_{k+3}}& =\hat
N_3 \end{array}\ee gives rises to a new operator $\hat N_{3}$ etc.

\subsection{Introduction of $u^2$}

Let us manifestly restore in the formulas the dependence on the
deformation parameter $u$.

The terms of order $u^0$ are underlined in the table. Those to the
right come with higher powers of $u^2$. Those to the left come with
powers of $1/u^2$, but they are eliminated in linear combinations of
$z^k$, needed to produce $v_k$.

\subsection{Commutation relations\label{trM}}

For \be \hat {\cal M}_{-1} = \hat L_{-1} - \frac{1}{24}u^2, \ee and
\be 2\hat {\cal M}_0 = 2\hat L_0 + u^2(\alpha_{01} \hat L_1-\nu_{01}
\hat N_1) + u^4(\alpha_{02} \hat L_2 - \nu_{02} \hat N_2) + \ldots
\ee one has \be \left[\hat {\cal M}_0, 2\hat{\cal M}_{-1}\right] =
2\hat L_{-1} + u^2\Big(2\alpha_{01}\hat L_0 - \nu_{01}[\hat N_1,\hat
L_{-1}]\Big) + u^4\Big(3\alpha_{02}\hat L_1 - \nu_{02}[\hat N_2,\hat
L_{-1}]\Big) + \ldots \label{0-1comm} \ee Now the point is that,
since \be \left[\hat N_1,\hat L_{-1}\right] = \sum_{k=0}^\infty
(2k+1)T_k\frac{\partial}{\partial T_k} = 2\hat L_0 - \frac{1}{8} \ee
the $u^2$ term at the r.h.s. of (\ref{0-1comm}) is actually
$u^2\left(\frac{\nu_{01}}{8} + 2(\alpha_{01}-\nu_{01})\hat
L_0\right)$, i.e. one may expect the r.h.s. of (\ref{0-1comm}) is
actually such that \be 2\left[\hat {\cal M}_0, \hat{\cal
M}_{-1}\right] = 2\underbrace{\left(\hat  L_{-1} -
\frac{1}{24}u^2\right)}_{\hat{\cal M}_{-1}} + {\rm const}\cdot
u^2\hat {\cal M}_0 + {\rm const}\cdot u^4 \hat{\cal M}_1 + \ldots
\ee and the operators $\hat {\cal M}_m$ form a closed algebra. This
requires a conspiracy of the coefficients, say, \be \frac{1}{8}\nu_{01} =
-\frac{2}{24}, \ \ \ {\rm i.e.}\ \ \nu_{01} = -\frac{2}{3} \ee which
is indeed the case.

\subsection{From ${\cal M}_n$ to Virasoro algebra}

An important property of this closed algebra of operators ${\cal
M}_k$ is that it can be converted, with a triangular transformation,
into the Virasoro algebra: \be \hat {\cal L}_{-1} = \hat {\cal
M}_{-1}
= \hat L_{-1} - \frac{u^2}{24}, \nn \\
\hat {\cal L}_0 = \hat {\cal M}_0 +\frac{2u^2}{45}\hat {\cal M}_1 +
0\cdot u^4 \left(\frac{8}{7}\hat L_2 - \hat N_2\right) +
 \ldots =
\hat L_0 - \frac{u^2}{3}(\hat L_1 - \hat N_1),\nn \\
\ldots \ee Then \be \left[\hat {\cal L}_0, \hat {\cal L}_{-1}\right]
= \hat L_{-1} - \frac{2u^2}{3}\hat L_0 + 2\left(\hat
L_0-\frac{1}{16}\right) + \ldots = \hat L_{-1} - \frac{u^2}{24} =
\hat {\cal L}_{-1} \ee Furthermore, it looks like \be \hat {\cal
L}_m = \hat U \hat L_m \hat U^{-1} \ee where \be \hat U = \exp
\left\{-\frac{u^2}{3}(\hat L_1^K-\hat N_1)+O(u^6)\right\} =
\exp\left\{-\frac{u^2}{12}\left(\sum_{k=0}^\infty \tilde
T_k\frac{\partial}{\partial T_{k+1}} + \frac{1}{2}
\frac{\partial^2}{\partial T_0^2}\right)+O(u^6)\right\} \ee Indeed,
\be \hat U \hat L_{-1} \hat U^{-1} = \hat L_{-1} -
\frac{u^2}{3}[(\hat L_1-\hat N_1),\hat L_{-1}] +
\frac{u^4}{18}\Big[(\hat L_1-\hat N_1),
[(\hat L_1-\hat N_1),\hat L_{-1}]\Big] + \ldots = \nn \\
= \hat L_{-1} - \frac{u^2}{3}\left(2\hat L_0 - 2\Big(\hat
L_0-\frac{1}{16}\Big)\right) + 0 = \hat L_{-1} - \frac{u^2}{24} =
\hat{\cal L}_{-1} \ee -- since the first commutator is a $c$-number,
all multiple commutators automatically vanish.

Similarly, \be \hat U \hat L_{0} \hat U^{-1} = \hat L_{0} -
\frac{u^2}{3}[(\hat L_1-\hat N_1),\hat L_{0}] +
\frac{u^4}{18}\Big[(\hat L_1-\hat N_1),
[(\hat L_1-\hat N_1),\hat L_{0}]\Big] + \ldots = \nn \\
= \hat L_{0} - \frac{u^2}{3}(\hat L_1 - \hat N_1) + 0 = \hat L_{-1}
- \frac{u^2}{3}(\hat L_1 - \hat N_1) = \hat{\cal L}_0 \ee -- again
the form of the first commutator implies the vanishing of all
multiple commutators.

Then, since the Kontsevich partition function $Z_K$ is annihilated
by $\hat L_{m\geq -1}$, while its Kontsevich-Hurwitz deformation
${\cal Z}$ by ${\cal M}_m$, one has \be {\cal Z}(u) = \hat U Z_K \ee
This implies a series of relations. Indeed, \be {\cal Z} = \left(1 -
\frac{u^2}{3}(\hat L_1^K-\hat N_1) + \frac{u^4}{18}(\hat L_1^K-\hat
N_1)^2 + \ldots\right)Z_K = Z_K + \frac{u^2}{3}\hat N_1 Z_K +
\frac{u^4}{18}\left(\frac{2}{3}\hat N_2 + \hat N_1^2\right)Z_K +
\ldots \ee We used here the fact that $\hat L_{m\geq -1}$ annihilate
$Z_K$. In other words, one should have \be
F_1 = \frac{1}{3}\hat N_1 F_0,\\
F_2 +\frac{1}{2}F_1^2 =
\frac{1}{18}\left(\frac{2}{3}\hat N_2F_0 + \hat N_1^2 F_0 + (\hat
N_1 F_0)^2\right),  \ \ \ {\rm i.e.} \ \ \ F_2 = \frac{1}{18}
\left(\frac{2}{3}\hat N_2F_0 + \hat N_1^2 F_0\right)
\nn \\
\ldots \ee In particular, all low-genus contributions should vanish.
For instance, one can check that \be \hat N_1 F^{(0)}_0 = \hat N_2
F^{(0)}_0 = 0 \label{annF00} \ee and \be \left(\frac{2}{3}\hat N_2 +
\hat N_1^2\right)F^{(1)}_0 = 0 \ee
\be F_1^{(1)} = \frac{1}{3}\hat N_1F_0^{(1)} \ee Since at the same
time \be F^{(1)}_1 = -\frac{1}{24}\partial^2_0F^{(0)}_0 \ee we
obtain an identity relating different genera of the Kontsevich
partition function: \be \hat N_1 F_0^{(1)} =
-\frac{1}{8}\partial^2_0 F^{(0)}_0 \ee which supplements the first
one in (\ref{annF00}).

The next similar relations are \be \hat
N_1 F_1^{(1)} = \frac{1}{12}\partial^2_{01} F^{(0)}_0 \ee \be
F^{(2)}_2 =
\frac{1}{18} \left(\frac{2}{3}\hat N_2 + \hat N_1^2\right)F^{(2)}_0
\ee etc.

Note that action of the operator $\hat {\cal L}_0 = \hat  L_0 -
\frac{u^2}{3}(\hat L_1-\hat N_1)$ on ${\cal Z} = \hat U Z_K$ can be
represented by action of the operator \be \hat L_0 + u^2\partial_{u^2}
\ee
which automatically guarantees that
\be
\left[\hat L_0+u^2\partial_{u^2},\hat L_{-1}-{u^2\over 24}\right]=
\hat L_{-1}-{u^2\over 24}
\ee
However, {\it such} representation does not continue to higher $\hat
L_{m\geq 1}$.

\subsection{Annihilators of $F^{(0)}_0$}

Relation (\ref{annF00}) implies that there is a whole family of
operators annihilating $F^{(0)}_0$. First of all, for each $m\geq 0$
there is a linear (without second derivatives) operator $\hat {\cal
N}_m$: $\hat {\cal N}_0 = \hat L_0^K$, $\hat {\cal N}_1 = \hat N_1$,
$\hat {\cal N}_2 = \hat N_2$, $\hat {\cal N}_3 = [\hat N_2,\hat N_1]
=2 \sum_k (k+1)(k+2)^2(k+3)\tilde T_k\partial_{k+3}$ and so on. They
all begin from \be \hat {\cal N}_m \sim T_0\partial_m +
(m+3)(T_1-1)\partial_{m+1} + \ldots \label{firt} \ee To illustrate
how the other potential linear annihilators disappear, let us
consider the level $3$: There are three other annihilators of
$F_0^{(0)}$ at this level: \be 1\cdot 8 \cdot [\hat N_1,\hat L_2] =
-2\sum_k (2k+3)(2k+5)(k^2+4k+1)\tilde T_k \partial_{k+3}
- 3\Big(\partial_1^2 + 4\partial^2_{02}\Big), \nn \\
2\cdot 4\cdot [\hat N_2, \hat L_1] = 2\sum_k
(2k+3)(2k+5)(k+2)^2\tilde T_k \partial_{k+3}
- 6\partial^2_{02}, \nn \\
16\hat L_3 = \sum_k (2k+1)(2k+3)(2k+5)(2k+7) \tilde T_k
\partial_{k+3} + \frac{9}{2}\partial_1^2 + 15\partial^2_{02} \ee
Quadratic derivatives cancel in a certain linear combination of
these three lines, at the same time the linear part contains a
factor $(2k+3)(2k+5)$, which can not be made consistent with
(\ref{firt}). This implies that this linear combination should
vanish identically. Indeed, \be -12[\hat N_1,\hat L_2] + 4[\hat
N_2,\hat L_1] -16\hat L_3
= 0 \ee

When operator contains second time-derivatives, it acts on
$F^{(0)}_0$ non-linearly: $\partial^2 \rightarrow \partial F
\partial F$. Adding such non-linear annihilators we obtain $1+{\rm
entier}\left(\frac{m+1}{2}\right)$ annihilators of $F^{(0)}_0$ at
each level $m$. For $m=1$ these are $\hat N_1$ and $\hat L_1$, for
$m=2$: $\hat N_2$ and $\hat L_2$, for $m=3$: $\hat{\cal N}_3\sim
[\hat N_1,\hat N_2]$, $[\hat L_1,\hat N_2]$ and $\hat L_3$ (the
fourth potential candidate, $[\hat L_2, \hat N_1]$ is a linear
combination of the last two, as we already know).

\section{Conclusion}
\setcounter{equation}{0}

To conclude, we demonstrated, at the level of convincing
evidence rather than a rigorous proof,
that the Kontsevich-Hurwitz partition function is annihilated
by the Virasoro generators (\ref{defvirco}),
which differ from the continuous Virasoro constraints by
a conjugation.

Therefore, the KH partition function ${\cal Z}$ is now known to
possess the following properties:

${\bf A.}$
It is a generating function for the Hodge integrals \cite{G1,Hodge}
\be
I_q^{(p)}(k_1,\ldots,k_m) \sim \int_{\overline{{\cal M}}_{p,m}}
\lambda_q\psi_{k_1}\ldots\psi_{k_m}
\ee

${\bf B.}$
A change of time-variables $T \rightarrow T(p)$
\be
T_k = u^{2k+1}\sum_{n=1}^\infty \frac{n^{n+k}}{n!}u^{3n} p_n
\label{Tpc}
\ee
converts ${\cal Z}$ into the Hurwitz partition function:
\be
{\cal Z}\Big(T(p)\Big) =
e^{H(p)} = e^{u^3\hat W_0}e^{p_1},
\label{Hpc}
\ee
where
\be
\hat W_0 = \sum_{m=0}^\infty p_m \hat V_m
= \frac{1}{2}\sum_{i,j\geq 1}\left(
(i+j)p_ip_j\frac{\partial}{\partial p_{i+j}}
+ ijp_{i+j}\frac{\partial^2}{\partial p_i\partial p_j}\right)
\ee
and $\hat V_m$ are the ``discrete Virasoro" operators ($p_k=kt_k$)
\be
\hat V_m = \sum_{k=0}^\infty
(k+m)p_k\frac{\partial}{\partial p_{k+m}}
+ \sum_{i+j=m} ij\frac{\partial^2}{\partial p_i\partial p_j}
\ee

${\bf C.}$
For any fixed $u$ and $g$, the partition function ${\cal Z}$
is a KP $\tau$-function.

${\bf D.}$
Associated multidensities (the generating functions of
certain subsets of coefficients in ${\cal F}$) satisfy the
AMM-Eynard equations on the Lambert spectral curve
$x = (z+1)e^{-z}$.

${\bf E.}$ ${\cal Z}$ is obtained from $Z_0 = e^{F_0}$ by the
explicit $u^2$-dependent transformation \be {\cal Z}(u,g) = \hat
U(u,g) Z_K(g) \label{calZvsZ} \ee where \be \hat U =
\exp\left\{\frac{u^2}{3} \Big(\hat N_1 - \hat
L_1\Big)+O(u^6)\right\} = \exp\left\{\frac{u^2}{12}\left(
\sum_{k=0}^\infty T_k\frac{\partial}{\partial T_{k+1}} -
\frac{g^2}{2}\frac{\partial^2}{\partial T_0^2}\right)+O(u^6)
\right\} \ee and, consequently, it satisfies the Virasoro
constraints \be \hat {\cal L}_{m\geq -1}{\cal Z} = 0 \ee with \be
\hat {\cal L}_m = \hat U \hat L_m \hat U^{-1} \label{calLvsL} \ee
where $\hat L_{m\geq -1}$ are the usual ''continuous Virasoro''
operators \cite{covirco}, annihilating the Kontsevich partition
function $Z_K$. This can be considered as a deformation of the
Virasoro sub-algebra induced by the constant shift of the lowest
$\hat L_{-1}$: \be \hat{\cal L}_{-1} = \hat L_{-1} - \frac{u^2}{24}
\ee and generated by the new important operator \be \hat N_1 =
\sum_{k=0}^\infty (k+1)^2\tilde T_k \frac{\partial}{\partial
T_{k+1}} \ee which annihilates the genus-zero Kontsevich free
energy, \be \hat N_1 F_0^{(0)} = 0 \ee

\bigskip

The property {\bf A} is the original problem, addressed
by E.Witten and M.Kontsevich (at $q=0$). It enters \cite{LaKa,Kaz}
through the celebrated ELSV formulas.

The property {\bf B} refers to representation of $H(p))$
through the action of "cut-and-join"
operator $\hat W_0$, which was found in
\cite{G}. The relevant change of variables $T(p)$ is described
in \cite{Kaz}, see also \cite{G1}.

The property {\bf C} for the original Kontsevich
model (at $u^2=0$, when the $\tau$-function actually
belongs to a narrow KdV class) was proved in \cite{Ko,MMM,GKM}
and was later studied by
numerous different methods.
For arbitrary $u^2$ it was proved by M.Kazarian.
This is a non-trivial generalization from the $u^2=0$ case,
in particular, the number of time variables in KP
$\tau$-function is effectively doubled as compared to
the KdV one, and appropriate time-variables for $u^2\neq 0$
are actually different from $T$ (they are called $q$-variables
in \cite{Kaz}).
In fact, as we explained in s.4.3.4 above,
once {\bf B} is known, {\bf C}
is a simple direct corollary of the old theory of
equivalent hierarchies \cite{equivhi,Kh}.

The property {\bf D} was conjectured by
V.Bouchard and M.Marino in \cite{BM}. They conjectured that the constraints are
indeed quadratic (and thus reduced to the Virasoro, but not to some $W$-algebra)
and associated with the Lambert curve, they also introduced the basis of
$\zeta$-differentials.

The property {\bf E} has been our main concern in this paper.
Note that the very fact
that the generic Hodge integrals are somehow expressed through the
intersection numbers, i.e. that the Kontsevich-Hurwitz partition
function should be expressed through the Kontsevich one,
is well known since \cite{Fab}
(based on earlier results due to D.Mumford). Our goal was to
make this relation as explicit as possible.

We demonstrated in s.3 that the twisting (\ref{defpartfun})-(\ref{Uop})
of the Kontsevich partition function {\it a la} \cite{AMM.IM}
immediately reproduces (\ref{F0})-(\ref{F4})
and explained in ss.4 and 5 how this fact is related to the previous
works \cite{LaKa,Kaz,BM}.
We do not provide rigorous proofs in this paper
and concentrate instead on decisive evidence in support of
(\ref{calZvsZ}) and (\ref{calLvsL}).
The reason for this is that once these
relations are accepted, they can be used as the
{\it better definition} of the Kontsevich-Hurwitz partition function.
As explained in the Introduction, the definition provided
by such reformulation is more fundamental than the original
ones, as a generating function either of the Hodge integrals or of the
Hurwitz numbers.
Therefore, the detailed proof of (\ref{defvirco}) starting
from the old definitions is, in fact, a problem of a rather limited
interest, concerning the properties of moduli spaces or ramified coverings
more than
the theory of integrability and partition functions.
Most of properties of the Kontsevich-Hurwitz partition function,
including {\bf A-D}, should now be derived directly
from (\ref{defvirco}).
We explained that parts of the relevant statements are already
available in the matrix-model literature, still a complete derivation of
{\bf A-D} provides a set of important open problems.

Of greatest interest is the search for one more property
{\bf F}: an integral representation of ${\cal Z}$ --
an appropriate $u$-dependent deformation of the Kontsevich
matrix integral
$$Z_K = S(\Lambda)
\int dX \exp \left\{-\frac{2g^2}{27}{\rm tr} X^3 +
{\rm tr}\Lambda^2 X\right\}, \ \ \ \ \
T_k = \frac{(2k+1)!!}{2^k}\tau _k =
\frac{(2k+1)!!}{2^k k}{\rm tr}\Lambda^{-2k-1}$$
for which the relation to $H(p)$ should arise
as a character expansion {\it a la} \cite{MMS} and
Virasoro constraints (\ref{defvirco}) should be the Ward
identities, following from the reparametrization of integration
variables {\it a la} \cite{virc}.

Far more straightforward should be three other exercises.

First, one can investigate Virasoro constraints for
the Hurwitz function $\exp \Big(H(p)\Big)$ directly
in terms of the $p$-variables and relate them to our
$\hat{\cal L}_m$ through a change of variables -- in
the spirit of \cite{MMMM}.

Second,
of certain interest are generalizations to multi-Hurwitz free
energies, which enumerate coverings of the Riemann sphere with several
non-simple critical points (some results are already available on the
combinatorial side for the case of two non-simple points). This
research direction should be related to the celebrated conjecture
about the Mumford measure on the universal moduli space made in the
last chapter of \cite{Kn}.

Third, one can find the $\hat U$-operator,
associated with the family $$X=(1-Z)Z^f\ \ \ \ {\rm or}\ \ \ \
\left(1-\frac{1}{f}\right)^f\!\!x=
(1+z)\left(1-\frac{1+z}{f}\right)^f$$
of spectral curves, of which the Lambert curve
$x = (1+z)e^{-z}$ is the $f=\infty$ limit.
This family is important for applications
\cite{ADKMV} and the relevant
AMM-Eynard equations are already suggested in \cite{BM}.
It remains "only" to repeat the consideration of our section
\ref{BMap}.

In fact, as explained in \cite{AMM.IM},
one expects that the continuous Virasoro algebra
is relevant in the vicinity of any quadratic
ramification point on a spectral curve,
only the twisting operator $\hat U$
should be appropriately adjusted, and it is
natural to expect that the quadratic AMM-Eynard equations
on an arbitrary spectral curve describe some set
of Virasoro constraints.
Thus, the same formalism should work in many more
cases.
We understand that the same attitude is expressed in \cite{EO}
(see, for example, the discussion of Mirzakhani
relations in terms of the Virasoro algebra on
the Weyl-Petersson curve
$y = \frac{1}{2\pi}\sin(2\pi\sqrt{x})$
in these wonderful papers).
Somewhat unexpected to us is a mysteriously simple
form of the twisting operator $\hat U$ in the
case of the Lambert curve, it would be interesting
to see if this property persists in other
important examples.

\section*{Acknowledgements}

We are grateful to M.Kazarian for the discussion.
Our work is partly supported by Russian Federal Nuclear
Energy Agency, by the joint grant 06-01-92059-CE,  by NWO project
047.011.2004.026, by INTAS grant 05-1000008-7865, by
ANR-05-BLAN-0029-01 project and by the Russian President's Grant of
Support for the Scientific Schools NSh-3035.2008.2, by RFBR grants
07-02-00878 (A.Mir.) and 07-02-00645 (A.Mor.).

\end{document}